\documentclass[preprint,tightenlines,amsmath,amssymb]{revtex4-2}

\usepackage{graphicx}
\usepackage{dcolumn}
\usepackage{bm}
\usepackage{braket}
\usepackage{booktabs}
\usepackage{siunitx}
\usepackage{multirow}
\usepackage{array}
\usepackage{colortbl}

\newcommand{\yb}{$^{171}\mathrm{Yb}^+$ }
\newcommand{\renyientropy}{\text{R\'enyi entropy}}

\begin{document}

\title{Observation of a topological edge state among localized bulk states in the anisotropic quantum Rabi model}

\author{Sungjoo Lim, Chanyang Im}

\affiliation{SKKU Advanced Institute of Nanotechnology \& Department of Nano Science and Technology (SAINT), Sungkyunkwan University, Suwon 16419, Republic of Korea
}
\author{Christopher G. Yale, Brian K. McFarland, Edward C. Tortorici, Daniel S. Lobser, Melissa C. Revelle, Susan M. Clark}
\affiliation{
Sandia National Laboratories, Albuquerque, New Mexico 87123, USA}
\author{Mahn-Soo Choi}
\affiliation{
Department of Physics, Korea University, Seoul 02841, Republic of Korea
}
\author{Junki Kim}
\email{junki.kim.q@skku.edu}
\affiliation{Department of Quantum Information Engineering, Sungkyunkwan University, Suwon 16419, Republic of Korea\\
}
\affiliation{SKKU Advanced Institute of Nanotechnology \& Department of Nano Science and Technology (SAINT), Sungkyunkwan University, Suwon 16419, Republic of Korea
}
\begin{abstract}
Topological phases are governed by discrete symmetries that protect boundary modes against local perturbations.
When translational periodicity is absent, the bulk states also become localized, so that a topological edge state can no longer be distinguished from them by spatial localization alone.
Here, we investigate the topological edge state (TES) and bulk eigenstates of the anisotropic quantum Rabi model (AQRM) in a trapped-ion quantum simulator.
The AQRM hosts a topological phase in a one-dimensional synthetic lattice, whose translational symmetry is broken by the non-uniform couplings scaling with the site index.
While both the TES and bulk states show localized distributions, we find that the TES exhibits well-defined chirality and near-complete spin--boson separability as signatures of the topological phase, in contrast to the bulk states.
Phase-space tomography further reveals that the bosonic component of the TES is a squeezed vacuum state, with squeezing up to 6.45 dB.
These results identify the TES through its intrinsic topological signatures and establish eigenstate-level characterization as a route to probing topological phenomena.

\end{abstract}

\maketitle

\section{Introduction}
\label{sec:Introduction}

Since the discovery of the quantum Hall effect\cite{klitzing_new_1980}, topological materials have emerged as a central topic in modern physics, exhibiting novel phases of matter such as topological insulators and topological superconductors\cite{hasan_colloquium_2010, moore_birth_2010, sato_topological_2017, qi_topological_nodate, kane_quantum_2005, bernevig_quantum_2006, konig_quantum_2007}.
In these materials, discrete symmetries determine the topological classification and protect boundary states whose properties qualitatively differ from those of the bulk\cite{chiu_classification_2016}.
The robustness of such boundary states against local perturbations underlies a wide range of applications from spintronics\cite{he_topological_2022} to fault-tolerant quantum information processing\cite{nayak_non-abelian_2008}.
Such topological physics has recently been extended beyond crystalline solids to synthetic quantum platforms, including optical lattices\cite{goldman_topological_2016, cooper_topological_2019}, photonic systems\cite{ozawa_topological_2019, lu_topological_2014}, superconducting circuits\cite{malz_topological_2021, google_quantum_ai_and_collaborators_non-abelian_2023}, neutral atoms\cite{de_leseleuc_observation_2019}, and trapped ions\cite{dumitrescu_dynamical_2022, iqbal_non-abelian_2024, katz_floquet_2025}.

Boundary states are localized at edges, in contrast to typical bulk states that extend over the system when translational symmetry is present.
When this symmetry is broken, however, bulk states can also become spatially localized, while topological boundary states can persist as long as the protecting discrete symmetries remain intact, as exemplified by topological Anderson insulators\cite{li_topological_2009, groth_theory_2009, meier_observation_2018} and amorphous topological phases\cite{agarwala_topological_2017, mitchell_amorphous_2018}.
In such systems, distinguishing boundary and bulk eigenstates requires topological signatures beyond spatial localization.

The anisotropic quantum Rabi model (AQRM)\cite{xie_anisotropic_2014, tomka_exceptional_2014, shen_ground_2014, zhang_analytical_2015} is one of the simplest spin--boson models with unequal rotating and counter-rotating couplings, yet it already encodes a broad range of critical, topological, and quantum-information phenomena\cite{liu_universal_2017, shen_quantum_2017, wang_quantum_2018, zhu_criticality_2023, ying_from_2022, ying_hidden_2022, ying_nodes_2023}.
Interactions in the AQRM form semi-infinite one-dimensional synthetic lattices of spin--Fock states that are expected to host topological phases\cite{deng_observing_2022,saugmann_fock-state-lattice_2023, lee_phase-space_2026}, closely resembling the Su--Schrieffer--Heeger (SSH) model\cite{su_solitons_1979, heeger_solitons_1988}.
A key distinction is that the nearest-neighbor couplings scale as $\sqrt{n}$ with the bosonic Fock index, so translational symmetry is intrinsically broken, unlike in the SSH model.
Observing the edge state in the AQRM therefore requires more than spatial localization. It is instead identified through direct topological properties that differ from those of the bulk states.

Trapped ions are a leading platform for quantum information processing and quantum simulation, offering high-fidelity control\cite{ballance_high-fidelity_2016,wang_high-fidelity_2020} and long coherence times\cite{wang_single-qubit_2017,wang_single_2021}.
Their internal spin and external motional degrees of freedom, or phonons, provide a natural setting for spin--boson physics\cite{leibfried_quantum_2003, wineland_quantum_2003}.
Laser-induced spin--phonon couplings enable tunable interactions, while established tomography of both spin and phonon modes allows thorough characterization of the simulated states\cite{leibfried_quantum_2003,fluhmann_direct_2020}.
These capabilities have been used extensively to simulate spin--boson models, including the quantum Rabi model\cite{pedernales_quantum_2015, lv_quantum_2018, cai_observation_2021, zhao_experimental_2025}, Hubbard-like models\cite{dutta_nonequilibrium_2012, mei_experimental_2022}, and vibronic models relevant to quantum chemistry\cite{sun_quantum_2025,so_trapped-ion_2024}.

In this work, we experimentally study the AQRM in a trapped-ion quantum simulator and characterize the topological properties of its eigenstates.
We adiabatically prepare the topological edge state (TES) as well as bulk eigenstates and compare their properties.
The TES exhibits clear topological signatures, including well-defined chirality and near-complete spin--phonon separability, even though spatial localization alone does not distinguish it from the bulk states.
Furthermore, phase-space measurements reveal that the bosonic component of the TES is a squeezed vacuum, showing highly non-classical behavior.

This article is organized as follows. In Sec.~\ref{sec:AQRM}, we introduce the AQRM and analyze its topological structure, deriving the TES and bulk eigenstates and identifying the signatures that distinguish them. 
In Sec.~\ref{sec:experiment}, we describe the trapped-ion implementation of the AQRM and the adiabatic protocol used to prepare each eigenstate.
Sections~\ref{sec:population}, \ref{sec:separability}, and \ref{sec:phasespace} present the experimental characterization of the prepared states through three complementary measurements: the joint spin--Fock populations and chirality (Sec.~\ref{sec:population}), the spin--boson separability quantified by the second-order Rényi entropy (Sec.~\ref{sec:separability}), and the bosonic phase-space distribution probed via the characteristic function (Sec.~\ref{sec:phasespace}).
Across all three measurements, the TES exhibits signatures distinct from those of the bulk eigenstates.

\section{The Anisotropic Rabi Model}
\label{sec:AQRM}

\begin{figure}
    \centering
    \includegraphics[width=1\linewidth]{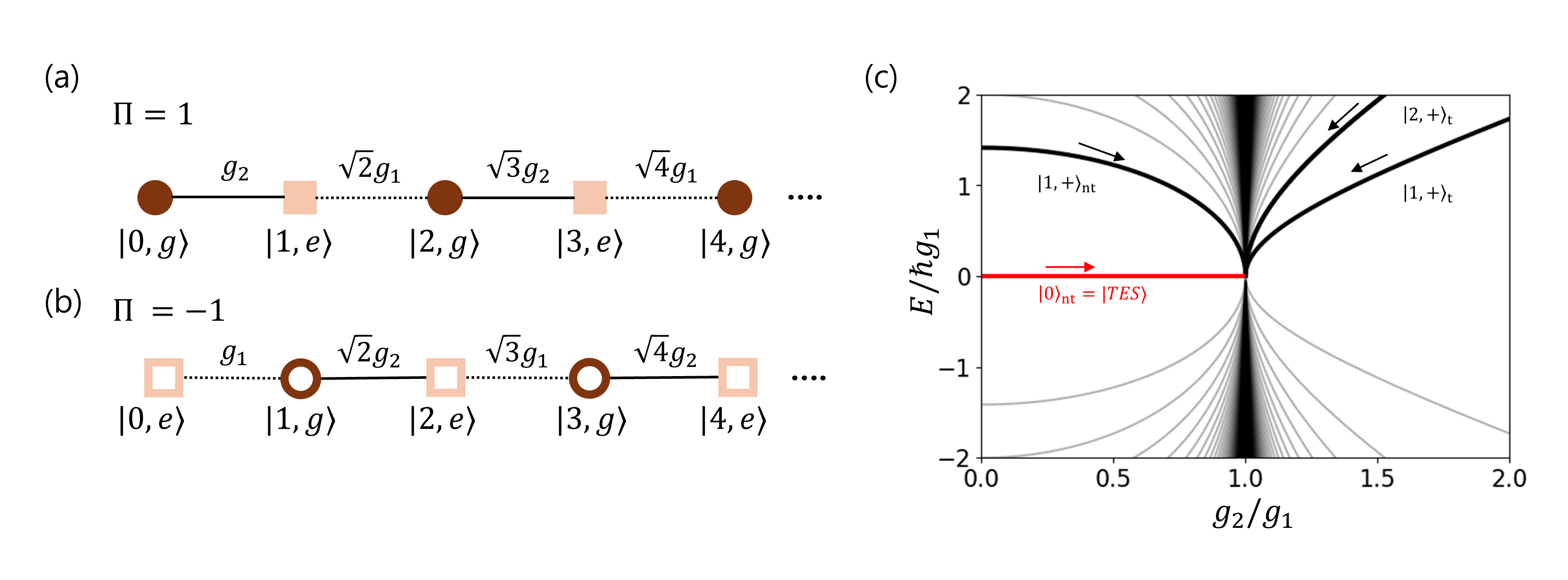}
    \caption{Parity chains and subspace energy spectrum of the AQRM. 
    (a), (b)  Parity-resolved mapping of the AQRM Hilbert space onto two decoupled one-dimensional subspaces.
    Each site corresponds to a spin--Fock state $\lvert n,g(e)\rangle$, with nearest-neighbor couplings $\propto \sqrt{n}\,g_{1,2}$ arising from the Jaynes--Cummings (JC) and anti-Jaynes--Cummings (AJC) interactions.
    The resulting bipartite lattice is analogous to that of the SSH model but with $\sqrt{n}$-dependent couplings breaking translational symmetry.
    (c) Energy spectrum as a function of the coupling ratio $g_2/g_1$ in the even-parity subspace, with energies in units of $\hbar g_1$.
    In the regime $g_2<g_1$, the TES ($|0\rangle_{\mathrm{nt}}$, red) appears at zero energy inside the bulk gap, serving as the main spectral signature of the topological phase.
    For $g_2>g_1$, the zero-energy state disappears and only bulk states (black) remain, indicating a transition to a topologically trivial phase.
    }
    \label{fig:chain_spectrum}
\end{figure}

The AQRM is a generalized form of the quantum Rabi model, in which the Jaynes--Cummings (JC) coupling strength $g_1$ and the anti-Jaynes--Cummings (AJC) coupling strength $g_2$ are independently controlled.
We consider the resonant limit of the AQRM, in which both the spin~$(\omega_{0})$ and bosonic~$(\omega_x)$ frequencies are set to zero.
The system Hamiltonian is given by
\begin{equation}
    \hat{H}
    = \hbar g_1(\hat{\sigma}_+ \hat{a} + \hat{\sigma}_- \hat{a}^\dagger)
    + \hbar g_2(\hat{\sigma}_+ \hat{a}^\dagger + \hat{\sigma}_- \hat{a}),
     \label{eq:AQRMHamiltonian}
\end{equation}
where $\hat{a}$ ($\hat{a}^\dagger$) is the annihilation (creation) operator of the bosonic mode and $\hat{\sigma}_\pm$ are the spin raising and lowering operators, with $\hat{\sigma}_z = |e\rangle\langle e| - |g\rangle\langle g|$.

The AQRM conserves a $\mathbb{Z}_2$ symmetry generated by
$\hat{\Pi} = -\exp(i\pi \hat{a}^\dagger \hat{a}) \hat{\sigma}_z$,
which allows the full Hilbert space to be decomposed into decoupled even and odd parity subspaces~\cite{casanova_deep_2010}.
By interpreting each spin--Fock state as a lattice site, each parity subspace can be mapped onto an effective semi-infinite one-dimensional lattice (see Fig.~\ref{fig:chain_spectrum}(a,b)).
Alternating JC and AJC couplings then generate a bipartite chain of the SSH type~\cite{su_solitons_1979, batra_physics_2020}.
Unlike the uniform SSH chain, the hopping amplitudes here grow as $\sqrt{n}$ with the bosonic Fock index, so the lattice is intrinsically inhomogeneous.

The AQRM Hamiltonian also exhibits chiral symmetry, as expressed by the anti-commutation relation $\{\hat{\Xi}, \hat{H}\} = 0$ with the chiral operator $\hat{\Xi} = -\hat{\sigma}_z$.
Since $\hat{\Xi}$ commutes with $\hat{\Pi}$ and thus preserves the parity, the chiral symmetry holds within each parity subspace as well.
This symmetry enforces spectral symmetry about zero energy and permits the existence of a topologically protected zero-energy state.
The corresponding topological invariant is the edge index of each parity chain, which counts zero-energy states weighted by their chirality.
It is given by the Fredholm index of the chiral block of the chain Hamiltonian and remains well defined in the absence of translational symmetry~\cite{thiang_topological_2023,Essin11a}.
The index in the even-parity sector is $1$ for $g_1>g_2$ and $0$ for $g_2>g_1$, reflecting the presence or absence of the single zero-energy state (Appendix~\ref{Appendix:AQRMTheory}).
The assignment is reversed in the odd-parity sector, such that the topological classification is defined within each parity sector.
In this work, we focus on the even-parity subspace, and Fig.~\ref{fig:chain_spectrum}(c) shows the corresponding AQRM energy spectrum.

For $g_1 > g_2$, corresponding to the topologically non-trivial phase, the TES is denoted by $|0\rangle_{\mathrm{nt}}$ and the bulk eigenstates by $|n_b,\pm\rangle_{\mathrm{nt}}$, where $n_b=1,2,\ldots$ is the Fock index of a Bogoliubov-transformed bosonic mode (Appendix~\ref{Appendix:AQRMTheory}) and $\pm$ labels the positive- and negative-energy branches. Their eigenenergies are given by
\begin{equation}
    E_{n_b} =
    \begin{cases}
        0, & n_b = 0 \\
        \pm E_{\mathrm{gap}}\sqrt{2n_b}, & n_b = 1,2,\ldots
    \end{cases}
\end{equation}
In this phase, a zero-energy eigenstate emerges at $n_b=0$, corresponding to the TES of a one-dimensional chiral system.
In contrast, in the regime $g_2 > g_1$, which corresponds to the trivial phase, the eigenenergies are given by
\begin{equation}
    E_{n_b} = \pm E_{\mathrm{gap}}\sqrt{2n_b-1}, \qquad n_b = 1,2,3,\ldots
\end{equation}
with eigenstates $|n_b,\pm\rangle_{\mathrm{t}}$, where no zero-energy eigenstate exists and the spectrum is separated by a finite bulk energy gap (see Appendix~\ref{Appendix:AQRMTheory} for the detailed derivation).
As a result, the energy spectrum exhibits a topological-insulator-like behavior, with trivial and non-trivial phases determined by the coupling ratio $g_2/g_1$.
We define the relative coupling ratio $g_{\mathrm{r}}$ as $g_2/g_1$ in the non-trivial phase and $g_1/g_2$ in the trivial phase.
In both phases, the energy gap follows $E_{\mathrm{gap}} = \hbar\sqrt{|g_1^2 - g_2^2|}\propto \hbar\sqrt{1 - g_{\mathrm{r}}^2}$, showing the gap closing near $g_{\mathrm{r}} = 1$.

The TES is simultaneously an eigenstate of both the Hamiltonian and the chiral operator and belongs to the right-handed (or positive) chirality sector ($\Xi=+1$) in the even-parity subspace.
Owing to this single-valued chirality, the TES occupies only the even-Fock states, and hence it remains fully separable between the spin and boson degrees of freedom.
By contrast, bulk eigenstates are not chiral eigenstates and occupy equal populations for both chiral sectors, exhibiting strong spin--boson entanglement.

The bosonic part of the TES is a squeezed vacuum state $\hat{S}(r)\lvert 0 \rangle$, a distinctly non-classical state.
The squeezing parameter $r$ is fully specified by the coupling ratio, satisfying $\tanh r = g_{\mathrm{r}}$.
The population distribution of the TES shows strong localization near the origin, faster than a purely exponential envelope, originating from the non-uniform, $\sqrt{n}$-coupling strengths in the AQRM (see Appendix~\ref{Appendix:AQRMTheory} for details).

\section{Experimental implementation}
\label{sec:experiment}

In the experiment, quantum simulation of the AQRM is performed using \yb ions trapped in a radio-frequency~(RF) surface electrode trap~\cite{revelle_phoenix_2020} in the Quantum Scientific Computing Open User Testbed (QSCOUT) at Sandia National Laboratories~\cite{clark_engineering_2021}.
The spin degree of freedom in the AQRM is encoded in the two hyperfine states of the electronic ground state manifold $^{2}S_{1/2}$ of an \yb ion, denoted as $|g\rangle \equiv |F=0, m_F=0\rangle$ and $|e\rangle \equiv |F=1, m_F=0\rangle$, while the bosonic counterpart is encoded in the ion motional states, which can be well described as a harmonic oscillator.
We trap a chain of two \yb ions, designating one as the target ion for simulation and the other as a spectator ion for bosonic-state characterization.
We use the radial tilt mode as the bosonic mode because it experiences less heating than the center-of-mass mode.
Coherent control of the hyperfine spin and the spin--motion interaction is achieved using two counter-propagating Raman lasers.
The JC and AJC interactions in the AQRM Hamiltonian are mapped onto resonant red- and blue-sideband transitions of the ion's spin--motion interaction~\cite{leibfried_quantum_2003, haljan_spin-dependent_2005, haljan_entanglement_2005}.
The corresponding coupling strengths $g_1$ and $g_2$ are proportional to the driving-field amplitudes of the red and blue sidebands, respectively, which can be controlled and modulated by acousto-optic modulators driven by a custom RF-system-on-chip device, \textsc{Octet}~\cite{clark_engineering_2021, lobser_jaqalpaw_2023}.

The experimental sequence consists of adiabatic state preparation followed by characterization of the prepared states.
At each extreme of relative coupling strength $g_{\mathrm{r}}$ (\textit{i.e.}, $g_2/g_1 = 0$ or $g_1/g_2 = 0$), the energy eigenstates of the AQRM can be expressed as $\lvert 0, g \rangle$ for the TES and as equal superpositions of neighboring states for the bulk states.
Adopting these states as the initial states of the state-preparation stage, we adiabatically adjust $g_{\mathrm{r}}$ up to the desired value (red and black arrows in Fig.~\ref{fig:chain_spectrum}(c)) to prepare the target TES or bulk eigenstates.
Note that as $g_{\mathrm{r}}$ approaches unity, the energy gap between neighboring states closes, making adiabatic state transfer increasingly difficult due to the breakdown of the adiabatic approximation~\cite{born_beweis_1928}.

After adiabatic preparation of the TES and bulk states, we characterize the resulting states by separately addressing their spin and bosonic subsystems.
In the spin subsystem, we perform projective $Z$ measurements, with additional basis rotations to access $X$- and $Y$-basis measurements.
For the bosonic subsystem, we extract phonon-number statistics from blue-sideband Rabi flopping of the spectator ion~\cite{wineland_quantum_2003, lv_quantum_2018}.
Combined with spin-$Z$ readout, this yields the joint spin--Fock-basis population distribution.
Additionally, for the TES, we probe the bosonic phase-space distribution by measuring the characteristic function via spin-dependent kicks~\cite{fluhmann_direct_2020}.

Further details about the experimental methods are provided in Appendix~\ref{Appendix:ExperimentalDetails}.

\section{Spin--Fock populations and chirality}
\label{sec:population}

\begin{figure}
    \centering
    \includegraphics[width=1\linewidth]{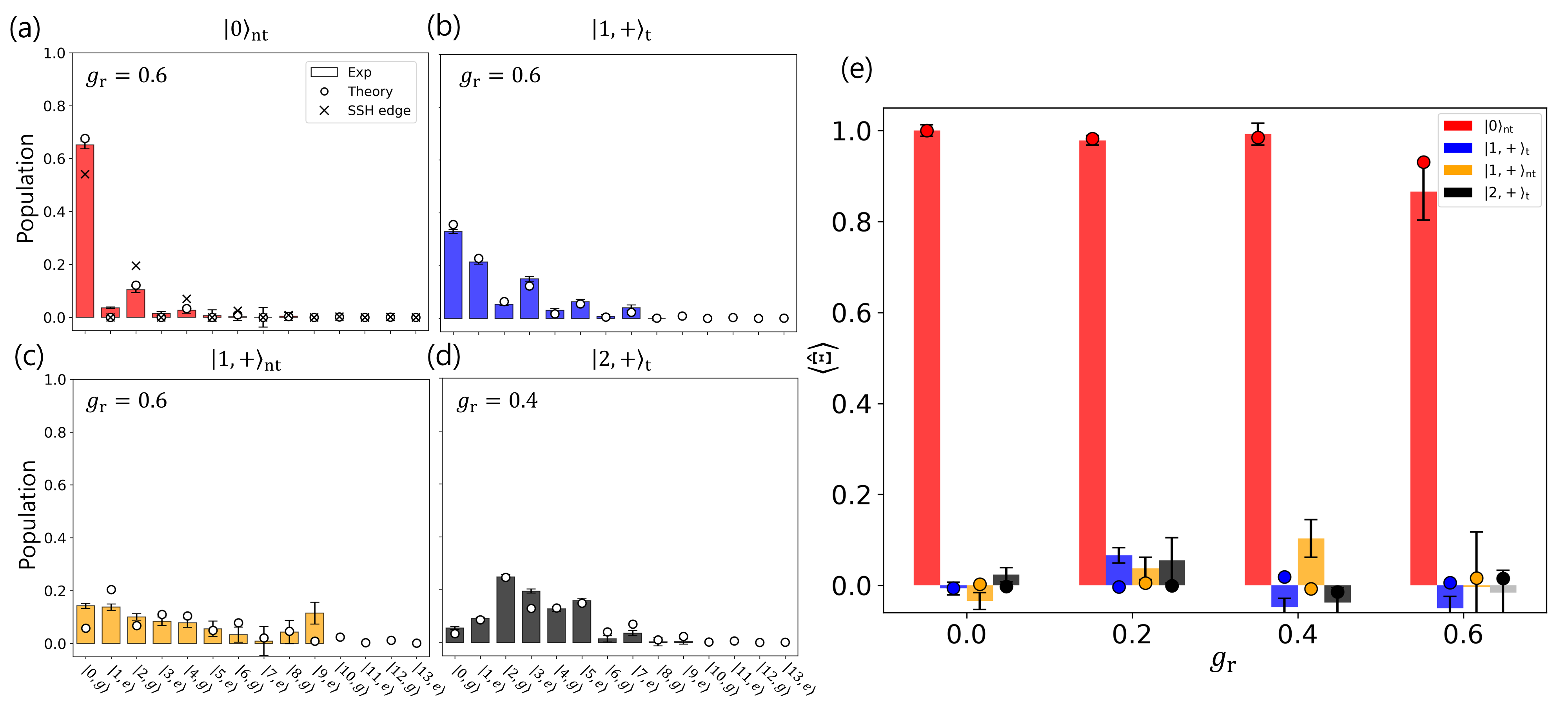}
    \caption{Measured joint spin--Fock populations and chirality of the TES and three representative bulk states $\lvert 1, +\rangle_{\mathrm{t}}$, $\lvert 1, +\rangle_{\mathrm{nt}}$, and $\lvert 2, +\rangle_{\mathrm{t}}$.
    (a--d) Joint spin--Fock populations of the TES $\lvert 0\rangle_{\mathrm{nt}}$ at $g_{\mathrm{r}} = 0.6$ (a), $\ket{1,+}_{\mathrm{t}}$ at $g_{\mathrm{r}} = 0.6$ (b), $\ket{1,+}_{\mathrm{nt}}$ at $g_{\mathrm{r}} = 0.6$ (c), and $\ket{2,+}_{\mathrm{t}}$ at $g_{\mathrm{r}} = 0.4$ (d).
    The $\ket{2,+}_{\mathrm{t}}$ state is shown at $g_{\mathrm{r}} = 0.4$, as its simulated preparation fidelity at $g_{\mathrm{r}} = 0.6$ is expected to fall below 0.5 (see Appendix~\ref{appendix:LindbladEquation}).
    Bars and circles denote experimental data and theoretical predictions.
    In (a), crosses indicate the SSH edge-state profile for comparison.
    The TES shows strong localization near the origin, faster than the exponential decay of the SSH case, and exclusively occupies the even-Fock states, whereas the bulk states exhibit moderate localization, with the mean phonon number increasing with the state excitation, and comparable populations in the even- and odd-Fock states.    
    (e) Expectation value of the chiral operator $\langle\hat{\Xi}\rangle$ as a function of relative coupling strength $g_{\mathrm{r}}$.
    Bars with error bars are experimental results and dots are simulation results.
    The faded bar indicates the $\ket{2,+}_{\mathrm{t}}$ data at $g_{\mathrm{r}} = 0.6$ with limited preparation fidelity.
    The $\langle\hat{\Xi}\rangle$ of the TES remains close to +1, indicating right-handed chirality, whereas the bulk states lie near zero, consistent with mixed chirality.
    Additional data for other $g_{\mathrm{r}}$ values are provided in Appendix~\ref{Appendix:SpinBosonPopulation}.
    }
    \label{fig:Figure2}
\end{figure}

We characterize the prepared TES ($\lvert 0\rangle_{\mathrm{nt}}$) and three representative bulk states, $\lvert 1, +\rangle_{\mathrm{t}}$, $\lvert 1, +\rangle_{\mathrm{nt}}$, and $\lvert 2, +\rangle_{\mathrm{t}}$, through their joint spin--Fock populations and chirality (Fig.~\ref{fig:Figure2}).
Figure~\ref{fig:Figure2}(a--d) shows the measured joint spin--Fock populations of those states within the even-parity subspace.
While imperfect ground-state cooling leaves a residual odd-parity population, the two parity subspaces do not couple to each other, and joint spin--boson measurements can be used to extract the even-parity population distribution (see Appendix~\ref{Appendix:SpinBosonPopulation} for details).
The TES distribution is sharply localized near the edge, closely resembling the exponentially decaying profile of an SSH edge state (cross symbols in Fig.~\ref{fig:Figure2}(a)).
However, unlike a purely exponential decay, the TES population exhibits an additional $1/\sqrt{n}$ suppression with increasing site index $n$, reflecting the non-uniform couplings of the AQRM.

The bulk state population distributions are also affected by the non-uniform coupling of the AQRM and therefore differ from those of SSH bulk states.
Unlike the globally delocalized SSH bulk distributions, all three bulk states in Fig.~\ref{fig:Figure2}(b-d) remain moderately localized.
In particular, the $\lvert 1, +\rangle_{\mathrm{t}}$ state, the lowest bulk state of the trivial phase, also shows a population distribution concentrated near the edge, only slightly broader than that of the TES.
With increasing bulk-state excitation, the mean phonon number shifts to higher values, while increasing $g_{\mathrm{r}}$ broadens the corresponding population distributions (see Appendix~\ref{Appendix:SpinBosonPopulation} for additional data).

From the extracted joint spin--Fock distributions, we compute the expectation value of the chiral operator $\langle\hat{\Xi}\rangle$ for each state.
As shown in Fig.~\ref{fig:Figure2}(e), the TES remains close to $+1$ ($\langle\hat{\Xi}\rangle=1.00\pm0.01$, $0.98\pm0.01$, $0.99\pm0.02$, and $0.87\pm0.06$ at $g_{\mathrm{r}}=0.0$, $0.2$, $0.4$, and $0.6$, respectively), indicating well-defined right-handed chirality of the TES.
On the other hand, all three bulk states remain clustered near zero, with maximum deviations of $0.10\pm0.04$, reflecting nearly balanced sublattice occupations.
These differences in chirality between the TES and bulk states provide a clear signature of the topological properties of the AQRM.
As $g_{\mathrm{r}}$ increases, we observe a slight reduction of the TES chirality, which is attributed to proximity to the energy-gap closing and to phonon dephasing during the longer preparation sequence, in agreement with our numerical simulations using experimental parameters (see Appendix~\ref{appendix:LindbladEquation}).
Nonetheless, the TES remains well distinguishable from the bulk states due to its chirality.

\section{Spin--boson separability} \label{sec:separability}

Another notable property of the TES is the separability between the spin and boson subsystems.
Although the AQRM Hamiltonian contains only interaction terms between the spin and bosonic subsystems, the TES remains fully separable, in contrast to the bulk states, which generally exhibit strong spin--boson entanglement.

We quantify bipartite correlations using the second-order \renyientropy~of the spin subsystem~\cite{islam_measuring_2015}, $S_{2,\mathrm{spin}} = -\ln \mathrm{Tr}[\rho_{\mathrm{spin}}^2]$.
Zero entropy indicates a pure and separable spin subsystem, while a higher entropy indicates reduced spin purity arising from spin--boson correlations, with $\ln 2$ corresponding to a maximally correlated spin--boson state.
In the experiments, imperfect ground-state cooling leaves a residual odd-parity population, so measured Pauli X- and Y-basis spin expectation values include unwanted odd-parity contributions, whereas the Z-basis spin expectation value can be extracted from joint spin--phonon populations.
Given this limitation, we estimate the feasible upper and lower bounds on $S_{2,\mathrm{spin}}$ for the TES and bulk eigenstates as
\begin{equation}
    -\ln \left[ \frac{1}{2}\left(1 + \langle \sigma_x \rangle_{\mathrm{even, max}}^2 + \langle \sigma_y \rangle_{\mathrm{even, max}}^2 + \langle  \sigma_z \rangle_{\mathrm{even}}^2 \right) \right] \leq S_{2,\mathrm{spin}} \leq -\ln \left[ \frac{1}{2}\left(1 + \langle \sigma_z \rangle_{\mathrm{even}}^2 \right) \right],
\end{equation}
\noindent where $\langle \sigma_i \rangle_{\mathrm{even}}$ denotes the spin $i$-basis expectation value in the even-parity subspace, and $\langle \sigma_i \rangle_{\mathrm{even, max}}$ denotes the maximum expectation value inferred from the projective Pauli $i$-basis measurements (see Appendix~\ref{appendix:entropybound} for details).

\begin{figure}
    \centering
    \includegraphics[width=0.8\linewidth]{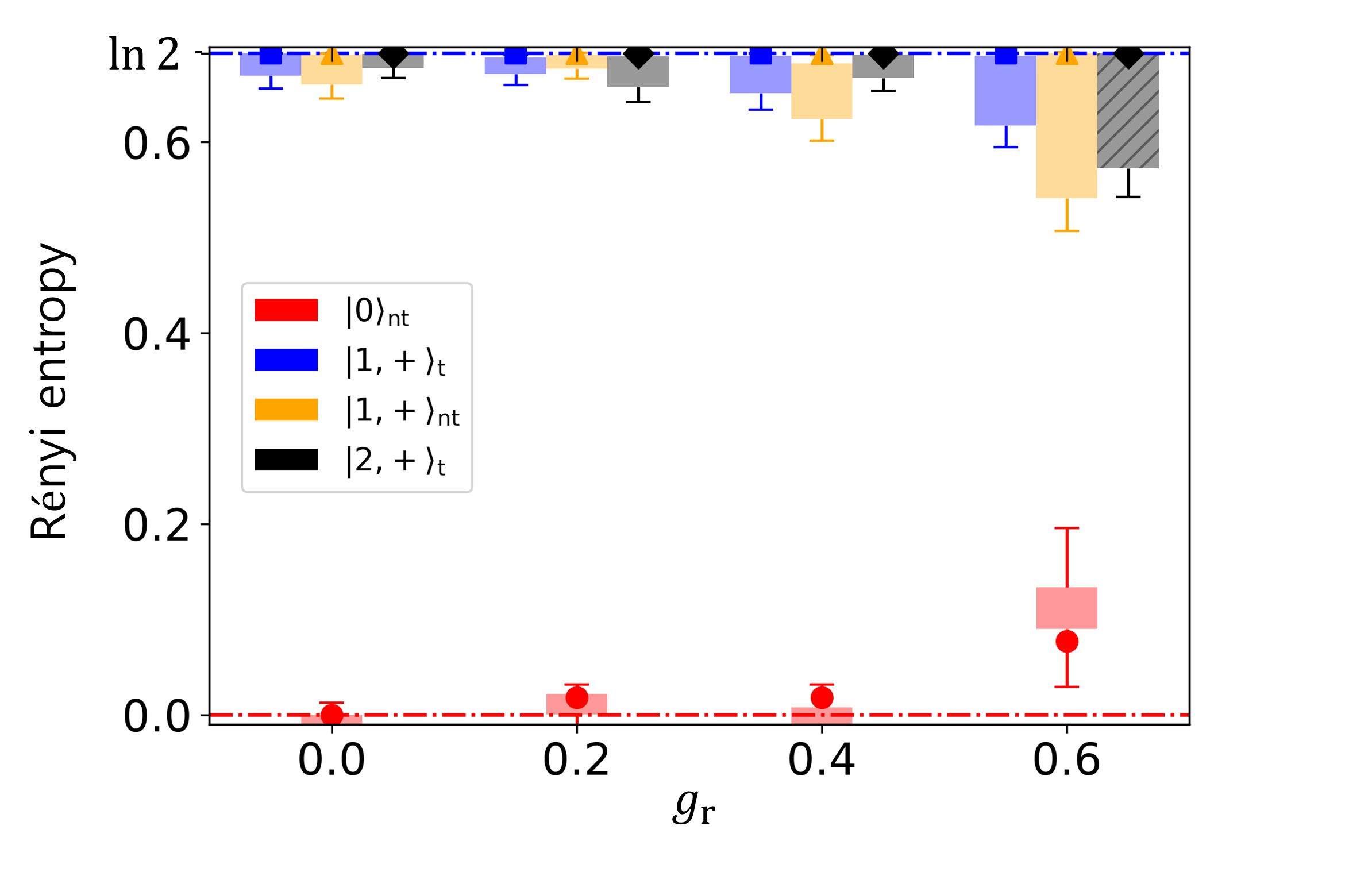}
    \caption{Spin--boson correlations in the TES and bulk states.
    The shaded bars indicate the experimentally estimated bounds on the second-order Rényi entropy, with error bars indicating uncertainties arising from binomial sampling and fitting. 
    The symbols show the corresponding numerical simulation results.
    The hatched bar indicates the $\ket{2,+}_{\mathrm{t}}$ data at $g_{\mathrm{r}} = 0.6$ with limited preparation fidelity (see Appendix~\ref{appendix:LindbladEquation}).
    The TES remains near zero, indicating a nearly separable spin subsystem, whereas the bulk states lie close to $\ln 2$, reflecting strong spin--boson correlations.
    }
    \label{fig:Figure3}
\end{figure}

Figure~\ref{fig:Figure3} presents the experimentally determined lower and upper bounds of $S_{2,\mathrm{spin}}$ for each state.
For the TES, the upper bound remains negligibly small, taking values of $0.00\pm0.01$, $0.02\pm0.01$, and $0.01\pm0.02$ for $g_{\mathrm{r}} = 0.0$, $0.2$, and $0.4$, respectively, indicating that the spin state of the TES is largely decoupled from its bosonic counterpart.
At $g_{\mathrm{r}} = 0.6$, the upper bound increases to $0.13\pm0.06$, primarily due to imperfect state preparation arising from non-adiabaticity and phonon dephasing.
We calculated the lower bounds for the bulk states, finding that $S_{2,\mathrm{spin}} \ge 0.62\pm0.02$ up to $g_{\mathrm{r}} = 0.4$, with the bound decreasing to $0.54\pm0.03$ at $g_{\mathrm{r}}=0.6$.
These results indicate that the spin--boson system in the bulk states is highly correlated, exhibiting near-maximal correlations.
In summary, the TES exhibits a high degree of spin--boson separability, making it distinguishable from the bulk eigenstates, which exhibit strong spin--boson correlations.

\section{Bosonic Phase-Space Distribution}
\label{sec:phasespace}

The bosonic subsystem of the TES forms a squeezed vacuum state, with the squeezing parameter $r$ determined by the relative coupling strength $g_{\mathrm{r}}$.
As $g_{\mathrm{r}}$ increases, the squeezing parameter also increases, leading to squeezing along the X quadrature and corresponding anti-squeezing along the P quadrature, revealing its non-classical phase-space structure.
As the bosonic subsystem of the TES is fully separable from its spin counterpart, direct phase-space tomography of the bosonic subsystem is sufficient to characterize the TES.

\begin{figure}
    \centering
    \includegraphics[width=1.0\linewidth]{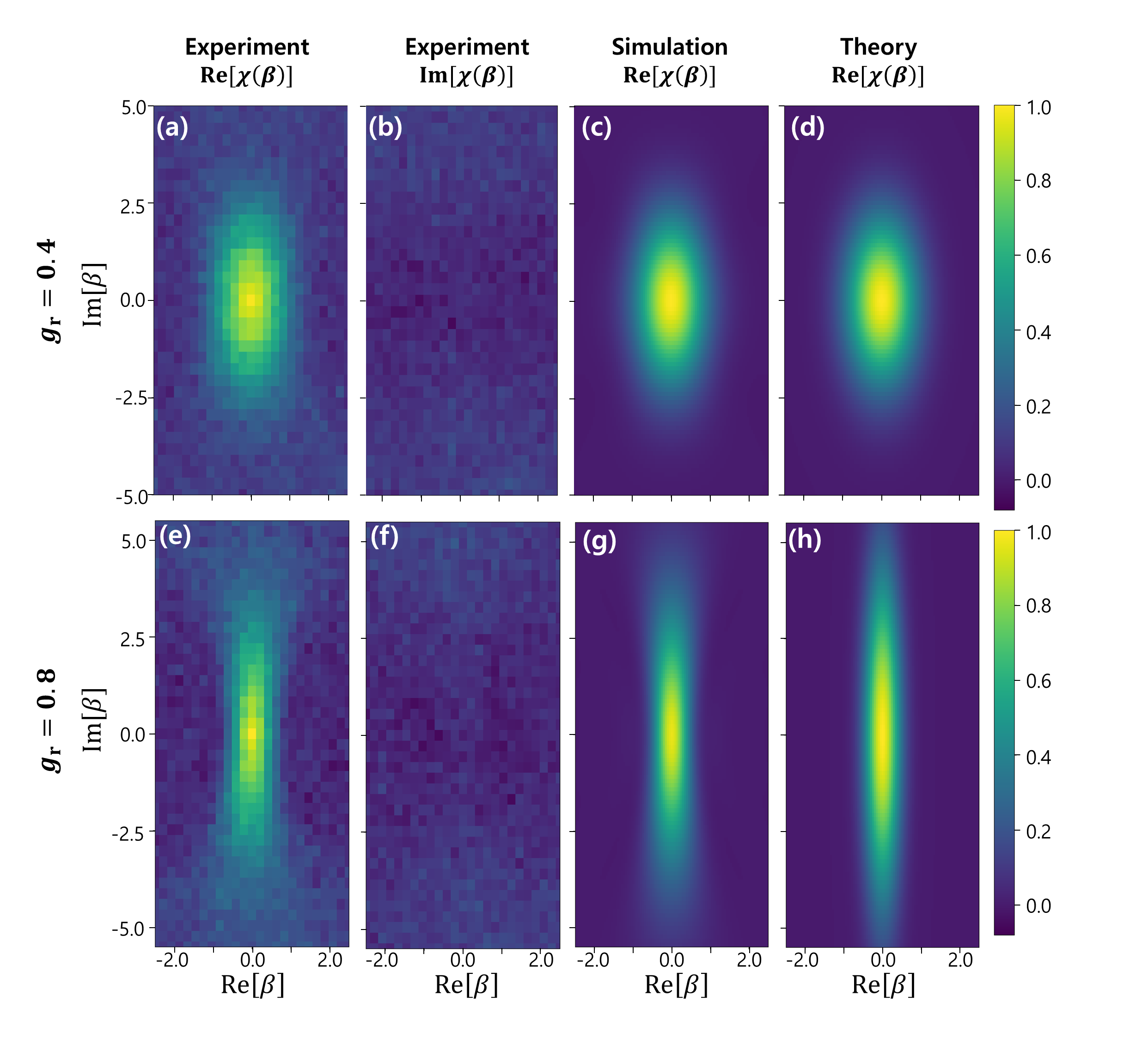}
    \caption{The characteristic function $\chi(\beta)$ of the bosonic subpart of the TES at $g_{\mathrm{r}}=0.4$ (a--d) and $0.8$ (e--h). (a,e) and (b,f) show the experimentally reconstructed $\mathrm{Re}[\chi(\beta)]$ and $\mathrm{Im}[\chi(\beta)]$, respectively. (c,g) show the corresponding numerical simulations of $\mathrm{Re}[\chi(\beta)]$, while (d,h) show the theoretical results.
    The anisotropic shape of $\mathrm{Re}[\chi(\beta)]$, compressed along $\mathrm{Re}[\beta]$ and expanded along $\mathrm{Im}[\beta]$, reveals anti-squeezing in the $P$ quadrature and squeezing in the $X$ quadrature.
    The measured $\mathrm{Im}[\chi(\beta)]$ remains close to zero for all $\beta$.
    }
    \label{fig:figure4}
\end{figure}

Figure~\ref{fig:figure4} shows the two-dimensional characteristic function $\chi(\beta)$ of the bosonic subsystem of the TES at $g_{\mathrm{r}}=0.4$ and $0.8$, compared with the simulation and theory.
The measured real part $\mathrm{Re}[\chi(\beta)]$ reveals an anisotropic Gaussian shape, compressed along the real axis and expanded along the imaginary axis, indicating anti-squeezing in the momentum quadrature and squeezing in the position quadrature, respectively.
The measured imaginary part $\mathrm{Im}[\chi(\beta)]$ remains close to zero for all $\beta$, as expected for an ideal squeezed vacuum state.

Using the measured characteristic function, we reconstructed the bosonic density matrix via maximum-likelihood estimation and evaluated its fidelity with the ideal TES bosonic state $\hat{S}(\tanh^{-1}g_{\mathrm{r}})\lvert0\rangle$.
At $g_{\mathrm{r}}=0.4$, the reconstructed state reaches a fidelity of $F=0.94$ with the ideal squeezed vacuum, confirming that the TES bosonic state is well prepared.
At $g_{\mathrm{r}}=0.8$, by contrast, the fidelity falls to $F=0.73$, as the phonon number increases and the state becomes more susceptible to phonon dephasing.
The $\mathrm{Re}[\chi(\beta)]$ distribution develops a bow-tie-like distortion in the $\beta$ plane, showing angular broadening at large $|\beta|$ due to phonon dephasing, which is well captured in the numerical simulations as well (Fig.~\ref{fig:figure4}(g)).
Nevertheless, the characteristic function still exhibits a distribution broader than that of the vacuum along the imaginary axis, indicating a non-classical quadrature distribution narrower than the standard quantum limit.

\begin{figure}
    \centering
    \includegraphics[width=1\linewidth]{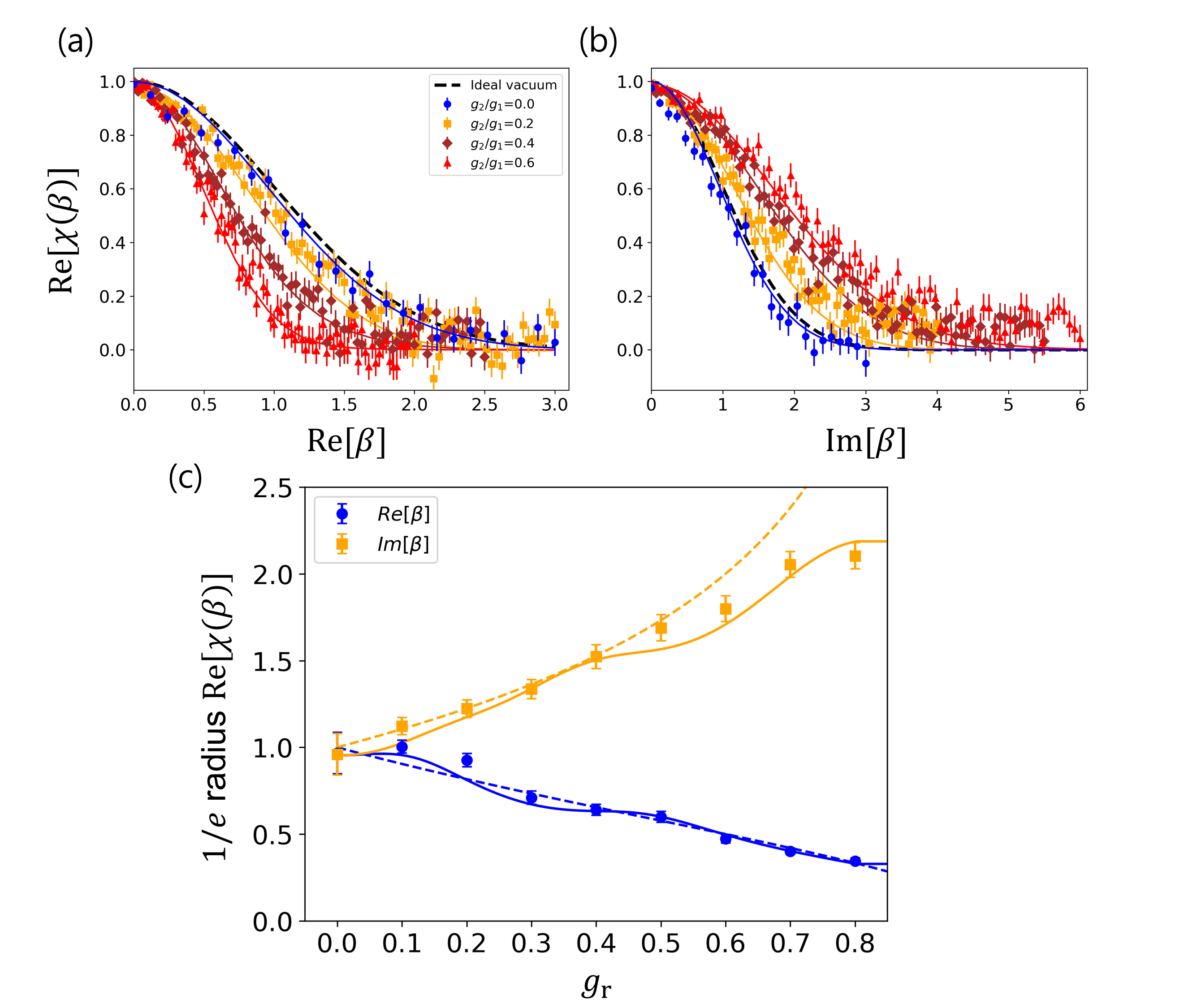}
    \caption{
    (a,b) Cross-sectional profiles of $\mathrm{Re}[\chi(\beta)]$ along the real (a) and imaginary (b) axes for various $g_{\mathrm{r}}$.
    Symbols represent the experimental measurements, with error bars in (a) indicating uncertainties arising from shot noise.
    The solid colored curves show the corresponding numerical simulations.
    The distribution shows compression (expansion) along the real (imaginary) axis compared to the reference vacuum-state profile (black dashed line).
    (c) Extracted $1/e$ radius of $\mathrm{Re}[\chi(\beta)]$ along the real and imaginary axes, normalized to that of the vacuum state, where the error bars reflect the step size in scanning $\beta$.
    As $g_{\mathrm{r}}$ increases, the TES exhibits stronger squeezing along the position quadrature, and the largest squeezing of 6.45 dB is observed at $g_{\mathrm{r}}=0.8$.
    In (c), the dashed lines indicate the theoretical predictions for the ideal TES, while the solid lines show the numerical simulations with measured experimental parameters.
    }
    \label{fig:Figure5}
\end{figure}

To evaluate the bosonic squeezing as a function of the relative coupling strength, we performed one-dimensional scans of $\mathrm{Re}[\chi(\beta)]$ along the $\mathrm{Re}[\beta]$ and $\mathrm{Im}[\beta]$ axes for various $g_{\mathrm{r}}$, as shown in Fig.~\ref{fig:Figure5}(a,b).
For all measured coupling ratios except $g_{\mathrm{r}}=0$, the $\mathrm{Re}[\chi(\beta)]$ profiles are narrower along $\mathrm{Re}[\beta]$ and broader along $\mathrm{Im}[\beta]$ than the corresponding vacuum-state profiles.
Increasing $g_{\mathrm{r}}$ systematically enhances both narrowing along $\mathrm{Re}[\beta]$ and widening along $\mathrm{Im}[\beta]$, consistent with progressively stronger squeezing of the bosonic mode.

From the one-dimensional scans, we extract the $1/e$ radius of $\mathrm{Re}[\chi(\beta)]$ along the $\mathrm{Re}[\beta]$ and $\mathrm{Im}[\beta]$ axes, normalized to that of the vacuum state, and plot it as a function of $g_{\mathrm{r}}$, as shown in Fig.~\ref{fig:Figure5}(c).
For $g_{\mathrm{r}}\le 0.4$, the extracted radius closely follows the theoretical prediction of $\exp(\pm r)$, where the squeezing parameter is given by $r = \tanh^{-1}g_{\mathrm{r}}$.
As $g_{\mathrm{r}}$ increases further, the larger phonon number renders the state more susceptible to phonon dephasing, reducing the measured radius along the imaginary axis relative to the theoretical expectation.
At the largest $g_{\mathrm{r}} = 0.8$, the radius along the imaginary axis is measured to be $2.10\pm0.08$, indicating a maximum position-quadrature squeezing of 6.45 dB, equivalent to a squeezing parameter $r=0.743$.
Across all $g_{\mathrm{r}}$ values, the experimental data agree well with the numerical simulations with measured phonon coherence time and residual phonon number after state preparation.

\section{Conclusion and Outlook}
\label{sec:Conclusion}

In this work, we experimentally prepared and characterized the TES and representative bulk eigenstates of the resonant AQRM using a trapped-ion quantum simulator. 
The spin--Fock states form a synthetic lattice that preserves chiral symmetry within each conserved parity sector, despite intrinsically non-uniform couplings and the absence of translational symmetry.
Adiabatic preparation enabled us to characterize the TES through a direct comparison with bulk eigenstates in the even-parity sector.
The TES exhibits near-definite chirality and a nearly pure reduced spin state, consistent with approximate spin--boson separability.
In contrast, the bulk states show nearly balanced populations of the two chiral sectors and substantially reduced spin purity.
The bosonic component of the TES exhibits a squeezed-vacuum structure, as revealed by phase-space tomography.
Together, these observations distinguish the TES from localized bulk states through its internal structure rather than localization alone.

The present work could be extended by incorporating additional spins or motional modes to realize synthetic lattices with tunable dimensionality and connectivity \cite{wang_realizing_2024, zhang_synthetic_2025, saugmann_fock-state-lattice_2023}.
Engineered dissipation could also provide a route to stabilizing selected topological states, provided that the dissipative dynamics are designed to preserve the relevant state structure \cite{kienzler_quantum_2015, so_trapped-ion_2024}.
These extensions would broaden the range of synthetic topological models accessible to controlled state preparation and characterization.

\begin{acknowledgments}
We thank Youngkuk Kim for helpful discussions and comments. This work is supported by the National Research Foundation of Korea (NRF) grant funded by the Korea government (MSIT) (RS-2022-NR068814, RS-2023-NR119931, RS-2023-00302576, RS-2024-00466865, RS-2023-NR068116, 2023R1A2C1005588) and Institute of Information \& Communications Technology Planning \& Evaluation (IITP) grant funded by the Korean government (MSIT) (RS-2022-II221040).

This material is supported in part by the US Department of Energy, Office of Science, Office of Advanced Scientific Computing Research under its Quantum Testbed Program. Sandia National Laboratories is a multimission laboratory managed and operated by National Technology \& Engineering Solutions of Sandia, LLC, a wholly owned subsidiary of Honeywell International Inc., for the US Department of Energy’s National Nuclear Security Administration under contract DE-NA0003525. This paper describes objective technical results and analysis. Any subjective views or opinions that might be expressed in the paper do not necessarily represent the views of the US Department of Energy or the United States Government. SAND2026-26092O.

S.L.\ designed the experimental implementation, analyzed the data, performed the numerical simulations, and wrote the original draft.
C.I.\ provided the experimental control code.
C.G.Y., B.K.M., E.C.T., D.S.L., M.C.R., and S.M.C.\ carried out the experiments and hardware calibration at the Quantum Scientific Computing Open User Testbed (QSCOUT).
M.-S.C.\ provided the theoretical concept.
J.K.\ supervised the project and revised the manuscript.
All authors reviewed the manuscript and provided comments.

\end{acknowledgments}

\appendix   

\section{Topological index and eigenstates of the AQRM}
\label{Appendix:AQRMTheory}

Within the even-parity subspace, the chiral operator $\hat{\Xi}$ separates the states into $\mathcal{H}_{+} = \{|2n\rangle|g\rangle\}$ and $\mathcal{H}_{-} = \{|2n+1\rangle|e\rangle\}$, where $n=0,1,2,\ldots$ and $|n\rangle$ denotes a Fock state of the bosonic mode $\hat{a}$.
Chiral symmetry forbids matrix elements within each sector, so the Hamiltonian is block off-diagonal,
\begin{equation}
    \hat{H} =
    \begin{pmatrix}
        0 & \hat{H}_{+-} \\
        \hat{H}_{-+} & 0
    \end{pmatrix},
    \label{eq:BlockOffDiagonal}
\end{equation}
where $\hat{H}_{-+} = \hbar(g_1 \hat{a} + g_2 \hat{a}^\dagger)$ maps $\mathcal{H}_{+}$ into $\mathcal{H}_{-}$ and $\hat{H}_{+-} = \hat{H}_{-+}^\dagger$ maps $\mathcal{H}_{-}$ into $\mathcal{H}_{+}$.
The topological invariant is the edge index, defined as the number of zero-energy states with positive chirality minus that with negative chirality~\cite{thiang_topological_2023,Essin11a}.
Since the zero-energy states of positive (negative) chirality span the kernel of $\hat{H}_{-+}$ ($\hat{H}_{+-}$), the edge index $\mathrm{Ind}_{e}\hat{H}$ equals the Fredholm index of the chiral block,
\begin{equation}
    \mathrm{Ind}_{e}\,\hat{H} = \dim\ker \hat{H}_{-+} - \dim\ker \hat{H}_{+-}.
    \label{eq:FredholmIndex}
\end{equation}

The chiral blocks take a particularly simple form after a Bogoliubov transformation~\cite{balian_nonunitary_1969}.
For the non-trivial phase $g_1 > g_2$, we define
\begin{equation}
    \hat{b} = \frac{g_1\, \hat{a} + g_2\, \hat{a}^\dagger}{\sqrt{g_1^2 - g_2^2}}
    = \hat{S}(r)\, \hat{a} \, \hat{S}^\dagger(r),
    \label{eq:Bogoliubov_nt}
\end{equation}
where $\hat{S}(r)=\exp\!\left[\frac{r}{2}\left(\hat{a}^2 - \hat{a}^{\dagger 2}\right)\right]$ is the squeezing operator with the squeezing parameter $r=\tanh^{-1}(g_2/g_1)$.
The operator $\hat{b}$ satisfies the canonical commutation relation $[\hat{b}, \hat{b}^\dagger]=1$, and its Fock states are related to those of $\hat{a}$ by $|n\rangle_b = \hat{S}(r) |n\rangle_a$.
Since $\hat{S}(r)$ commutes with boson parity $\exp(i\pi\hat{a}^\dagger\hat{a})$ and leaves the spin unchanged, it preserves both the parity and chirality sectors. In the transformed basis, the parity operator remains $\hat{\Pi} = -\exp\!\left(i \pi \hat{b}^\dagger \hat{b}\right)\hat{\sigma}_z$, and the chirality sectors can equally be written as $\mathcal{H}_{+} = \{|2n\rangle_b|g\rangle\}$ and $\mathcal{H}_{-} = \{|2n+1\rangle_b|e\rangle\}$.
In terms of $\hat{b}$, the Hamiltonian reduces to
\begin{equation}
    \hat{H} = E_{\mathrm{gap}}
    \left(
    \hat{b}^\dagger \hat{\sigma}_- + \hat{b} \hat{\sigma}_+
    \right),
    \qquad
    E_{\mathrm{gap}} = \hbar\sqrt{g_1^2 - g_2^2},
    \label{eq:JCHamiltonian_b}
\end{equation}
which contains only rotating terms and is therefore equivalent to the JC Hamiltonian.
The chiral blocks are read off as $\hat{H}_{-+} = E_{\mathrm{gap}}\,\hat{b}$ and $\hat{H}_{+-} = E_{\mathrm{gap}}\,\hat{b}^\dagger$.
Only $\hat{H}_{-+}$ has a kernel, spanned by the single state $|0\rangle_b|g\rangle$, so the index is $1$ and this zero mode is the TES.

For the trivial phase $g_2 > g_1$, we instead define
\begin{equation}
    \hat{b} = \frac{g_2\, \hat{a} + g_1\, \hat{a}^\dagger}{\sqrt{g_2^2 - g_1^2}}
    = \hat{S}(r)\, \hat{a} \, \hat{S}^\dagger(r),
    \label{eq:Bogoliubov_t}
\end{equation}
with $r=\tanh^{-1}(g_1/g_2)$, under which the Hamiltonian becomes
\begin{equation}
    \hat{H} = E_{\mathrm{gap}}
    (\hat{b}^\dagger \hat{\sigma}_+ + \hat{b} \hat{\sigma}_-),
    \qquad
    E_{\mathrm{gap}} = \hbar\sqrt{g_2^2 - g_1^2},
    \label{eq:AJCHamiltonian_b}
\end{equation}
which contains only the AJC interaction.
The roles of $\hat{b}$ and $\hat{b}^\dagger$ are exchanged in the chiral blocks, $\hat{H}_{-+} = E_{\mathrm{gap}}\,\hat{b}^\dagger$ and $\hat{H}_{+-} = E_{\mathrm{gap}}\,\hat{b}$.
Within the even-parity subspace, $\hat{b}^\dagger$ has no kernel and $\hat{b}$ acting on $\mathcal{H}_{-}$ annihilates no state because $\mathcal{H}_{-}$ contains no $\hat{b}$-vacuum.
Both kernels are therefore empty and the index vanishes, consistent with the absence of a zero-energy state.

The same analysis can be carried out for the odd-parity subspace, where the chirality sectors are $\mathcal{H}_{+} = \{|2n+1\rangle|g\rangle\}$ and $\mathcal{H}_{-} = \{|2n\rangle|e\rangle\}$.
The $\hat{b}$-vacuum now lies in $\mathcal{H}_{-}$, so the zero mode appears for $g_2 > g_1$ and carries negative chirality, giving the index $-1$.
Table~\ref{tab:FredholmIndex} summarizes the kernel dimensions and the resulting indices for both parity subspaces.

\begin{table}[t]
\centering
\caption{Kernel dimensions of the chiral blocks and the resulting Fredholm indices in each parity subspace.}
\label{tab:FredholmIndex}
\renewcommand{\arraystretch}{1.25}
\setlength{\tabcolsep}{8pt}
\begin{tabular}{ccccc}
\toprule
Parity & Phase & $\dim\ker\hat{H}_{-+}$ & $\dim\ker\hat{H}_{+-}$ & $\mathrm{Ind}_{e}\,\hat{H}$ \\
\midrule
Even & $g_1>g_2$ & 1 & 0 & 1 \\
Even & $g_2>g_1$ & 0 & 0 & 0 \\
Odd  & $g_1>g_2$ & 0 & 0 & 0 \\
Odd  & $g_2>g_1$ & 0 & 1 & $-1$ \\
\bottomrule
\end{tabular}
\end{table}

The Bogoliubov-transformed Hamiltonians also yield the full set of eigenstates.
In the non-trivial phase, the even-parity eigenstates of Eq.~(\ref{eq:JCHamiltonian_b}) are
\begin{equation}
    \begin{aligned}
        |0\rangle_{\mathrm{nt}} &= |0\rangle_b |g\rangle = \hat{S}(r)|0\rangle_a |g\rangle \\
        |n, \pm \rangle_{\mathrm{nt}} &= \frac{1}{\sqrt{2}} \left( |2n\rangle_b |g\rangle \pm |2n-1\rangle_b |e\rangle\right) \\
        &= \frac{1}{\sqrt{2}} \hat{S}(r) \left( |2n\rangle_a |g\rangle \pm |2n-1\rangle_a |e\rangle\right), \quad n=1,2,\ldots
    \end{aligned}
\end{equation}
with energy eigenvalues $E_n = \pm E_{\mathrm{gap}}\sqrt{2n}$.
The zero-energy eigenstate $|0\rangle_{\mathrm{nt}}$ is the TES identified above as the kernel of $\hat{H}_{-+}$, and it is a product of the ground spin state and a squeezed vacuum bosonic state.
The corresponding bosonic statistics is
\begin{equation}
    P_{2n}=\frac{1}{\cosh{r}}\frac{(2n)!}{2^{2n}(n!)^2}\,\tanh^{2n} r,
\end{equation}
with all odd-Fock populations vanishing by chiral symmetry.
The ratio of successive populations, $P_{2n}/P_{2n-2} = \tanh^2 r\,(2n-1)/(2n)$, approaches the constant $\tanh^2 r$ only asymptotically, and the extra factor $(2n-1)/(2n)$ accumulates into the $1/\sqrt{n}$ suppression of the population on top of the exponential envelope discussed in the main text.

In the trivial phase, the even-parity eigenstates of Eq.~(\ref{eq:AJCHamiltonian_b}) are
\begin{equation}
    \begin{aligned}
        |n, \pm \rangle_{\mathrm{t}} &= \frac{1}{\sqrt{2}} \left( |2n-2\rangle_b |g\rangle \pm |2n-1\rangle_b |e\rangle\right) \\
        &= \frac{1}{\sqrt{2}} \hat{S}(r) \left( |2n-2\rangle_a |g\rangle \pm |2n-1\rangle_a |e\rangle\right), \quad n=1,2,\ldots
    \end{aligned}
\end{equation}
with eigenvalues $E_n = \pm E_{\mathrm{gap}} \sqrt{2n-1}$.

The same transformations can be used to derive the odd-parity-subspace eigenstates, where the phase assignment is simply reversed, with $g_1>g_2$ corresponding to the trivial phase and $g_2>g_1$ to the non-trivial phase. This subspace is not explored in this work.

\section{Detailed Experimental Methods}
\label{Appendix:ExperimentalDetails}

Coherent control of both the hyperfine spin transition and the spin--motion interaction is achieved using stimulated Raman transitions driven by a pulsed 355 nm laser.
The pulsed laser has a frequency-comb structure, in which pairs of comb teeth bridge the hyperfine splitting of the $^2S_{1/2}$ ground-state manifold, with frequency splitting of $\omega_{\mathrm{0}}/2\pi = 12.642\,812~\mathrm{GHz} + (310.8~\mathrm{Hz/G^2})\,B^2$~\cite{hayes_entanglement_2010, olmschenk_manipulation_2007}.
The Raman transitions are implemented using a multi-channel acousto-optic modulator (AOM) system, which enables individual addressing of the ion as well as precise control of the frequency and amplitude of each Raman beam~\cite{lim_design_2025,debnath_demonstration_2016,wright_benchmarking_2019}. 

In the Lamb--Dicke regime, $\eta\sqrt{2\bar{n}+1}\ll1$, a counter-propagating Raman beam pair yields sideband interactions near the red and blue resonances. When both sidebands are simultaneously applied, the interaction-picture Hamiltonian becomes
\begin{equation}
    \hat{H}_{\mathrm{int}}
    =
    \left(
    \frac{\hbar\Omega_{\mathrm{red}}}{2}\, \hat{a}\, \hat{\sigma}_+ e^{i(\delta_{\mathrm{red}} t + \phi_{\mathrm{red}})}
    +
    \frac{\hbar\Omega_{\mathrm{blue}}}{2}\, \hat{a}^\dagger\, \hat{\sigma}_+ e^{i(\delta_{\mathrm{blue}} t + \phi_{\mathrm{blue}})}
    \right)
    + \mathrm{H.c.},
\end{equation}
where $\Omega$, $\delta$, and $\phi$ denote the effective Rabi frequency, detuning, and phase of each Raman beam, respectively. With both detunings and phases set to zero, \textit{i.e.}, $\delta_{\mathrm{red}}= \delta_{\mathrm{blue}}= 0$ and $\phi_{\mathrm{red}}= \phi_{\mathrm{blue}}= 0$, the interaction Hamiltonian becomes the AQRM Hamiltonian (Eq.~\ref{eq:AQRMHamiltonian}) where $g_1$ and $g_2$ are given by $\Omega_{\mathrm{red}}/2$ and $\Omega_{\mathrm{blue}}/2$, respectively.

For the adiabatic state preparation in the non-trivial phase ($g_1>g_2$), we first turn on $g_1$ to a fixed value of $g_1^{(0)}=2\pi\times 8.33~\mathrm{kHz}$, and then linearly increase $g_2$ to the desired value.
The slope of the $g_2$ ramp is set to $\nu=2\pi\times 39.79~\mathrm{kHz/ms}$, and the final value of $g_2$ is tuned by changing the total ramping time.
For the trivial phase ($g_2>g_1$), we follow the same protocol with the roles of the two couplings exchanged, holding $g_2$ at $g_2^{(0)}=2\pi\times 8.33~\mathrm{kHz}$ while ramping up $g_1$ at the same slope $\nu$.

During the ramp, the ion hyperfine levels experience a light shift due to off-resonant comb components of the 355 nm Raman laser, which depends on the intensity of the Raman beams ($|\Omega_{\mathrm{red}}|^2$ and $|\Omega_{\mathrm{blue}}|^2$)~\cite{lee_engineering_2016}.
To keep both sidebands resonant throughout the state preparation process, we dynamically update the red- and blue-sideband laser frequencies and maintain target detunings $\delta_{\mathrm{red}}$ and $\delta_{\mathrm{blue}}$ in real time~\cite{yale_realization_2025}.
This compensation is implemented through RF-frequency modulation of the AOM drives using the \textsc{Octet} control hardware system.

\section{Numerical simulation and state preparation fidelity}
\label{appendix:LindbladEquation}

To account for various experimental imperfections, we numerically solve the Lindblad master equation using the \texttt{QuTiP} Python package\cite{johansson_qutip_2013}.
The dynamics of the density matrix $\rho(t)$ are governed by
\begin{equation}
    \dot{\rho}(t)
    =
    -\frac{i}{\hbar}\,[\hat{H}, \rho(t)]
    +
    \sum_k \mathcal{L}[\hat{O}_k]\,\rho(t),
\end{equation}
where the Lindblad superoperator is defined as
\begin{equation}
    \mathcal{L}[\hat{O}]\,\rho
    \equiv
    \hat{O}\rho\hat{O}^\dagger
    -
    \frac{1}{2}\hat{O}^\dagger\hat{O}\rho
    -
    \frac{1}{2}\rho\hat{O}^\dagger\hat{O}.
\end{equation}
Here, $\hat{H}$ is the AQRM Hamiltonian (Eq.~\ref{eq:AQRMHamiltonian}) and $\hat{O}_k$ are the Lindblad jump operators.

In our experiment, the main source of decoherence was motional dephasing, with a dephasing time of $\tau = 2.5~\mathrm{ms}$.
We model phonon dephasing by including the Lindblad jump operator $\hat{O}=\sqrt{2\Gamma}\,\hat{a}^\dagger\hat{a}$ into the Lindblad equation, where $\Gamma = 1/\tau$ is the phonon dephasing rate.
Spin dephasing ($T_2^* \approx 30~\mathrm{ms}$) and motional heating (3 quanta per second for tilt mode) are negligible in the timescale of our experiment and were not included in the numerical simulations.

In addition, imperfect sideband cooling leaves a residual thermal population in the motional mode.
We model this by taking the initial motional state to be a thermal state with a mean phonon number of approximately $0.10$, as measured in our system\cite{leibfried_quantum_2003}.

While we could not directly access the state preparation fidelity in our experiment, we estimated the quality of the adiabatic state preparation through the fidelity between the simulated prepared state and the ideal target eigenstate using the numerical model described above.
Table~\ref{tab:prepfidelity} lists these estimated preparation fidelities for the TES and the three bulk states at different coupling ratios $g_{\mathrm{r}}$.

\begin{table}[t]
\centering
\caption{Simulated preparation fidelities for various coupling ratios $g_{\mathrm{r}}$.}
\label{tab:prepfidelity}
\renewcommand{\arraystretch}{1.25}
\setlength{\tabcolsep}{10pt}
\begin{tabular}{c S[table-format=1.2] S[table-format=1.2] S[table-format=1.2] S[table-format=1.2]}
\toprule
{$g_{\mathrm{r}}$}
& {$\lvert 0\rangle_{\mathrm{nt}}$}
& {$\lvert 1,+\rangle_{\mathrm{t}}$}
& {$\lvert 1,+\rangle_{\mathrm{nt}}$}
& {$\lvert 2,+\rangle_{\mathrm{t}}$} \\
\midrule
0.2 & 0.98 & 0.96 & 0.91 & 0.83 \\
0.4 & 0.98 & 0.90 & 0.77 & 0.56 \\
0.6 & 0.94 & 0.83 & 0.62 & 0.39 \\
0.8 & 0.83 & 0.62 & 0.38 & 0.20 \\
\bottomrule
\end{tabular}
\end{table}

The edge state is expected to be prepared with the highest fidelity across all coupling ratios, whereas the preparation fidelity decreases both as $g_{\mathrm{r}}$ increases and for the higher excited states, through distinct mechanisms.
As $g_{\mathrm{r}}$ approaches unity, the energy gap between neighboring eigenstates closes, making the adiabatic state transfer increasingly difficult.
The higher excited states are more fragile because their energy-level spacings are inherently smaller and their larger mean phonon number makes them more susceptible to phonon dephasing.
Although increasing both $g_1$ and $g_2$ would increase the energy gap, the fidelity is limited by off-resonant excitation of spectator motional modes.
We focus the experimental characterization on states with a simulated preparation fidelity above 0.5.
Accordingly, the joint spin--Fock populations of $|2,+\rangle_{\mathrm{t}}$ are shown at $g_{\mathrm{r}}=0.4$ rather than $0.6$ in Fig.~\ref{fig:Figure2}(d).
Its chirality and entropy bound at $g_{\mathrm{r}}=0.6$ are additionally presented as the faded bar in Fig.~\ref{fig:Figure2}(e) and the hatched bar in Fig.~\ref{fig:Figure3}.
We note, however, that fidelity is a relatively stringent metric for states spanning a large Hilbert space, so a fidelity below this threshold does not imply that the target-state component is absent.
This rapid decrease of the preparation fidelity for higher excited states also limited our exploration to bulk states up to $|2,+\rangle_{\mathrm{t}}$.

\section{Joint spin--boson state population measurement}
\label{Appendix:SpinBosonPopulation}

\begin{figure}
    \centering
    \includegraphics[width=1\linewidth]{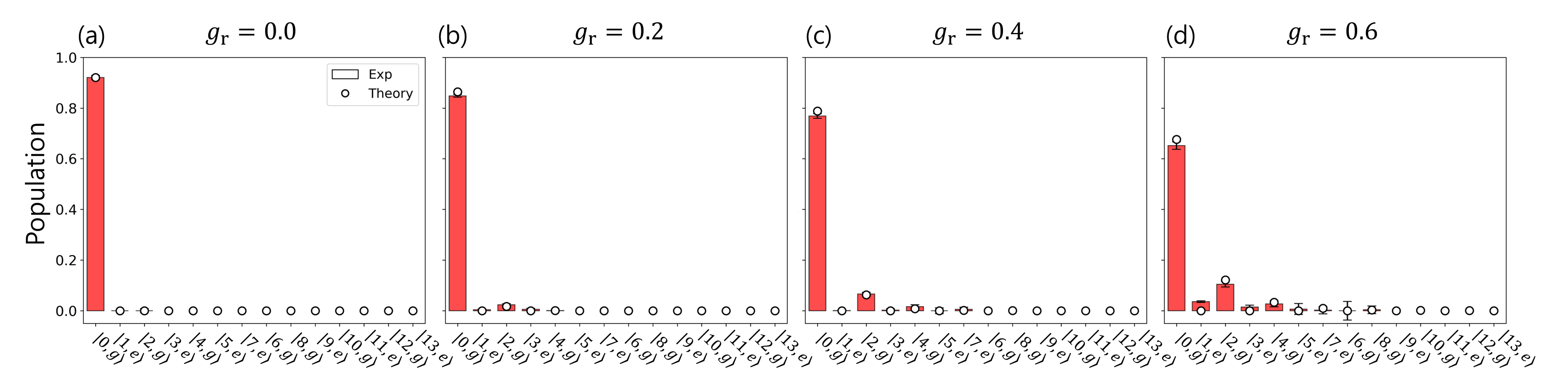}
    \caption{Spin--Fock populations for the TES on the even-parity subspace.
    Bars with error bars show the experimentally extracted populations, while circles indicate the theoretical predictions.
    (a)--(d) correspond to the TES at $g_2/g_1 = 0.0,\,0.2,\,0.4,\,0.6$, respectively. 
    The TES predominantly occupies the even-Fock states, with population of the odd-Fock states being forbidden across the full range of relative coupling strengths.
    }
    \label{fig:Appendix2}
\end{figure}

\begin{figure}
    \centering
    \includegraphics[width=1\linewidth]{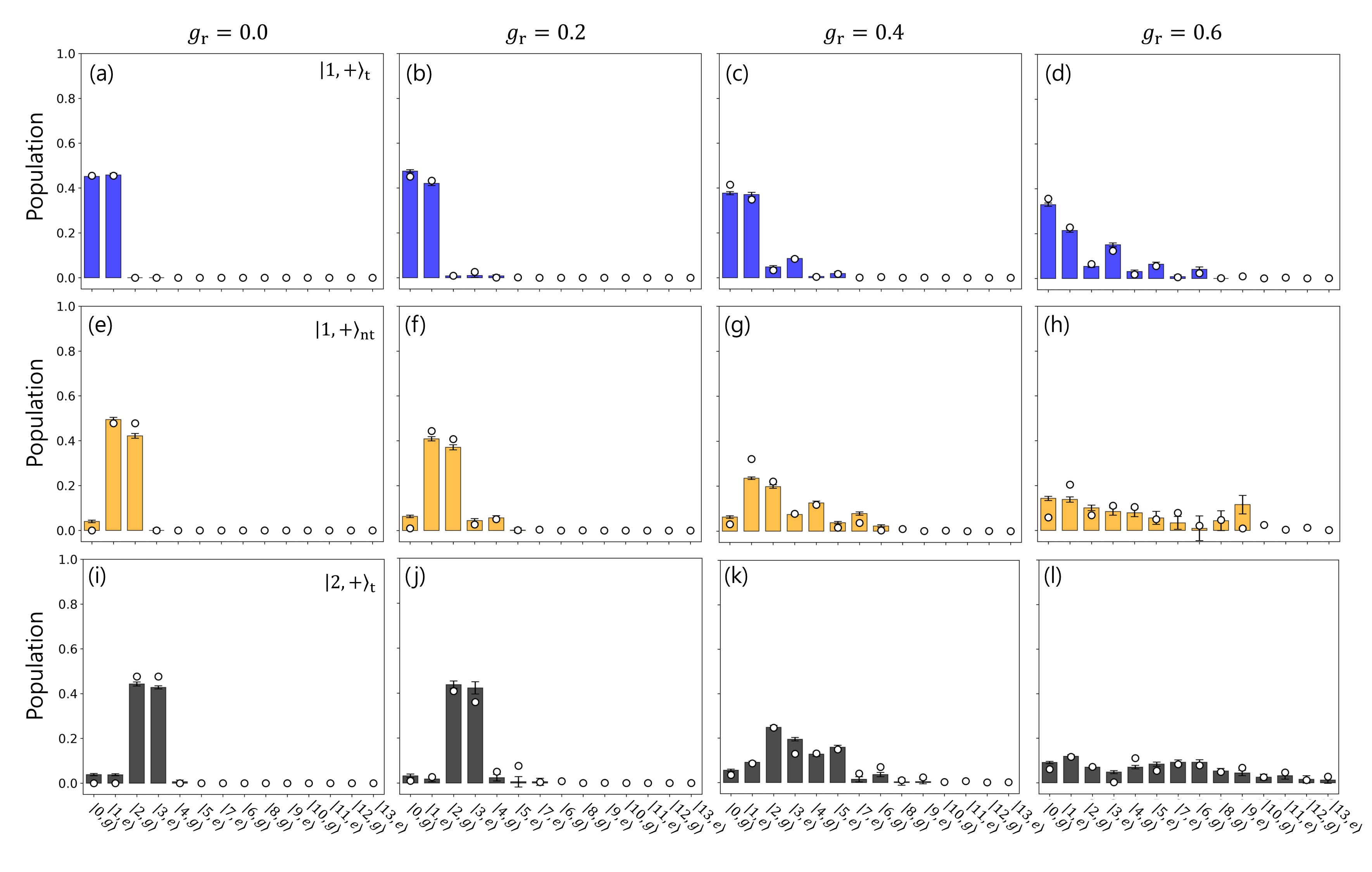}
    \caption{
    Spin--Fock-state populations of representative bulk states in the even-parity subspace for different relative coupling strengths. 
    Bars with error bars denote the experimentally extracted populations, while circles indicate the theoretical predictions. 
    (a)--(d) correspond to the bulk state $\lvert 1,+\rangle_{\mathrm{t}}$, (e)--(h) to $\lvert 1,+\rangle_{\mathrm{nt}}$, and (i)--(l) to $\lvert 2,+\rangle_{\mathrm{t}}$.
    In contrast to the TES, the bulk states occupy both even- and odd-Fock states and are not confined to a single chiral sector.
    }
    \label{fig:Appendix3}
\end{figure}

To obtain the joint spin--boson state population of the prepared states, we measured the blue-sideband Rabi oscillations with the spectator ion~\cite{leibfried_quantum_2003, lo_spinmotion_2015}.
After a blue-sideband pulse of duration $\tau$ is applied to the spectator ion, we measured both the spectator- and target-ion spin states and combined the measurement results to obtain two spectator ion excitation probability trajectories $P_{e/g}(\tau)$ conditioned on the target ion's spin state.
Each trajectory contains the information of phonon statistics conditioned on the target ion's spin state, which can be extracted by fitting with the model below
\begin{equation}
    P_{e/g}(\tau)=\frac{1}{2}[1-\sum_{n=0}^{n_{\max}}p^{\mathrm{phonon}}_{e/g}(n) e^{-\gamma_n\tau}\cos(\Omega_{n,n+1}\tau)]
\label{eq:phononfittingfunction}
\end{equation}
where $p^{\mathrm{phonon}}_{e/g}(n)$ denotes the occupation probability of the phonon number state $n$ conditioned on the target ion's spin in the $e/g$ state, $\gamma_n$ is an empirical decay rate satisfying $\gamma_n \propto (n+1)^{0.7}$, $\Omega_{n,n+1}=\sqrt{n+1}\Omega_{\mathrm{blue}}$ is the blue-sideband Rabi frequency, and $n_{\max}$ denotes the phonon-number cutoff.
The cutoff $n_{\max}$ was defined as the smallest phonon number $n$ satisfying $p^{\mathrm{phonon}}_{e/g}(n)< 1\%$ in the target state.
The extracted phonon number distribution was normalized to fulfill $\sum_{n=0}^{n_{\max}}p^{\mathrm{phonon}}_{e/g}(n)=1$.
Combined with the target ion's spin state population $p^{\mathrm{spin}}_{e/g}$, we can obtain the joint spin--Fock state population $p_{e/g,n}$ as
\begin{equation}
    p_{e/g,n}=p^{\mathrm{spin}}_{e/g}\cdot p^{\mathrm{phonon}}_{e/g}(n).
\end{equation}
Figure~\ref{fig:Appendix2} and Figure~\ref{fig:Appendix3} show the extracted joint spin--Fock state population for the TES and bulk states, respectively.

\begin{figure}
    \centering
    \includegraphics[width=0.5\linewidth]{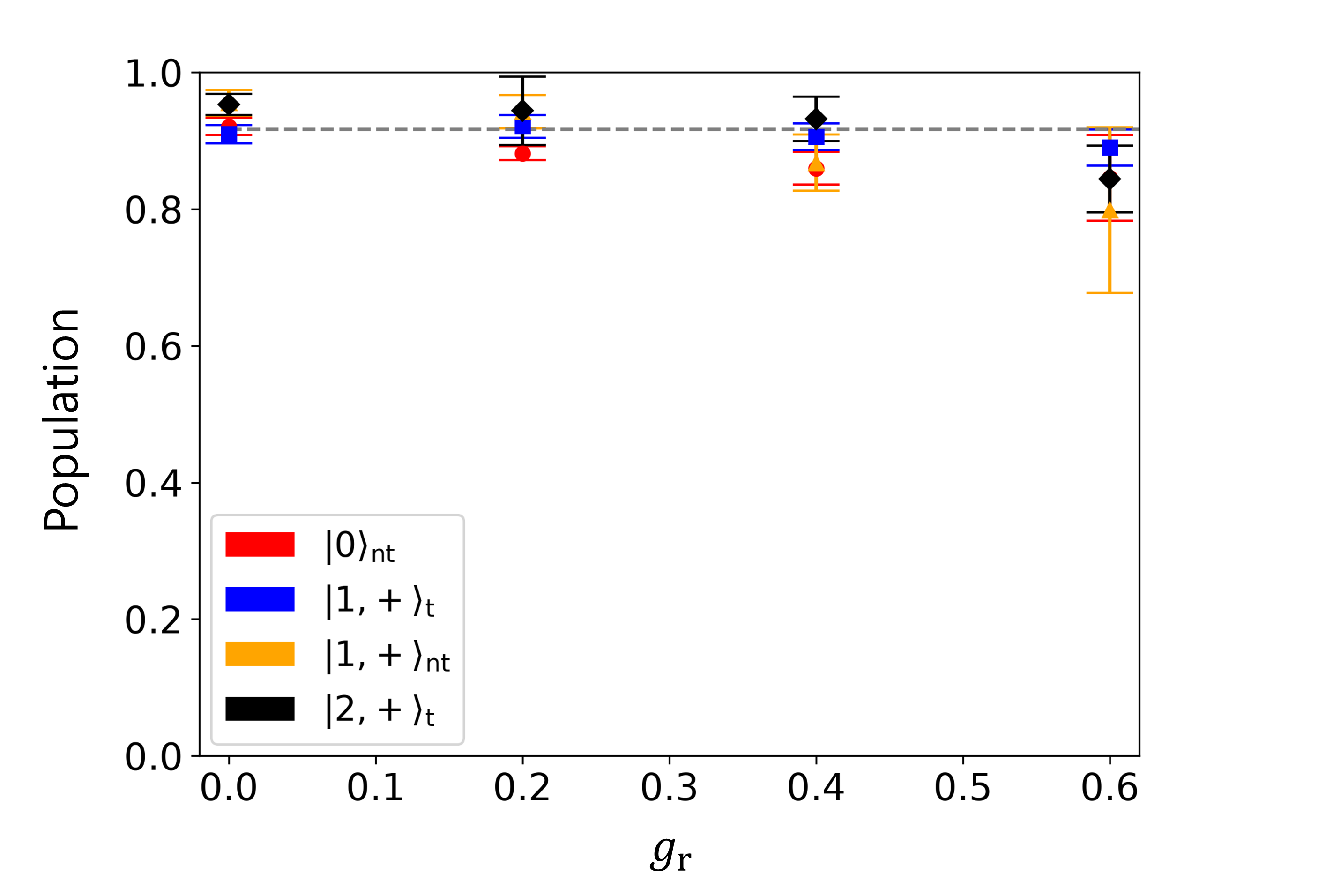}
    \caption{Measured even-parity subspace occupation for all prepared eigenstates. 
    The symbols denote the measured total populations of the TES and bulk states. The dashed gray line indicates the expected even-parity population for a residual phonon number of 0.1 after sideband cooling.
    The population in the even-parity subspace remains nearly constant for all eigenstates throughout the adiabatic evolution.
    }
    \label{fig:Appendix1}
\end{figure}

With the extracted joint spin--Fock state population, we can obtain the state occupation in even-parity states $p_{\textrm{even}} = \sum_{\Pi=+1}p_{e/g, n}$ and odd-parity states $p_{\textrm{odd}} = \sum_{\Pi=-1}p_{e/g, n}$.
Ideally, the state occupation in the even-parity subspace should be unity if the initial state is well prepared in the subspace, but imperfect phonon ground-state cooling leaves a residual phonon number of $\bar{n}\sim0.1$, which leads to around 10\% residual population in the odd-parity subspace.
Figure~\ref{fig:Appendix1} shows the measured even-parity subspace occupation $p_{\textrm{even}}$ for all prepared eigenstates.
Apart from the initial state-preparation leakage of 10\%, the even-parity occupation is well preserved during the adiabatic state transfer, showing that the Hamiltonian commutes with the parity operator.

\section{Upper and lower bound of the second-order Rényi entropy}
\label{appendix:entropybound}

The second-order Rényi entropy can be calculated from the Pauli expectation values of the spin state, given by
\begin{equation}
    \begin{aligned}
    S_{2,\mathrm{spin}} &= -\ln \mathrm{Tr}[\rho_{\mathrm{spin}}^2] \\
    &= -\ln \left[ \frac{1}{2}\left(1 + \langle \sigma_x \rangle^2 + \langle \sigma_y \rangle^2 + \langle \sigma_z \rangle^2 \right) \right].
    \end{aligned}
\end{equation}
In the experiment, the prepared states contain unwanted odd-parity populations due to imperfect state preparation, which should be subtracted from the analysis.
For Z-basis measurement, the odd-parity contribution can be subtracted by using joint spin--Fock state population measurement, described in Appendix~\ref{Appendix:SpinBosonPopulation}, leading to $\langle \sigma_z \rangle_{\mathrm{even}} = \frac{1}{p_{\textrm{even}}}\sum_{\Pi=+1} \left( p_{e, n} - p_{g,n} \right)$.

For X- and Y-basis measurement, the parity-resolved measurement is not accessible, so we derive the boundary of each Pauli expectation value from measured spin expectation values.
First, we consider our prepared state as a mixture of even- and odd-parity states, $\rho = p_{\textrm{even}}\rho_{\textrm{even}} + (1-p_{\textrm{even}})\rho_{\textrm{odd}}$.
Then, the Pauli measurement should give $\langle \sigma_i \rangle = p_{\textrm{even}}\langle \sigma_i \rangle_{\textrm{even}} + (1-p_{\textrm{even}})\langle \sigma_i \rangle_{\textrm{odd}}$.
Since $\langle \sigma_i \rangle_{\textrm{odd}}$ for $i = x, y$ is bounded in the range of $\left[-1, 1\right]$, we can derive the boundary of $\langle \sigma_i \rangle_{\textrm{even}}$ as
\begin{equation}
\frac{1}{p}\langle \sigma_i \rangle - \frac{1-p}{p}
\;\le\;
\langle \sigma_i \rangle_{\mathrm{even}}
\;\le\;
\frac{1}{p}\langle \sigma_i \rangle + \frac{1-p}{p},
\end{equation}
where $p \equiv p_{\textrm{even}}$ denotes the even-parity occupation.
As a result, we can derive the upper and lower bounds of the squared Pauli expectation value in the even-parity subspace as
\begin{equation}
    0 \leq \langle \sigma_i \rangle_{\mathrm{even}}^{2}
    \leq \langle\sigma_i \rangle_{\mathrm{even,max}}^{2}
    =
    \min\!\left\{
    1,\,
    \max\!\left[
    \left(
    \frac{1}{p}\langle \sigma_i \rangle-\frac{1-p}{p}
    \right)^2,
    \left(
    \frac{1}{p}\langle \sigma_i \rangle + \frac{1-p}{p}
    \right)^{2}
    \right]
    \right\}.
\end{equation}
resulting in the upper and lower bounds of the second-order Rényi entropy as
\begin{equation}
    -\ln \left[ \frac{1}{2}\left(1 + \langle \sigma_x \rangle_{\mathrm{even, max}}^2 + \langle \sigma_y \rangle_{\mathrm{even, max}}^2 + \langle  \sigma_z \rangle_{\mathrm{even}}^2 \right) \right] \leq S_{2,\mathrm{spin}} \leq -\ln \left[ \frac{1}{2}\left(1 + \langle \sigma_z \rangle_{\mathrm{even}}^2 \right) \right].
\end{equation}


\begin{thebibliography}{81}%
\makeatletter
\providecommand \@ifxundefined [1]{%
 \@ifx{#1\undefined}
}%
\providecommand \@ifnum [1]{%
 \ifnum #1\expandafter \@firstoftwo
 \else \expandafter \@secondoftwo
 \fi
}%
\providecommand \@ifx [1]{%
 \ifx #1\expandafter \@firstoftwo
 \else \expandafter \@secondoftwo
 \fi
}%
\providecommand \natexlab [1]{#1}%
\providecommand \enquote  [1]{``#1''}%
\providecommand \bibnamefont  [1]{#1}%
\providecommand \bibfnamefont [1]{#1}%
\providecommand \citenamefont [1]{#1}%
\providecommand \href@noop [0]{\@secondoftwo}%
\providecommand \href [0]{\begingroup \@sanitize@url \@href}%
\providecommand \@href[1]{\@@startlink{#1}\@@href}%
\providecommand \@@href[1]{\endgroup#1\@@endlink}%
\providecommand \@sanitize@url [0]{\catcode `\\12\catcode `\$12\catcode
  `\&12\catcode `\#12\catcode `\^12\catcode `\_12\catcode `\%12\relax}%
\providecommand \@@startlink[1]{}%
\providecommand \@@endlink[0]{}%
\providecommand \url  [0]{\begingroup\@sanitize@url \@url }%
\providecommand \@url [1]{\endgroup\@href {#1}{\urlprefix }}%
\providecommand \urlprefix  [0]{URL }%
\providecommand \Eprint [0]{\href }%
\providecommand \doibase [0]{https://doi.org/}%
\providecommand \selectlanguage [0]{\@gobble}%
\providecommand \bibinfo  [0]{\@secondoftwo}%
\providecommand \bibfield  [0]{\@secondoftwo}%
\providecommand \translation [1]{[#1]}%
\providecommand \BibitemOpen [0]{}%
\providecommand \bibitemStop [0]{}%
\providecommand \bibitemNoStop [0]{.\EOS\space}%
\providecommand \EOS [0]{\spacefactor3000\relax}%
\providecommand \BibitemShut  [1]{\csname bibitem#1\endcsname}%
\let\auto@bib@innerbib\@empty
%</preamble>
\bibitem [{\citenamefont {v.~Klitzing}\ \emph {et~al.}(1980)\citenamefont
  {v.~Klitzing}, \citenamefont {Dorda},\ and\ \citenamefont
  {Pepper}}]{klitzing_new_1980}%
  \BibitemOpen
  \bibfield  {author} {\bibinfo {author} {\bibfnamefont {K.}~\bibnamefont
  {v.~Klitzing}}, \bibinfo {author} {\bibfnamefont {G.}~\bibnamefont {Dorda}},\
  and\ \bibinfo {author} {\bibfnamefont {M.}~\bibnamefont {Pepper}},\
  }\bibfield  {title} {\bibinfo {title} {New {Method} for {High}-{Accuracy}
  {Determination} of the {Fine}-{Structure} {Constant} {Based} on {Quantized}
  {Hall} {Resistance}},\ }\href {https://doi.org/10.1103/PhysRevLett.45.494}
  {\bibfield  {journal} {\bibinfo  {journal} {Phys. Rev. Lett.}\ }\textbf
  {\bibinfo {volume} {45}},\ \bibinfo {pages} {494} (\bibinfo {year}
  {1980})}\BibitemShut {NoStop}%
\bibitem [{\citenamefont {Hasan}\ and\ \citenamefont
  {Kane}(2010)}]{hasan_colloquium_2010}%
  \BibitemOpen
  \bibfield  {author} {\bibinfo {author} {\bibfnamefont {M.~Z.}\ \bibnamefont
  {Hasan}}\ and\ \bibinfo {author} {\bibfnamefont {C.~L.}\ \bibnamefont
  {Kane}},\ }\bibfield  {title} {\bibinfo {title} {Colloquium: {Topological}
  insulators},\ }\href {https://doi.org/10.1103/RevModPhys.82.3045} {\bibfield
  {journal} {\bibinfo  {journal} {Rev. Mod. Phys.}\ }\textbf {\bibinfo {volume}
  {82}},\ \bibinfo {pages} {3045} (\bibinfo {year} {2010})}\BibitemShut
  {NoStop}%
\bibitem [{\citenamefont {Moore}(2010)}]{moore_birth_2010}%
  \BibitemOpen
  \bibfield  {author} {\bibinfo {author} {\bibfnamefont {J.~E.}\ \bibnamefont
  {Moore}},\ }\bibfield  {title} {\bibinfo {title} {The birth of topological
  insulators},\ }\href {https://doi.org/10.1038/nature08916} {\bibfield
  {journal} {\bibinfo  {journal} {Nature}\ }\textbf {\bibinfo {volume} {464}},\
  \bibinfo {pages} {194} (\bibinfo {year} {2010})}\BibitemShut {NoStop}%
\bibitem [{\citenamefont {Sato}\ and\ \citenamefont
  {Ando}(2017)}]{sato_topological_2017}%
  \BibitemOpen
  \bibfield  {author} {\bibinfo {author} {\bibfnamefont {M.}~\bibnamefont
  {Sato}}\ and\ \bibinfo {author} {\bibfnamefont {Y.}~\bibnamefont {Ando}},\
  }\bibfield  {title} {\bibinfo {title} {Topological superconductors: a
  review},\ }\href {https://doi.org/10.1088/1361-6633/aa6ac7} {\bibfield
  {journal} {\bibinfo  {journal} {Rep. Prog. Phys.}\ }\textbf {\bibinfo
  {volume} {80}},\ \bibinfo {pages} {076501} (\bibinfo {year}
  {2017})}\BibitemShut {NoStop}%
\bibitem [{\citenamefont {Qi}\ and\ \citenamefont
  {Zhang}(2011)}]{qi_topological_nodate}%
  \BibitemOpen
  \bibfield  {author} {\bibinfo {author} {\bibfnamefont {X.-L.}\ \bibnamefont
  {Qi}}\ and\ \bibinfo {author} {\bibfnamefont {S.-C.}\ \bibnamefont {Zhang}},\
  }\bibfield  {title} {\bibinfo {title} {Topological insulators and
  superconductors},\ }\href {https://doi.org/10.1103/RevModPhys.83.1057}
  {\bibfield  {journal} {\bibinfo  {journal} {Rev. Mod. Phys.}\ }\textbf
  {\bibinfo {volume} {83}},\ \bibinfo {pages} {1057} (\bibinfo {year}
  {2011})}\BibitemShut {NoStop}%
\bibitem [{\citenamefont {Kane}\ and\ \citenamefont
  {Mele}(2005)}]{kane_quantum_2005}%
  \BibitemOpen
  \bibfield  {author} {\bibinfo {author} {\bibfnamefont {C.~L.}\ \bibnamefont
  {Kane}}\ and\ \bibinfo {author} {\bibfnamefont {E.~J.}\ \bibnamefont
  {Mele}},\ }\bibfield  {title} {\bibinfo {title} {Quantum {Spin} {Hall}
  {Effect} in {Graphene}},\ }\href
  {https://doi.org/10.1103/PhysRevLett.95.226801} {\bibfield  {journal}
  {\bibinfo  {journal} {Phys. Rev. Lett.}\ }\textbf {\bibinfo {volume} {95}},\
  \bibinfo {pages} {226801} (\bibinfo {year} {2005})}\BibitemShut {NoStop}%
\bibitem [{\citenamefont {Bernevig}\ and\ \citenamefont
  {Zhang}(2006)}]{bernevig_quantum_2006}%
  \BibitemOpen
  \bibfield  {author} {\bibinfo {author} {\bibfnamefont {B.~A.}\ \bibnamefont
  {Bernevig}}\ and\ \bibinfo {author} {\bibfnamefont {S.-C.}\ \bibnamefont
  {Zhang}},\ }\bibfield  {title} {\bibinfo {title} {Quantum {Spin} {Hall}
  {Effect}},\ }\href {https://doi.org/10.1103/PhysRevLett.96.106802} {\bibfield
   {journal} {\bibinfo  {journal} {Phys. Rev. Lett.}\ }\textbf {\bibinfo
  {volume} {96}},\ \bibinfo {pages} {106802} (\bibinfo {year}
  {2006})}\BibitemShut {NoStop}%
\bibitem [{\citenamefont {König}\ \emph {et~al.}(2007)\citenamefont {König},
  \citenamefont {Wiedmann}, \citenamefont {Brüne}, \citenamefont {Roth},
  \citenamefont {Buhmann}, \citenamefont {Molenkamp}, \citenamefont {Qi},\ and\
  \citenamefont {Zhang}}]{konig_quantum_2007}%
  \BibitemOpen
  \bibfield  {author} {\bibinfo {author} {\bibfnamefont {M.}~\bibnamefont
  {König}}, \bibinfo {author} {\bibfnamefont {S.}~\bibnamefont {Wiedmann}},
  \bibinfo {author} {\bibfnamefont {C.}~\bibnamefont {Brüne}}, \bibinfo
  {author} {\bibfnamefont {A.}~\bibnamefont {Roth}}, \bibinfo {author}
  {\bibfnamefont {H.}~\bibnamefont {Buhmann}}, \bibinfo {author} {\bibfnamefont
  {L.~W.}\ \bibnamefont {Molenkamp}}, \bibinfo {author} {\bibfnamefont {X.-L.}\
  \bibnamefont {Qi}},\ and\ \bibinfo {author} {\bibfnamefont {S.-C.}\
  \bibnamefont {Zhang}},\ }\bibfield  {title} {\bibinfo {title} {Quantum {Spin}
  {Hall} {Insulator} {State} in {HgTe} {Quantum} {Wells}},\ }\href
  {https://doi.org/10.1126/science.1148047} {\bibfield  {journal} {\bibinfo
  {journal} {Science}\ }\textbf {\bibinfo {volume} {318}},\ \bibinfo {pages}
  {766} (\bibinfo {year} {2007})}\BibitemShut {NoStop}%
\bibitem [{\citenamefont {Chiu}\ \emph {et~al.}(2016)\citenamefont {Chiu},
  \citenamefont {Teo}, \citenamefont {Schnyder},\ and\ \citenamefont
  {Ryu}}]{chiu_classification_2016}%
  \BibitemOpen
  \bibfield  {author} {\bibinfo {author} {\bibfnamefont {C.-K.}\ \bibnamefont
  {Chiu}}, \bibinfo {author} {\bibfnamefont {J.~C.~Y.}\ \bibnamefont {Teo}},
  \bibinfo {author} {\bibfnamefont {A.~P.}\ \bibnamefont {Schnyder}},\ and\
  \bibinfo {author} {\bibfnamefont {S.}~\bibnamefont {Ryu}},\ }\bibfield
  {title} {\bibinfo {title} {Classification of topological quantum matter with
  symmetries},\ }\href {https://doi.org/10.1103/RevModPhys.88.035005}
  {\bibfield  {journal} {\bibinfo  {journal} {Rev. Mod. Phys.}\ }\textbf
  {\bibinfo {volume} {88}},\ \bibinfo {pages} {035005} (\bibinfo {year}
  {2016})}\BibitemShut {NoStop}%
\bibitem [{\citenamefont {He}\ \emph {et~al.}(2022)\citenamefont {He},
  \citenamefont {Hughes}, \citenamefont {Armitage}, \citenamefont {Tokura},\
  and\ \citenamefont {Wang}}]{he_topological_2022}%
  \BibitemOpen
  \bibfield  {author} {\bibinfo {author} {\bibfnamefont {Q.~L.}\ \bibnamefont
  {He}}, \bibinfo {author} {\bibfnamefont {T.~L.}\ \bibnamefont {Hughes}},
  \bibinfo {author} {\bibfnamefont {N.~P.}\ \bibnamefont {Armitage}}, \bibinfo
  {author} {\bibfnamefont {Y.}~\bibnamefont {Tokura}},\ and\ \bibinfo {author}
  {\bibfnamefont {K.~L.}\ \bibnamefont {Wang}},\ }\bibfield  {title} {\bibinfo
  {title} {Topological spintronics and magnetoelectronics},\ }\href
  {https://doi.org/10.1038/s41563-021-01138-5} {\bibfield  {journal} {\bibinfo
  {journal} {Nat. Mater.}\ }\textbf {\bibinfo {volume} {21}},\ \bibinfo {pages}
  {15} (\bibinfo {year} {2022})}\BibitemShut {NoStop}%
\bibitem [{\citenamefont {Nayak}\ \emph {et~al.}(2008)\citenamefont {Nayak},
  \citenamefont {Simon}, \citenamefont {Stern}, \citenamefont {Freedman},\ and\
  \citenamefont {{Das Sarma}}}]{nayak_non-abelian_2008}%
  \BibitemOpen
  \bibfield  {author} {\bibinfo {author} {\bibfnamefont {C.}~\bibnamefont
  {Nayak}}, \bibinfo {author} {\bibfnamefont {S.~H.}\ \bibnamefont {Simon}},
  \bibinfo {author} {\bibfnamefont {A.}~\bibnamefont {Stern}}, \bibinfo
  {author} {\bibfnamefont {M.}~\bibnamefont {Freedman}},\ and\ \bibinfo
  {author} {\bibfnamefont {S.}~\bibnamefont {{Das Sarma}}},\ }\bibfield
  {title} {\bibinfo {title} {Non-{Abelian} {Anyons} and {Topological} {Quantum}
  {Computation}},\ }\href {https://doi.org/10.1103/RevModPhys.80.1083}
  {\bibfield  {journal} {\bibinfo  {journal} {Rev. Mod. Phys.}\ }\textbf
  {\bibinfo {volume} {80}},\ \bibinfo {pages} {1083} (\bibinfo {year}
  {2008})}\BibitemShut {NoStop}%
\bibitem [{\citenamefont {Goldman}\ \emph {et~al.}(2016)\citenamefont
  {Goldman}, \citenamefont {Budich},\ and\ \citenamefont
  {Zoller}}]{goldman_topological_2016}%
  \BibitemOpen
  \bibfield  {author} {\bibinfo {author} {\bibfnamefont {N.}~\bibnamefont
  {Goldman}}, \bibinfo {author} {\bibfnamefont {J.~C.}\ \bibnamefont
  {Budich}},\ and\ \bibinfo {author} {\bibfnamefont {P.}~\bibnamefont
  {Zoller}},\ }\bibfield  {title} {\bibinfo {title} {Topological quantum matter
  with ultracold gases in optical lattices},\ }\href
  {https://doi.org/10.1038/nphys3803} {\bibfield  {journal} {\bibinfo
  {journal} {Nat. Phys.}\ }\textbf {\bibinfo {volume} {12}},\ \bibinfo {pages}
  {639} (\bibinfo {year} {2016})}\BibitemShut {NoStop}%
\bibitem [{\citenamefont {Cooper}\ \emph {et~al.}(2019)\citenamefont {Cooper},
  \citenamefont {Dalibard},\ and\ \citenamefont
  {Spielman}}]{cooper_topological_2019}%
  \BibitemOpen
  \bibfield  {author} {\bibinfo {author} {\bibfnamefont {N.~R.}\ \bibnamefont
  {Cooper}}, \bibinfo {author} {\bibfnamefont {J.}~\bibnamefont {Dalibard}},\
  and\ \bibinfo {author} {\bibfnamefont {I.~B.}\ \bibnamefont {Spielman}},\
  }\bibfield  {title} {\bibinfo {title} {Topological bands for ultracold
  atoms},\ }\href {https://doi.org/10.1103/RevModPhys.91.015005} {\bibfield
  {journal} {\bibinfo  {journal} {Rev. Mod. Phys.}\ }\textbf {\bibinfo {volume}
  {91}},\ \bibinfo {pages} {015005} (\bibinfo {year} {2019})}\BibitemShut
  {NoStop}%
\bibitem [{\citenamefont {Ozawa}\ \emph {et~al.}(2019)\citenamefont {Ozawa},
  \citenamefont {Price}, \citenamefont {Amo}, \citenamefont {Goldman},
  \citenamefont {Hafezi}, \citenamefont {Lu}, \citenamefont {Rechtsman},
  \citenamefont {Schuster}, \citenamefont {Simon}, \citenamefont {Zilberberg},\
  and\ \citenamefont {Carusotto}}]{ozawa_topological_2019}%
  \BibitemOpen
  \bibfield  {author} {\bibinfo {author} {\bibfnamefont {T.}~\bibnamefont
  {Ozawa}}, \bibinfo {author} {\bibfnamefont {H.~M.}\ \bibnamefont {Price}},
  \bibinfo {author} {\bibfnamefont {A.}~\bibnamefont {Amo}}, \bibinfo {author}
  {\bibfnamefont {N.}~\bibnamefont {Goldman}}, \bibinfo {author} {\bibfnamefont
  {M.}~\bibnamefont {Hafezi}}, \bibinfo {author} {\bibfnamefont
  {L.}~\bibnamefont {Lu}}, \bibinfo {author} {\bibfnamefont {M.~C.}\
  \bibnamefont {Rechtsman}}, \bibinfo {author} {\bibfnamefont {D.}~\bibnamefont
  {Schuster}}, \bibinfo {author} {\bibfnamefont {J.}~\bibnamefont {Simon}},
  \bibinfo {author} {\bibfnamefont {O.}~\bibnamefont {Zilberberg}},\ and\
  \bibinfo {author} {\bibfnamefont {I.}~\bibnamefont {Carusotto}},\ }\bibfield
  {title} {\bibinfo {title} {Topological photonics},\ }\href
  {https://doi.org/10.1103/RevModPhys.91.015006} {\bibfield  {journal}
  {\bibinfo  {journal} {Rev. Mod. Phys.}\ }\textbf {\bibinfo {volume} {91}},\
  \bibinfo {pages} {015006} (\bibinfo {year} {2019})}\BibitemShut {NoStop}%
\bibitem [{\citenamefont {Lu}\ \emph {et~al.}(2014)\citenamefont {Lu},
  \citenamefont {Joannopoulos},\ and\ \citenamefont
  {Soljačić}}]{lu_topological_2014}%
  \BibitemOpen
  \bibfield  {author} {\bibinfo {author} {\bibfnamefont {L.}~\bibnamefont
  {Lu}}, \bibinfo {author} {\bibfnamefont {J.~D.}\ \bibnamefont
  {Joannopoulos}},\ and\ \bibinfo {author} {\bibfnamefont {M.}~\bibnamefont
  {Soljačić}},\ }\bibfield  {title} {\bibinfo {title} {Topological
  photonics},\ }\href {https://doi.org/10.1038/nphoton.2014.248} {\bibfield
  {journal} {\bibinfo  {journal} {Nat. Photonics}\ }\textbf {\bibinfo {volume}
  {8}},\ \bibinfo {pages} {821} (\bibinfo {year} {2014})}\BibitemShut {NoStop}%
\bibitem [{\citenamefont {Malz}\ and\ \citenamefont
  {Smith}(2021)}]{malz_topological_2021}%
  \BibitemOpen
  \bibfield  {author} {\bibinfo {author} {\bibfnamefont {D.}~\bibnamefont
  {Malz}}\ and\ \bibinfo {author} {\bibfnamefont {A.}~\bibnamefont {Smith}},\
  }\bibfield  {title} {\bibinfo {title} {Topological {Two}-{Dimensional}
  {Floquet} {Lattice} on a {Single} {Superconducting} {Qubit}},\ }\href
  {https://doi.org/10.1103/PhysRevLett.126.163602} {\bibfield  {journal}
  {\bibinfo  {journal} {Phys. Rev. Lett.}\ }\textbf {\bibinfo {volume} {126}},\
  \bibinfo {pages} {163602} (\bibinfo {year} {2021})}\BibitemShut {NoStop}%
\bibitem [{\citenamefont {{Google Quantum AI and Collaborators}}\ \emph
  {et~al.}(2023)\citenamefont {{Google Quantum AI and Collaborators}},
  \citenamefont {Andersen}, \citenamefont {Lensky}, \citenamefont {Kechedzhi},
  \citenamefont {Drozdov}, \citenamefont {Bengtsson}, \citenamefont {Hong},
  \citenamefont {Morvan}, \citenamefont {Mi}, \citenamefont {Opremcak},
  \citenamefont {Acharya}, \citenamefont {Allen}, \citenamefont {Ansmann},
  \citenamefont {Arute}, \citenamefont {Arya}, \citenamefont {Asfaw},
  \citenamefont {Atalaya}, \citenamefont {Babbush}, \citenamefont {Bacon},
  \citenamefont {Bardin}, \citenamefont {Bortoli}, \citenamefont {Bourassa},
  \citenamefont {Bovaird}, \citenamefont {Brill}, \citenamefont {Broughton},
  \citenamefont {Buckley}, \citenamefont {Buell}, \citenamefont {Burger},
  \citenamefont {Burkett}, \citenamefont {Bushnell}, \citenamefont {Chen},
  \citenamefont {Chiaro}, \citenamefont {Chik}, \citenamefont {Chou},
  \citenamefont {Cogan}, \citenamefont {Collins}, \citenamefont {Conner},
  \citenamefont {Courtney}, \citenamefont {Crook}, \citenamefont {Curtin},
  \citenamefont {Debroy}, \citenamefont {Del Toro~Barba}, \citenamefont
  {Demura}, \citenamefont {Dunsworth}, \citenamefont {Eppens}, \citenamefont
  {Erickson}, \citenamefont {Faoro}, \citenamefont {Farhi}, \citenamefont
  {Fatemi}, \citenamefont {Ferreira}, \citenamefont {Burgos}, \citenamefont
  {Forati}, \citenamefont {Fowler}, \citenamefont {Foxen}, \citenamefont
  {Giang}, \citenamefont {Gidney}, \citenamefont {Gilboa}, \citenamefont
  {Giustina}, \citenamefont {Gosula}, \citenamefont {Dau}, \citenamefont
  {Gross}, \citenamefont {Habegger}, \citenamefont {Hamilton}, \citenamefont
  {Hansen}, \citenamefont {Harrigan}, \citenamefont {Harrington}, \citenamefont
  {Heu}, \citenamefont {Hilton}, \citenamefont {Hoffmann}, \citenamefont
  {Huang}, \citenamefont {Huff}, \citenamefont {Huggins}, \citenamefont
  {Ioffe}, \citenamefont {Isakov}, \citenamefont {Iveland}, \citenamefont
  {Jeffrey}, \citenamefont {Jiang}, \citenamefont {Jones}, \citenamefont
  {Juhas}, \citenamefont {Kafri}, \citenamefont {Khattar}, \citenamefont
  {Khezri}, \citenamefont {Kieferová}, \citenamefont {Kim}, \citenamefont
  {Kitaev}, \citenamefont {Klimov}, \citenamefont {Klots}, \citenamefont
  {Korotkov}, \citenamefont {Kostritsa}, \citenamefont {Kreikebaum},
  \citenamefont {Landhuis}, \citenamefont {Laptev}, \citenamefont {Lau},
  \citenamefont {Laws}, \citenamefont {Lee}, \citenamefont {Lee}, \citenamefont
  {Lester}, \citenamefont {Lill}, \citenamefont {Liu}, \citenamefont
  {Locharla}, \citenamefont {Lucero}, \citenamefont {Malone}, \citenamefont
  {Martin}, \citenamefont {McClean}, \citenamefont {McCourt}, \citenamefont
  {McEwen}, \citenamefont {Miao}, \citenamefont {Mieszala}, \citenamefont
  {Mohseni}, \citenamefont {Montazeri}, \citenamefont {Mount}, \citenamefont
  {Movassagh}, \citenamefont {Mruczkiewicz}, \citenamefont {Naaman},
  \citenamefont {Neeley}, \citenamefont {Neill}, \citenamefont {Nersisyan},
  \citenamefont {Newman}, \citenamefont {Ng}, \citenamefont {Nguyen},
  \citenamefont {Nguyen}, \citenamefont {Niu}, \citenamefont {O’Brien},
  \citenamefont {Omonije}, \citenamefont {Petukhov}, \citenamefont {Potter},
  \citenamefont {Pryadko}, \citenamefont {Quintana}, \citenamefont {Rocque},
  \citenamefont {Rubin}, \citenamefont {Saei}, \citenamefont {Sank},
  \citenamefont {Sankaragomathi}, \citenamefont {Satzinger}, \citenamefont
  {Schurkus}, \citenamefont {Schuster}, \citenamefont {Shearn}, \citenamefont
  {Shorter}, \citenamefont {Shutty}, \citenamefont {Shvarts}, \citenamefont
  {Skruzny}, \citenamefont {Smith}, \citenamefont {Somma}, \citenamefont
  {Sterling}, \citenamefont {Strain}, \citenamefont {Szalay}, \citenamefont
  {Torres}, \citenamefont {Vidal}, \citenamefont {Villalonga}, \citenamefont
  {Heidweiller}, \citenamefont {White}, \citenamefont {Woo}, \citenamefont
  {Xing}, \citenamefont {Yao}, \citenamefont {Yeh}, \citenamefont {Yoo},
  \citenamefont {Young}, \citenamefont {Zalcman}, \citenamefont {Zhang},
  \citenamefont {Zhu}, \citenamefont {Zobrist}, \citenamefont {Neven},
  \citenamefont {Boixo}, \citenamefont {Megrant}, \citenamefont {Kelly},
  \citenamefont {Chen}, \citenamefont {Smelyanskiy}, \citenamefont {Kim},
  \citenamefont {Aleiner},\ and\ \citenamefont
  {Roushan}}]{google_quantum_ai_and_collaborators_non-abelian_2023}%
  \BibitemOpen
  \bibfield  {author} {\bibinfo {author} {\bibnamefont {{Google Quantum AI and
  Collaborators}}}, \bibinfo {author} {\bibfnamefont {T.~I.}\ \bibnamefont
  {Andersen}}, \bibinfo {author} {\bibfnamefont {Y.~D.}\ \bibnamefont
  {Lensky}}, \bibinfo {author} {\bibfnamefont {K.}~\bibnamefont {Kechedzhi}},
  \bibinfo {author} {\bibfnamefont {I.~K.}\ \bibnamefont {Drozdov}}, \bibinfo
  {author} {\bibfnamefont {A.}~\bibnamefont {Bengtsson}}, \bibinfo {author}
  {\bibfnamefont {S.}~\bibnamefont {Hong}}, \bibinfo {author} {\bibfnamefont
  {A.}~\bibnamefont {Morvan}}, \bibinfo {author} {\bibfnamefont
  {X.}~\bibnamefont {Mi}}, \bibinfo {author} {\bibfnamefont {A.}~\bibnamefont
  {Opremcak}}, \bibinfo {author} {\bibfnamefont {R.}~\bibnamefont {Acharya}},
  \bibinfo {author} {\bibfnamefont {R.}~\bibnamefont {Allen}}, \bibinfo
  {author} {\bibfnamefont {M.}~\bibnamefont {Ansmann}}, \bibinfo {author}
  {\bibfnamefont {F.}~\bibnamefont {Arute}}, \bibinfo {author} {\bibfnamefont
  {K.}~\bibnamefont {Arya}}, \bibinfo {author} {\bibfnamefont {A.}~\bibnamefont
  {Asfaw}}, \bibinfo {author} {\bibfnamefont {J.}~\bibnamefont {Atalaya}},
  \bibinfo {author} {\bibfnamefont {R.}~\bibnamefont {Babbush}}, \bibinfo
  {author} {\bibfnamefont {D.}~\bibnamefont {Bacon}}, \bibinfo {author}
  {\bibfnamefont {J.~C.}\ \bibnamefont {Bardin}}, \bibinfo {author}
  {\bibfnamefont {G.}~\bibnamefont {Bortoli}}, \bibinfo {author} {\bibfnamefont
  {A.}~\bibnamefont {Bourassa}}, \bibinfo {author} {\bibfnamefont
  {J.}~\bibnamefont {Bovaird}}, \bibinfo {author} {\bibfnamefont
  {L.}~\bibnamefont {Brill}}, \bibinfo {author} {\bibfnamefont
  {M.}~\bibnamefont {Broughton}}, \bibinfo {author} {\bibfnamefont {B.~B.}\
  \bibnamefont {Buckley}}, \bibinfo {author} {\bibfnamefont {D.~A.}\
  \bibnamefont {Buell}}, \bibinfo {author} {\bibfnamefont {T.}~\bibnamefont
  {Burger}}, \bibinfo {author} {\bibfnamefont {B.}~\bibnamefont {Burkett}},
  \bibinfo {author} {\bibfnamefont {N.}~\bibnamefont {Bushnell}}, \bibinfo
  {author} {\bibfnamefont {Z.}~\bibnamefont {Chen}}, \bibinfo {author}
  {\bibfnamefont {B.}~\bibnamefont {Chiaro}}, \bibinfo {author} {\bibfnamefont
  {D.}~\bibnamefont {Chik}}, \bibinfo {author} {\bibfnamefont {C.}~\bibnamefont
  {Chou}}, \bibinfo {author} {\bibfnamefont {J.}~\bibnamefont {Cogan}},
  \bibinfo {author} {\bibfnamefont {R.}~\bibnamefont {Collins}}, \bibinfo
  {author} {\bibfnamefont {P.}~\bibnamefont {Conner}}, \bibinfo {author}
  {\bibfnamefont {W.}~\bibnamefont {Courtney}}, \bibinfo {author}
  {\bibfnamefont {A.~L.}\ \bibnamefont {Crook}}, \bibinfo {author}
  {\bibfnamefont {B.}~\bibnamefont {Curtin}}, \bibinfo {author} {\bibfnamefont
  {D.~M.}\ \bibnamefont {Debroy}}, \bibinfo {author} {\bibfnamefont
  {A.}~\bibnamefont {Del Toro~Barba}}, \bibinfo {author} {\bibfnamefont
  {S.}~\bibnamefont {Demura}}, \bibinfo {author} {\bibfnamefont
  {A.}~\bibnamefont {Dunsworth}}, \bibinfo {author} {\bibfnamefont
  {D.}~\bibnamefont {Eppens}}, \bibinfo {author} {\bibfnamefont
  {C.}~\bibnamefont {Erickson}}, \bibinfo {author} {\bibfnamefont
  {L.}~\bibnamefont {Faoro}}, \bibinfo {author} {\bibfnamefont
  {E.}~\bibnamefont {Farhi}}, \bibinfo {author} {\bibfnamefont
  {R.}~\bibnamefont {Fatemi}}, \bibinfo {author} {\bibfnamefont {V.~S.}\
  \bibnamefont {Ferreira}}, \bibinfo {author} {\bibfnamefont {L.~F.}\
  \bibnamefont {Burgos}}, \bibinfo {author} {\bibfnamefont {E.}~\bibnamefont
  {Forati}}, \bibinfo {author} {\bibfnamefont {A.~G.}\ \bibnamefont {Fowler}},
  \bibinfo {author} {\bibfnamefont {B.}~\bibnamefont {Foxen}}, \bibinfo
  {author} {\bibfnamefont {W.}~\bibnamefont {Giang}}, \bibinfo {author}
  {\bibfnamefont {C.}~\bibnamefont {Gidney}}, \bibinfo {author} {\bibfnamefont
  {D.}~\bibnamefont {Gilboa}}, \bibinfo {author} {\bibfnamefont
  {M.}~\bibnamefont {Giustina}}, \bibinfo {author} {\bibfnamefont
  {R.}~\bibnamefont {Gosula}}, \bibinfo {author} {\bibfnamefont {A.~G.}\
  \bibnamefont {Dau}}, \bibinfo {author} {\bibfnamefont {J.~A.}\ \bibnamefont
  {Gross}}, \bibinfo {author} {\bibfnamefont {S.}~\bibnamefont {Habegger}},
  \bibinfo {author} {\bibfnamefont {M.~C.}\ \bibnamefont {Hamilton}}, \bibinfo
  {author} {\bibfnamefont {M.}~\bibnamefont {Hansen}}, \bibinfo {author}
  {\bibfnamefont {M.~P.}\ \bibnamefont {Harrigan}}, \bibinfo {author}
  {\bibfnamefont {S.~D.}\ \bibnamefont {Harrington}}, \bibinfo {author}
  {\bibfnamefont {P.}~\bibnamefont {Heu}}, \bibinfo {author} {\bibfnamefont
  {J.}~\bibnamefont {Hilton}}, \bibinfo {author} {\bibfnamefont {M.~R.}\
  \bibnamefont {Hoffmann}}, \bibinfo {author} {\bibfnamefont {T.}~\bibnamefont
  {Huang}}, \bibinfo {author} {\bibfnamefont {A.}~\bibnamefont {Huff}},
  \bibinfo {author} {\bibfnamefont {W.~J.}\ \bibnamefont {Huggins}}, \bibinfo
  {author} {\bibfnamefont {L.~B.}\ \bibnamefont {Ioffe}}, \bibinfo {author}
  {\bibfnamefont {S.~V.}\ \bibnamefont {Isakov}}, \bibinfo {author}
  {\bibfnamefont {J.}~\bibnamefont {Iveland}}, \bibinfo {author} {\bibfnamefont
  {E.}~\bibnamefont {Jeffrey}}, \bibinfo {author} {\bibfnamefont
  {Z.}~\bibnamefont {Jiang}}, \bibinfo {author} {\bibfnamefont
  {C.}~\bibnamefont {Jones}}, \bibinfo {author} {\bibfnamefont
  {P.}~\bibnamefont {Juhas}}, \bibinfo {author} {\bibfnamefont
  {D.}~\bibnamefont {Kafri}}, \bibinfo {author} {\bibfnamefont
  {T.}~\bibnamefont {Khattar}}, \bibinfo {author} {\bibfnamefont
  {M.}~\bibnamefont {Khezri}}, \bibinfo {author} {\bibfnamefont
  {M.}~\bibnamefont {Kieferová}}, \bibinfo {author} {\bibfnamefont
  {S.}~\bibnamefont {Kim}}, \bibinfo {author} {\bibfnamefont {A.}~\bibnamefont
  {Kitaev}}, \bibinfo {author} {\bibfnamefont {P.~V.}\ \bibnamefont {Klimov}},
  \bibinfo {author} {\bibfnamefont {A.~R.}\ \bibnamefont {Klots}}, \bibinfo
  {author} {\bibfnamefont {A.~N.}\ \bibnamefont {Korotkov}}, \bibinfo {author}
  {\bibfnamefont {F.}~\bibnamefont {Kostritsa}}, \bibinfo {author}
  {\bibfnamefont {J.~M.}\ \bibnamefont {Kreikebaum}}, \bibinfo {author}
  {\bibfnamefont {D.}~\bibnamefont {Landhuis}}, \bibinfo {author}
  {\bibfnamefont {P.}~\bibnamefont {Laptev}}, \bibinfo {author} {\bibfnamefont
  {K.-M.}\ \bibnamefont {Lau}}, \bibinfo {author} {\bibfnamefont
  {L.}~\bibnamefont {Laws}}, \bibinfo {author} {\bibfnamefont {J.}~\bibnamefont
  {Lee}}, \bibinfo {author} {\bibfnamefont {K.~W.}\ \bibnamefont {Lee}},
  \bibinfo {author} {\bibfnamefont {B.~J.}\ \bibnamefont {Lester}}, \bibinfo
  {author} {\bibfnamefont {A.~T.}\ \bibnamefont {Lill}}, \bibinfo {author}
  {\bibfnamefont {W.}~\bibnamefont {Liu}}, \bibinfo {author} {\bibfnamefont
  {A.}~\bibnamefont {Locharla}}, \bibinfo {author} {\bibfnamefont
  {E.}~\bibnamefont {Lucero}}, \bibinfo {author} {\bibfnamefont {F.~D.}\
  \bibnamefont {Malone}}, \bibinfo {author} {\bibfnamefont {O.}~\bibnamefont
  {Martin}}, \bibinfo {author} {\bibfnamefont {J.~R.}\ \bibnamefont {McClean}},
  \bibinfo {author} {\bibfnamefont {T.}~\bibnamefont {McCourt}}, \bibinfo
  {author} {\bibfnamefont {M.}~\bibnamefont {McEwen}}, \bibinfo {author}
  {\bibfnamefont {K.~C.}\ \bibnamefont {Miao}}, \bibinfo {author}
  {\bibfnamefont {A.}~\bibnamefont {Mieszala}}, \bibinfo {author}
  {\bibfnamefont {M.}~\bibnamefont {Mohseni}}, \bibinfo {author} {\bibfnamefont
  {S.}~\bibnamefont {Montazeri}}, \bibinfo {author} {\bibfnamefont
  {E.}~\bibnamefont {Mount}}, \bibinfo {author} {\bibfnamefont
  {R.}~\bibnamefont {Movassagh}}, \bibinfo {author} {\bibfnamefont
  {W.}~\bibnamefont {Mruczkiewicz}}, \bibinfo {author} {\bibfnamefont
  {O.}~\bibnamefont {Naaman}}, \bibinfo {author} {\bibfnamefont
  {M.}~\bibnamefont {Neeley}}, \bibinfo {author} {\bibfnamefont
  {C.}~\bibnamefont {Neill}}, \bibinfo {author} {\bibfnamefont
  {A.}~\bibnamefont {Nersisyan}}, \bibinfo {author} {\bibfnamefont
  {M.}~\bibnamefont {Newman}}, \bibinfo {author} {\bibfnamefont {J.~H.}\
  \bibnamefont {Ng}}, \bibinfo {author} {\bibfnamefont {A.}~\bibnamefont
  {Nguyen}}, \bibinfo {author} {\bibfnamefont {M.}~\bibnamefont {Nguyen}},
  \bibinfo {author} {\bibfnamefont {M.~Y.}\ \bibnamefont {Niu}}, \bibinfo
  {author} {\bibfnamefont {T.~E.}\ \bibnamefont {O’Brien}}, \bibinfo {author}
  {\bibfnamefont {S.}~\bibnamefont {Omonije}}, \bibinfo {author} {\bibfnamefont
  {A.}~\bibnamefont {Petukhov}}, \bibinfo {author} {\bibfnamefont
  {R.}~\bibnamefont {Potter}}, \bibinfo {author} {\bibfnamefont {L.~P.}\
  \bibnamefont {Pryadko}}, \bibinfo {author} {\bibfnamefont {C.}~\bibnamefont
  {Quintana}}, \bibinfo {author} {\bibfnamefont {C.}~\bibnamefont {Rocque}},
  \bibinfo {author} {\bibfnamefont {N.~C.}\ \bibnamefont {Rubin}}, \bibinfo
  {author} {\bibfnamefont {N.}~\bibnamefont {Saei}}, \bibinfo {author}
  {\bibfnamefont {D.}~\bibnamefont {Sank}}, \bibinfo {author} {\bibfnamefont
  {K.}~\bibnamefont {Sankaragomathi}}, \bibinfo {author} {\bibfnamefont
  {K.~J.}\ \bibnamefont {Satzinger}}, \bibinfo {author} {\bibfnamefont {H.~F.}\
  \bibnamefont {Schurkus}}, \bibinfo {author} {\bibfnamefont {C.}~\bibnamefont
  {Schuster}}, \bibinfo {author} {\bibfnamefont {M.~J.}\ \bibnamefont
  {Shearn}}, \bibinfo {author} {\bibfnamefont {A.}~\bibnamefont {Shorter}},
  \bibinfo {author} {\bibfnamefont {N.}~\bibnamefont {Shutty}}, \bibinfo
  {author} {\bibfnamefont {V.}~\bibnamefont {Shvarts}}, \bibinfo {author}
  {\bibfnamefont {J.}~\bibnamefont {Skruzny}}, \bibinfo {author} {\bibfnamefont
  {W.~C.}\ \bibnamefont {Smith}}, \bibinfo {author} {\bibfnamefont
  {R.}~\bibnamefont {Somma}}, \bibinfo {author} {\bibfnamefont
  {G.}~\bibnamefont {Sterling}}, \bibinfo {author} {\bibfnamefont
  {D.}~\bibnamefont {Strain}}, \bibinfo {author} {\bibfnamefont
  {M.}~\bibnamefont {Szalay}}, \bibinfo {author} {\bibfnamefont
  {A.}~\bibnamefont {Torres}}, \bibinfo {author} {\bibfnamefont
  {G.}~\bibnamefont {Vidal}}, \bibinfo {author} {\bibfnamefont
  {B.}~\bibnamefont {Villalonga}}, \bibinfo {author} {\bibfnamefont {C.~V.}\
  \bibnamefont {Heidweiller}}, \bibinfo {author} {\bibfnamefont
  {T.}~\bibnamefont {White}}, \bibinfo {author} {\bibfnamefont {B.~W.~K.}\
  \bibnamefont {Woo}}, \bibinfo {author} {\bibfnamefont {C.}~\bibnamefont
  {Xing}}, \bibinfo {author} {\bibfnamefont {Z.~J.}\ \bibnamefont {Yao}},
  \bibinfo {author} {\bibfnamefont {P.}~\bibnamefont {Yeh}}, \bibinfo {author}
  {\bibfnamefont {J.}~\bibnamefont {Yoo}}, \bibinfo {author} {\bibfnamefont
  {G.}~\bibnamefont {Young}}, \bibinfo {author} {\bibfnamefont
  {A.}~\bibnamefont {Zalcman}}, \bibinfo {author} {\bibfnamefont
  {Y.}~\bibnamefont {Zhang}}, \bibinfo {author} {\bibfnamefont
  {N.}~\bibnamefont {Zhu}}, \bibinfo {author} {\bibfnamefont {N.}~\bibnamefont
  {Zobrist}}, \bibinfo {author} {\bibfnamefont {H.}~\bibnamefont {Neven}},
  \bibinfo {author} {\bibfnamefont {S.}~\bibnamefont {Boixo}}, \bibinfo
  {author} {\bibfnamefont {A.}~\bibnamefont {Megrant}}, \bibinfo {author}
  {\bibfnamefont {J.}~\bibnamefont {Kelly}}, \bibinfo {author} {\bibfnamefont
  {Y.}~\bibnamefont {Chen}}, \bibinfo {author} {\bibfnamefont {V.}~\bibnamefont
  {Smelyanskiy}}, \bibinfo {author} {\bibfnamefont {E.-A.}\ \bibnamefont
  {Kim}}, \bibinfo {author} {\bibfnamefont {I.}~\bibnamefont {Aleiner}},\ and\
  \bibinfo {author} {\bibfnamefont {P.}~\bibnamefont {Roushan}},\ }\bibfield
  {title} {\bibinfo {title} {Non-{Abelian} braiding of graph vertices in a
  superconducting processor},\ }\href
  {https://doi.org/10.1038/s41586-023-05954-4} {\bibfield  {journal} {\bibinfo
  {journal} {Nature}\ }\textbf {\bibinfo {volume} {618}},\ \bibinfo {pages}
  {264} (\bibinfo {year} {2023})}\BibitemShut {NoStop}%
\bibitem [{\citenamefont {De~Léséleuc}\ \emph {et~al.}(2019)\citenamefont
  {De~Léséleuc}, \citenamefont {Lienhard}, \citenamefont {Scholl},
  \citenamefont {Barredo}, \citenamefont {Weber}, \citenamefont {Lang},
  \citenamefont {Büchler}, \citenamefont {Lahaye},\ and\ \citenamefont
  {Browaeys}}]{de_leseleuc_observation_2019}%
  \BibitemOpen
  \bibfield  {author} {\bibinfo {author} {\bibfnamefont {S.}~\bibnamefont
  {De~Léséleuc}}, \bibinfo {author} {\bibfnamefont {V.}~\bibnamefont
  {Lienhard}}, \bibinfo {author} {\bibfnamefont {P.}~\bibnamefont {Scholl}},
  \bibinfo {author} {\bibfnamefont {D.}~\bibnamefont {Barredo}}, \bibinfo
  {author} {\bibfnamefont {S.}~\bibnamefont {Weber}}, \bibinfo {author}
  {\bibfnamefont {N.}~\bibnamefont {Lang}}, \bibinfo {author} {\bibfnamefont
  {H.~P.}\ \bibnamefont {Büchler}}, \bibinfo {author} {\bibfnamefont
  {T.}~\bibnamefont {Lahaye}},\ and\ \bibinfo {author} {\bibfnamefont
  {A.}~\bibnamefont {Browaeys}},\ }\bibfield  {title} {\bibinfo {title}
  {Observation of a symmetry-protected topological phase of interacting bosons
  with {Rydberg} atoms},\ }\href {https://doi.org/10.1126/science.aav9105}
  {\bibfield  {journal} {\bibinfo  {journal} {Science}\ }\textbf {\bibinfo
  {volume} {365}},\ \bibinfo {pages} {775} (\bibinfo {year}
  {2019})}\BibitemShut {NoStop}%
\bibitem [{\citenamefont {Dumitrescu}\ \emph {et~al.}(2022)\citenamefont
  {Dumitrescu}, \citenamefont {Bohnet}, \citenamefont {Gaebler}, \citenamefont
  {Hankin}, \citenamefont {Hayes}, \citenamefont {Kumar}, \citenamefont
  {Neyenhuis}, \citenamefont {Vasseur},\ and\ \citenamefont
  {Potter}}]{dumitrescu_dynamical_2022}%
  \BibitemOpen
  \bibfield  {author} {\bibinfo {author} {\bibfnamefont {P.~T.}\ \bibnamefont
  {Dumitrescu}}, \bibinfo {author} {\bibfnamefont {J.~G.}\ \bibnamefont
  {Bohnet}}, \bibinfo {author} {\bibfnamefont {J.~P.}\ \bibnamefont {Gaebler}},
  \bibinfo {author} {\bibfnamefont {A.}~\bibnamefont {Hankin}}, \bibinfo
  {author} {\bibfnamefont {D.}~\bibnamefont {Hayes}}, \bibinfo {author}
  {\bibfnamefont {A.}~\bibnamefont {Kumar}}, \bibinfo {author} {\bibfnamefont
  {B.}~\bibnamefont {Neyenhuis}}, \bibinfo {author} {\bibfnamefont
  {R.}~\bibnamefont {Vasseur}},\ and\ \bibinfo {author} {\bibfnamefont {A.~C.}\
  \bibnamefont {Potter}},\ }\bibfield  {title} {\bibinfo {title} {Dynamical
  topological phase realized in a trapped-ion quantum simulator},\ }\href
  {https://doi.org/10.1038/s41586-022-04853-4} {\bibfield  {journal} {\bibinfo
  {journal} {Nature}\ }\textbf {\bibinfo {volume} {607}},\ \bibinfo {pages}
  {463} (\bibinfo {year} {2022})}\BibitemShut {NoStop}%
\bibitem [{\citenamefont {Iqbal}\ \emph {et~al.}(2024)\citenamefont {Iqbal},
  \citenamefont {Tantivasadakarn}, \citenamefont {Verresen}, \citenamefont
  {Campbell}, \citenamefont {Dreiling}, \citenamefont {Figgatt}, \citenamefont
  {Gaebler}, \citenamefont {Johansen}, \citenamefont {Mills}, \citenamefont
  {Moses}, \citenamefont {Pino}, \citenamefont {Ransford}, \citenamefont
  {Rowe}, \citenamefont {Siegfried}, \citenamefont {Stutz}, \citenamefont
  {Foss-Feig}, \citenamefont {Vishwanath},\ and\ \citenamefont
  {Dreyer}}]{iqbal_non-abelian_2024}%
  \BibitemOpen
  \bibfield  {author} {\bibinfo {author} {\bibfnamefont {M.}~\bibnamefont
  {Iqbal}}, \bibinfo {author} {\bibfnamefont {N.}~\bibnamefont
  {Tantivasadakarn}}, \bibinfo {author} {\bibfnamefont {R.}~\bibnamefont
  {Verresen}}, \bibinfo {author} {\bibfnamefont {S.~L.}\ \bibnamefont
  {Campbell}}, \bibinfo {author} {\bibfnamefont {J.~M.}\ \bibnamefont
  {Dreiling}}, \bibinfo {author} {\bibfnamefont {C.}~\bibnamefont {Figgatt}},
  \bibinfo {author} {\bibfnamefont {J.~P.}\ \bibnamefont {Gaebler}}, \bibinfo
  {author} {\bibfnamefont {J.}~\bibnamefont {Johansen}}, \bibinfo {author}
  {\bibfnamefont {M.}~\bibnamefont {Mills}}, \bibinfo {author} {\bibfnamefont
  {S.~A.}\ \bibnamefont {Moses}}, \bibinfo {author} {\bibfnamefont {J.~M.}\
  \bibnamefont {Pino}}, \bibinfo {author} {\bibfnamefont {A.}~\bibnamefont
  {Ransford}}, \bibinfo {author} {\bibfnamefont {M.}~\bibnamefont {Rowe}},
  \bibinfo {author} {\bibfnamefont {P.}~\bibnamefont {Siegfried}}, \bibinfo
  {author} {\bibfnamefont {R.~P.}\ \bibnamefont {Stutz}}, \bibinfo {author}
  {\bibfnamefont {M.}~\bibnamefont {Foss-Feig}}, \bibinfo {author}
  {\bibfnamefont {A.}~\bibnamefont {Vishwanath}},\ and\ \bibinfo {author}
  {\bibfnamefont {H.}~\bibnamefont {Dreyer}},\ }\bibfield  {title} {\bibinfo
  {title} {Non-{Abelian} topological order and anyons on a trapped-ion
  processor},\ }\href {https://doi.org/10.1038/s41586-023-06934-4} {\bibfield
  {journal} {\bibinfo  {journal} {Nature}\ }\textbf {\bibinfo {volume} {626}},\
  \bibinfo {pages} {505} (\bibinfo {year} {2024})}\BibitemShut {NoStop}%
\bibitem [{\citenamefont {Katz}\ \emph {et~al.}(2025)\citenamefont {Katz},
  \citenamefont {Feng}, \citenamefont {Porras},\ and\ \citenamefont
  {Monroe}}]{katz_floquet_2025}%
  \BibitemOpen
  \bibfield  {author} {\bibinfo {author} {\bibfnamefont {O.}~\bibnamefont
  {Katz}}, \bibinfo {author} {\bibfnamefont {L.}~\bibnamefont {Feng}}, \bibinfo
  {author} {\bibfnamefont {D.}~\bibnamefont {Porras}},\ and\ \bibinfo {author}
  {\bibfnamefont {C.}~\bibnamefont {Monroe}},\ }\bibfield  {title} {\bibinfo
  {title} {Floquet control of interactions and edge states in a programmable
  quantum simulator},\ }\href {https://doi.org/10.1038/s41467-025-62897-2}
  {\bibfield  {journal} {\bibinfo  {journal} {Nat. Commun.}\ }\textbf {\bibinfo
  {volume} {16}},\ \bibinfo {pages} {8815} (\bibinfo {year}
  {2025})}\BibitemShut {NoStop}%
\bibitem [{\citenamefont {Li}\ \emph {et~al.}(2009)\citenamefont {Li},
  \citenamefont {Chu}, \citenamefont {Jain},\ and\ \citenamefont
  {Shen}}]{li_topological_2009}%
  \BibitemOpen
  \bibfield  {author} {\bibinfo {author} {\bibfnamefont {J.}~\bibnamefont
  {Li}}, \bibinfo {author} {\bibfnamefont {R.-L.}\ \bibnamefont {Chu}},
  \bibinfo {author} {\bibfnamefont {J.~K.}\ \bibnamefont {Jain}},\ and\
  \bibinfo {author} {\bibfnamefont {S.-Q.}\ \bibnamefont {Shen}},\ }\bibfield
  {title} {\bibinfo {title} {Topological {Anderson} {Insulator}},\ }\href
  {https://doi.org/10.1103/PhysRevLett.102.136806} {\bibfield  {journal}
  {\bibinfo  {journal} {Phys. Rev. Lett.}\ }\textbf {\bibinfo {volume} {102}},\
  \bibinfo {pages} {136806} (\bibinfo {year} {2009})}\BibitemShut {NoStop}%
\bibitem [{\citenamefont {Groth}\ \emph {et~al.}(2009)\citenamefont {Groth},
  \citenamefont {Wimmer}, \citenamefont {Akhmerov}, \citenamefont
  {Tworzydło},\ and\ \citenamefont {Beenakker}}]{groth_theory_2009}%
  \BibitemOpen
  \bibfield  {author} {\bibinfo {author} {\bibfnamefont {C.~W.}\ \bibnamefont
  {Groth}}, \bibinfo {author} {\bibfnamefont {M.}~\bibnamefont {Wimmer}},
  \bibinfo {author} {\bibfnamefont {A.~R.}\ \bibnamefont {Akhmerov}}, \bibinfo
  {author} {\bibfnamefont {J.}~\bibnamefont {Tworzydło}},\ and\ \bibinfo
  {author} {\bibfnamefont {C.~W.~J.}\ \bibnamefont {Beenakker}},\ }\bibfield
  {title} {\bibinfo {title} {Theory of the {Topological} {Anderson}
  {Insulator}},\ }\href {https://doi.org/10.1103/PhysRevLett.103.196805}
  {\bibfield  {journal} {\bibinfo  {journal} {Phys. Rev. Lett.}\ }\textbf
  {\bibinfo {volume} {103}},\ \bibinfo {pages} {196805} (\bibinfo {year}
  {2009})}\BibitemShut {NoStop}%
\bibitem [{\citenamefont {Meier}\ \emph {et~al.}(2018)\citenamefont {Meier},
  \citenamefont {An}, \citenamefont {Dauphin}, \citenamefont {Maffei},
  \citenamefont {Massignan}, \citenamefont {Hughes},\ and\ \citenamefont
  {Gadway}}]{meier_observation_2018}%
  \BibitemOpen
  \bibfield  {author} {\bibinfo {author} {\bibfnamefont {E.~J.}\ \bibnamefont
  {Meier}}, \bibinfo {author} {\bibfnamefont {F.~A.}\ \bibnamefont {An}},
  \bibinfo {author} {\bibfnamefont {A.}~\bibnamefont {Dauphin}}, \bibinfo
  {author} {\bibfnamefont {M.}~\bibnamefont {Maffei}}, \bibinfo {author}
  {\bibfnamefont {P.}~\bibnamefont {Massignan}}, \bibinfo {author}
  {\bibfnamefont {T.~L.}\ \bibnamefont {Hughes}},\ and\ \bibinfo {author}
  {\bibfnamefont {B.}~\bibnamefont {Gadway}},\ }\bibfield  {title} {\bibinfo
  {title} {Observation of the topological {Anderson} insulator in disordered
  atomic wires},\ }\href {https://doi.org/10.1126/science.aat3406} {\bibfield
  {journal} {\bibinfo  {journal} {Science}\ }\textbf {\bibinfo {volume}
  {362}},\ \bibinfo {pages} {929} (\bibinfo {year} {2018})}\BibitemShut
  {NoStop}%
\bibitem [{\citenamefont {Agarwala}\ and\ \citenamefont
  {Shenoy}(2017)}]{agarwala_topological_2017}%
  \BibitemOpen
  \bibfield  {author} {\bibinfo {author} {\bibfnamefont {A.}~\bibnamefont
  {Agarwala}}\ and\ \bibinfo {author} {\bibfnamefont {V.~B.}\ \bibnamefont
  {Shenoy}},\ }\bibfield  {title} {\bibinfo {title} {Topological {Insulators}
  in {Amorphous} {Systems}},\ }\href
  {https://doi.org/10.1103/PhysRevLett.118.236402} {\bibfield  {journal}
  {\bibinfo  {journal} {Phys. Rev. Lett.}\ }\textbf {\bibinfo {volume} {118}},\
  \bibinfo {pages} {236402} (\bibinfo {year} {2017})}\BibitemShut {NoStop}%
\bibitem [{\citenamefont {Mitchell}\ \emph {et~al.}(2018)\citenamefont
  {Mitchell}, \citenamefont {Nash}, \citenamefont {Hexner}, \citenamefont
  {Turner},\ and\ \citenamefont {Irvine}}]{mitchell_amorphous_2018}%
  \BibitemOpen
  \bibfield  {author} {\bibinfo {author} {\bibfnamefont {N.~P.}\ \bibnamefont
  {Mitchell}}, \bibinfo {author} {\bibfnamefont {L.~M.}\ \bibnamefont {Nash}},
  \bibinfo {author} {\bibfnamefont {D.}~\bibnamefont {Hexner}}, \bibinfo
  {author} {\bibfnamefont {A.~M.}\ \bibnamefont {Turner}},\ and\ \bibinfo
  {author} {\bibfnamefont {W.~T.~M.}\ \bibnamefont {Irvine}},\ }\bibfield
  {title} {\bibinfo {title} {Amorphous topological insulators constructed from
  random point sets},\ }\href {https://doi.org/10.1038/s41567-017-0024-5}
  {\bibfield  {journal} {\bibinfo  {journal} {Nat. Phys.}\ }\textbf {\bibinfo
  {volume} {14}},\ \bibinfo {pages} {380} (\bibinfo {year} {2018})}\BibitemShut
  {NoStop}%
\bibitem [{\citenamefont {Xie}\ \emph {et~al.}(2014)\citenamefont {Xie},
  \citenamefont {Cui}, \citenamefont {Cao}, \citenamefont {Amico},\ and\
  \citenamefont {Fan}}]{xie_anisotropic_2014}%
  \BibitemOpen
  \bibfield  {author} {\bibinfo {author} {\bibfnamefont {Q.-T.}\ \bibnamefont
  {Xie}}, \bibinfo {author} {\bibfnamefont {S.}~\bibnamefont {Cui}}, \bibinfo
  {author} {\bibfnamefont {J.-P.}\ \bibnamefont {Cao}}, \bibinfo {author}
  {\bibfnamefont {L.}~\bibnamefont {Amico}},\ and\ \bibinfo {author}
  {\bibfnamefont {H.}~\bibnamefont {Fan}},\ }\bibfield  {title} {\bibinfo
  {title} {Anisotropic {Rabi} model},\ }\href
  {https://doi.org/10.1103/PhysRevX.4.021046} {\bibfield  {journal} {\bibinfo
  {journal} {Phys. Rev. X}\ }\textbf {\bibinfo {volume} {4}},\ \bibinfo {pages}
  {021046} (\bibinfo {year} {2014})}\BibitemShut {NoStop}%
\bibitem [{\citenamefont {Tomka}\ \emph {et~al.}(2014)\citenamefont {Tomka},
  \citenamefont {El~Araby}, \citenamefont {Pletyukhov},\ and\ \citenamefont
  {Gritsev}}]{tomka_exceptional_2014}%
  \BibitemOpen
  \bibfield  {author} {\bibinfo {author} {\bibfnamefont {M.}~\bibnamefont
  {Tomka}}, \bibinfo {author} {\bibfnamefont {O.}~\bibnamefont {El~Araby}},
  \bibinfo {author} {\bibfnamefont {M.}~\bibnamefont {Pletyukhov}},\ and\
  \bibinfo {author} {\bibfnamefont {V.}~\bibnamefont {Gritsev}},\ }\bibfield
  {title} {\bibinfo {title} {Exceptional and regular spectra of a generalized
  {Rabi} model},\ }\href {https://doi.org/10.1103/PhysRevA.90.063839}
  {\bibfield  {journal} {\bibinfo  {journal} {Phys. Rev. A}\ }\textbf {\bibinfo
  {volume} {90}},\ \bibinfo {pages} {063839} (\bibinfo {year}
  {2014})}\BibitemShut {NoStop}%
\bibitem [{\citenamefont {Shen}\ \emph {et~al.}(2014)\citenamefont {Shen},
  \citenamefont {Yang}, \citenamefont {Lu}, \citenamefont {Chen},\ and\
  \citenamefont {Wu}}]{shen_ground_2014}%
  \BibitemOpen
  \bibfield  {author} {\bibinfo {author} {\bibfnamefont {L.-T.}\ \bibnamefont
  {Shen}}, \bibinfo {author} {\bibfnamefont {Z.-B.}\ \bibnamefont {Yang}},
  \bibinfo {author} {\bibfnamefont {M.}~\bibnamefont {Lu}}, \bibinfo {author}
  {\bibfnamefont {R.-X.}\ \bibnamefont {Chen}},\ and\ \bibinfo {author}
  {\bibfnamefont {H.-Z.}\ \bibnamefont {Wu}},\ }\bibfield  {title} {\bibinfo
  {title} {Ground state of the asymmetric {Rabi} model in the ultrastrong
  coupling regime},\ }\href {https://doi.org/10.1007/s00340-014-5821-2}
  {\bibfield  {journal} {\bibinfo  {journal} {Appl. Phys. B}\ }\textbf
  {\bibinfo {volume} {117}},\ \bibinfo {pages} {195} (\bibinfo {year}
  {2014})}\BibitemShut {NoStop}%
\bibitem [{\citenamefont {Zhang}\ and\ \citenamefont
  {Zhu}(2015)}]{zhang_analytical_2015}%
  \BibitemOpen
  \bibfield  {author} {\bibinfo {author} {\bibfnamefont {G.}~\bibnamefont
  {Zhang}}\ and\ \bibinfo {author} {\bibfnamefont {H.}~\bibnamefont {Zhu}},\
  }\bibfield  {title} {\bibinfo {title} {Analytical {Solution} for the
  {Anisotropic} {Rabi} {Model}: {Effects} of {Counter}-{Rotating} {Terms}},\
  }\href {https://doi.org/10.1038/srep08756} {\bibfield  {journal} {\bibinfo
  {journal} {Sci. Rep.}\ }\textbf {\bibinfo {volume} {5}},\ \bibinfo {pages}
  {8756} (\bibinfo {year} {2015})}\BibitemShut {NoStop}%
\bibitem [{\citenamefont {Liu}\ \emph {et~al.}(2017)\citenamefont {Liu},
  \citenamefont {Chesi}, \citenamefont {Ying}, \citenamefont {Chen},
  \citenamefont {Luo},\ and\ \citenamefont {Lin}}]{liu_universal_2017}%
  \BibitemOpen
  \bibfield  {author} {\bibinfo {author} {\bibfnamefont {M.}~\bibnamefont
  {Liu}}, \bibinfo {author} {\bibfnamefont {S.}~\bibnamefont {Chesi}}, \bibinfo
  {author} {\bibfnamefont {Z.-J.}\ \bibnamefont {Ying}}, \bibinfo {author}
  {\bibfnamefont {X.}~\bibnamefont {Chen}}, \bibinfo {author} {\bibfnamefont
  {H.-G.}\ \bibnamefont {Luo}},\ and\ \bibinfo {author} {\bibfnamefont {H.-Q.}\
  \bibnamefont {Lin}},\ }\bibfield  {title} {\bibinfo {title} {Universal
  {Scaling} and {Critical} {Exponents} of the {Anisotropic} {Quantum} {Rabi}
  {Model}},\ }\href {https://doi.org/10.1103/PhysRevLett.119.220601} {\bibfield
   {journal} {\bibinfo  {journal} {Phys. Rev. Lett.}\ }\textbf {\bibinfo
  {volume} {119}},\ \bibinfo {pages} {220601} (\bibinfo {year}
  {2017})}\BibitemShut {NoStop}%
\bibitem [{\citenamefont {Shen}\ \emph {et~al.}(2017)\citenamefont {Shen},
  \citenamefont {Yang}, \citenamefont {Wu},\ and\ \citenamefont
  {Zheng}}]{shen_quantum_2017}%
  \BibitemOpen
  \bibfield  {author} {\bibinfo {author} {\bibfnamefont {L.-T.}\ \bibnamefont
  {Shen}}, \bibinfo {author} {\bibfnamefont {Z.-B.}\ \bibnamefont {Yang}},
  \bibinfo {author} {\bibfnamefont {H.-Z.}\ \bibnamefont {Wu}},\ and\ \bibinfo
  {author} {\bibfnamefont {S.-B.}\ \bibnamefont {Zheng}},\ }\bibfield  {title}
  {\bibinfo {title} {Quantum phase transition and quench dynamics in the
  anisotropic {Rabi} model},\ }\href
  {https://doi.org/10.1103/PhysRevA.95.013819} {\bibfield  {journal} {\bibinfo
  {journal} {Phys. Rev. A}\ }\textbf {\bibinfo {volume} {95}},\ \bibinfo
  {pages} {013819} (\bibinfo {year} {2017})}\BibitemShut {NoStop}%
\bibitem [{\citenamefont {Wang}\ \emph {et~al.}(2018)\citenamefont {Wang},
  \citenamefont {You}, \citenamefont {Liu}, \citenamefont {Dong}, \citenamefont
  {Luo}, \citenamefont {Romero},\ and\ \citenamefont
  {You}}]{wang_quantum_2018}%
  \BibitemOpen
  \bibfield  {author} {\bibinfo {author} {\bibfnamefont {Y.}~\bibnamefont
  {Wang}}, \bibinfo {author} {\bibfnamefont {W.-L.}\ \bibnamefont {You}},
  \bibinfo {author} {\bibfnamefont {M.}~\bibnamefont {Liu}}, \bibinfo {author}
  {\bibfnamefont {Y.-L.}\ \bibnamefont {Dong}}, \bibinfo {author}
  {\bibfnamefont {H.-G.}\ \bibnamefont {Luo}}, \bibinfo {author} {\bibfnamefont
  {G.}~\bibnamefont {Romero}},\ and\ \bibinfo {author} {\bibfnamefont {J.~Q.}\
  \bibnamefont {You}},\ }\bibfield  {title} {\bibinfo {title} {Quantum
  criticality and state engineering in the simulated anisotropic quantum {Rabi}
  model},\ }\href {https://doi.org/10.1088/1367-2630/aac5b5} {\bibfield
  {journal} {\bibinfo  {journal} {New J. Phys.}\ }\textbf {\bibinfo {volume}
  {20}},\ \bibinfo {pages} {053061} (\bibinfo {year} {2018})}\BibitemShut
  {NoStop}%
\bibitem [{\citenamefont {Zhu}\ \emph {et~al.}(2023)\citenamefont {Zhu},
  \citenamefont {Lü}, \citenamefont {Ning}, \citenamefont {Wu}, \citenamefont
  {Shen}, \citenamefont {Yang},\ and\ \citenamefont
  {Zheng}}]{zhu_criticality_2023}%
  \BibitemOpen
  \bibfield  {author} {\bibinfo {author} {\bibfnamefont {X.}~\bibnamefont
  {Zhu}}, \bibinfo {author} {\bibfnamefont {J.-H.}\ \bibnamefont {Lü}},
  \bibinfo {author} {\bibfnamefont {W.}~\bibnamefont {Ning}}, \bibinfo {author}
  {\bibfnamefont {F.}~\bibnamefont {Wu}}, \bibinfo {author} {\bibfnamefont
  {L.-T.}\ \bibnamefont {Shen}}, \bibinfo {author} {\bibfnamefont {Z.-B.}\
  \bibnamefont {Yang}},\ and\ \bibinfo {author} {\bibfnamefont {S.-B.}\
  \bibnamefont {Zheng}},\ }\bibfield  {title} {\bibinfo {title}
  {Criticality-enhanced quantum sensing in the anisotropic quantum {Rabi}
  model},\ }\href {https://doi.org/10.1007/s11433-022-2073-9} {\bibfield
  {journal} {\bibinfo  {journal} {Sci. China Phys. Mech. Astron.}\ }\textbf
  {\bibinfo {volume} {66}},\ \bibinfo {pages} {250313} (\bibinfo {year}
  {2023})}\BibitemShut {NoStop}%
\bibitem [{\citenamefont {Ying}(2022{\natexlab{a}})}]{ying_from_2022}%
  \BibitemOpen
  \bibfield  {author} {\bibinfo {author} {\bibfnamefont {Z.-J.}\ \bibnamefont
  {Ying}},\ }\bibfield  {title} {\bibinfo {title} {From {Quantum} {Rabi}
  {Model} to {Jaynes}--{Cummings} {Model}: {Symmetry}-{Breaking} {Quantum}
  {Phase} {Transitions}, {Symmetry}-{Protected} {Topological} {Transitions} and
  {Multicriticality}},\ }\href {https://doi.org/10.1002/qute.202100088}
  {\bibfield  {journal} {\bibinfo  {journal} {Adv. Quantum Technol.}\ }\textbf
  {\bibinfo {volume} {5}},\ \bibinfo {pages} {2100088} (\bibinfo {year}
  {2022}{\natexlab{a}})}\BibitemShut {NoStop}%
\bibitem [{\citenamefont {Ying}(2022{\natexlab{b}})}]{ying_hidden_2022}%
  \BibitemOpen
  \bibfield  {author} {\bibinfo {author} {\bibfnamefont {Z.-J.}\ \bibnamefont
  {Ying}},\ }\bibfield  {title} {\bibinfo {title} {Hidden {Single}-{Qubit}
  {Topological} {Phase} {Transition} without {Gap} {Closing} in {Anisotropic}
  {Light}-{Matter} {Interactions}},\ }\href
  {https://doi.org/10.1002/qute.202100165} {\bibfield  {journal} {\bibinfo
  {journal} {Adv. Quantum Technol.}\ }\textbf {\bibinfo {volume} {5}},\
  \bibinfo {pages} {2100165} (\bibinfo {year}
  {2022}{\natexlab{b}})}\BibitemShut {NoStop}%
\bibitem [{\citenamefont {Ying}(2023)}]{ying_nodes_2023}%
  \BibitemOpen
  \bibfield  {author} {\bibinfo {author} {\bibfnamefont {Z.-J.}\ \bibnamefont
  {Ying}},\ }\bibfield  {title} {\bibinfo {title} {Nodes and {Spin} {Windings}
  for {Topological} {Transitions} in {Light}--{Matter} {Interactions}},\ }\href
  {https://doi.org/10.1002/qute.202200177} {\bibfield  {journal} {\bibinfo
  {journal} {Adv. Quantum Technol.}\ }\textbf {\bibinfo {volume} {6}},\
  \bibinfo {pages} {2200177} (\bibinfo {year} {2023})}\BibitemShut {NoStop}%
\bibitem [{\citenamefont {Deng}\ \emph {et~al.}(2022)\citenamefont {Deng},
  \citenamefont {Dong}, \citenamefont {Zhang}, \citenamefont {Wu},
  \citenamefont {Yuan}, \citenamefont {Zhu}, \citenamefont {Jin}, \citenamefont
  {Li}, \citenamefont {Wang}, \citenamefont {Cai}, \citenamefont {Song},
  \citenamefont {Wang}, \citenamefont {You},\ and\ \citenamefont
  {Wang}}]{deng_observing_2022}%
  \BibitemOpen
  \bibfield  {author} {\bibinfo {author} {\bibfnamefont {J.}~\bibnamefont
  {Deng}}, \bibinfo {author} {\bibfnamefont {H.}~\bibnamefont {Dong}}, \bibinfo
  {author} {\bibfnamefont {C.}~\bibnamefont {Zhang}}, \bibinfo {author}
  {\bibfnamefont {Y.}~\bibnamefont {Wu}}, \bibinfo {author} {\bibfnamefont
  {J.}~\bibnamefont {Yuan}}, \bibinfo {author} {\bibfnamefont {X.}~\bibnamefont
  {Zhu}}, \bibinfo {author} {\bibfnamefont {F.}~\bibnamefont {Jin}}, \bibinfo
  {author} {\bibfnamefont {H.}~\bibnamefont {Li}}, \bibinfo {author}
  {\bibfnamefont {Z.}~\bibnamefont {Wang}}, \bibinfo {author} {\bibfnamefont
  {H.}~\bibnamefont {Cai}}, \bibinfo {author} {\bibfnamefont {C.}~\bibnamefont
  {Song}}, \bibinfo {author} {\bibfnamefont {H.}~\bibnamefont {Wang}}, \bibinfo
  {author} {\bibfnamefont {J.~Q.}\ \bibnamefont {You}},\ and\ \bibinfo {author}
  {\bibfnamefont {D.-W.}\ \bibnamefont {Wang}},\ }\bibfield  {title} {\bibinfo
  {title} {Observing the quantum topology of light},\ }\href
  {https://doi.org/10.1126/science.ade6219} {\bibfield  {journal} {\bibinfo
  {journal} {Science}\ }\textbf {\bibinfo {volume} {378}},\ \bibinfo {pages}
  {966} (\bibinfo {year} {2022})}\BibitemShut {NoStop}%
\bibitem [{\citenamefont {Saugmann}\ and\ \citenamefont
  {Larson}(2023)}]{saugmann_fock-state-lattice_2023}%
  \BibitemOpen
  \bibfield  {author} {\bibinfo {author} {\bibfnamefont {P.}~\bibnamefont
  {Saugmann}}\ and\ \bibinfo {author} {\bibfnamefont {J.}~\bibnamefont
  {Larson}},\ }\bibfield  {title} {\bibinfo {title} {Fock-state-lattice
  approach to quantum optics},\ }\href
  {https://doi.org/10.1103/PhysRevA.108.033721} {\bibfield  {journal} {\bibinfo
   {journal} {Phys. Rev. A}\ }\textbf {\bibinfo {volume} {108}},\ \bibinfo
  {pages} {033721} (\bibinfo {year} {2023})}\BibitemShut {NoStop}%
\bibitem [{\citenamefont {Lee}\ \emph {et~al.}(2026)\citenamefont {Lee},
  \citenamefont {Yu}, \citenamefont {Kang}, \citenamefont {Yu}, \citenamefont
  {Choi}, \citenamefont {Chung}, \citenamefont {Park},\ and\ \citenamefont
  {Kim}}]{lee_phase-space_2026}%
  \BibitemOpen
  \bibfield  {author} {\bibinfo {author} {\bibfnamefont {K.}~\bibnamefont
  {Lee}}, \bibinfo {author} {\bibfnamefont {S.}~\bibnamefont {Yu}}, \bibinfo
  {author} {\bibfnamefont {J.}~\bibnamefont {Kang}}, \bibinfo {author}
  {\bibfnamefont {S.}~\bibnamefont {Yu}}, \bibinfo {author} {\bibfnamefont
  {W.}~\bibnamefont {Choi}}, \bibinfo {author} {\bibfnamefont {D.}~\bibnamefont
  {Chung}}, \bibinfo {author} {\bibfnamefont {S.}~\bibnamefont {Park}},\ and\
  \bibinfo {author} {\bibfnamefont {T.}~\bibnamefont {Kim}},\ }\bibfield
  {title} {\bibinfo {title} {Phase-space topology in a single-atom synthetic
  dimension},\ }\href {https://doi.org/10.1103/xgym-bqy5} {\bibfield  {journal}
  {\bibinfo  {journal} {Phys. Rev. A}\ }\textbf {\bibinfo {volume} {113}},\
  \bibinfo {pages} {L010401} (\bibinfo {year} {2026})}\BibitemShut {NoStop}%
\bibitem [{\citenamefont {Su}\ \emph {et~al.}(1979)\citenamefont {Su},
  \citenamefont {Schrieffer},\ and\ \citenamefont {Heeger}}]{su_solitons_1979}%
  \BibitemOpen
  \bibfield  {author} {\bibinfo {author} {\bibfnamefont {W.~P.}\ \bibnamefont
  {Su}}, \bibinfo {author} {\bibfnamefont {J.~R.}\ \bibnamefont {Schrieffer}},\
  and\ \bibinfo {author} {\bibfnamefont {A.~J.}\ \bibnamefont {Heeger}},\
  }\bibfield  {title} {\bibinfo {title} {Solitons in {Polyacetylene}},\ }\href
  {https://doi.org/10.1103/PhysRevLett.42.1698} {\bibfield  {journal} {\bibinfo
   {journal} {Phys. Rev. Lett.}\ }\textbf {\bibinfo {volume} {42}},\ \bibinfo
  {pages} {1698} (\bibinfo {year} {1979})}\BibitemShut {NoStop}%
\bibitem [{\citenamefont {Heeger}\ \emph {et~al.}(1988)\citenamefont {Heeger},
  \citenamefont {Kivelson}, \citenamefont {Schrieffer},\ and\ \citenamefont
  {Su}}]{heeger_solitons_1988}%
  \BibitemOpen
  \bibfield  {author} {\bibinfo {author} {\bibfnamefont {A.~J.}\ \bibnamefont
  {Heeger}}, \bibinfo {author} {\bibfnamefont {S.}~\bibnamefont {Kivelson}},
  \bibinfo {author} {\bibfnamefont {J.~R.}\ \bibnamefont {Schrieffer}},\ and\
  \bibinfo {author} {\bibfnamefont {W.-P.}\ \bibnamefont {Su}},\ }\bibfield
  {title} {\bibinfo {title} {Solitons in conducting polymers},\ }\href
  {https://doi.org/10.1103/RevModPhys.60.781} {\bibfield  {journal} {\bibinfo
  {journal} {Rev. Mod. Phys.}\ }\textbf {\bibinfo {volume} {60}},\ \bibinfo
  {pages} {781} (\bibinfo {year} {1988})}\BibitemShut {NoStop}%
\bibitem [{\citenamefont {Ballance}\ \emph {et~al.}(2016)\citenamefont
  {Ballance}, \citenamefont {Harty}, \citenamefont {Linke}, \citenamefont
  {Sepiol},\ and\ \citenamefont {Lucas}}]{ballance_high-fidelity_2016}%
  \BibitemOpen
  \bibfield  {author} {\bibinfo {author} {\bibfnamefont {C.~J.}\ \bibnamefont
  {Ballance}}, \bibinfo {author} {\bibfnamefont {T.~P.}\ \bibnamefont {Harty}},
  \bibinfo {author} {\bibfnamefont {N.~M.}\ \bibnamefont {Linke}}, \bibinfo
  {author} {\bibfnamefont {M.~A.}\ \bibnamefont {Sepiol}},\ and\ \bibinfo
  {author} {\bibfnamefont {D.~M.}\ \bibnamefont {Lucas}},\ }\bibfield  {title}
  {\bibinfo {title} {High-{Fidelity} {Quantum} {Logic} {Gates} {Using}
  {Trapped}-{Ion} {Hyperfine} {Qubits}},\ }\href
  {https://doi.org/10.1103/PhysRevLett.117.060504} {\bibfield  {journal}
  {\bibinfo  {journal} {Phys. Rev. Lett.}\ }\textbf {\bibinfo {volume} {117}},\
  \bibinfo {pages} {060504} (\bibinfo {year} {2016})}\BibitemShut {NoStop}%
\bibitem [{\citenamefont {Wang}\ \emph {et~al.}(2020)\citenamefont {Wang},
  \citenamefont {Crain}, \citenamefont {Fang}, \citenamefont {Zhang},
  \citenamefont {Huang}, \citenamefont {Liang}, \citenamefont {Leung},
  \citenamefont {Brown},\ and\ \citenamefont {Kim}}]{wang_high-fidelity_2020}%
  \BibitemOpen
  \bibfield  {author} {\bibinfo {author} {\bibfnamefont {Y.}~\bibnamefont
  {Wang}}, \bibinfo {author} {\bibfnamefont {S.}~\bibnamefont {Crain}},
  \bibinfo {author} {\bibfnamefont {C.}~\bibnamefont {Fang}}, \bibinfo {author}
  {\bibfnamefont {B.}~\bibnamefont {Zhang}}, \bibinfo {author} {\bibfnamefont
  {S.}~\bibnamefont {Huang}}, \bibinfo {author} {\bibfnamefont
  {Q.}~\bibnamefont {Liang}}, \bibinfo {author} {\bibfnamefont {P.~H.}\
  \bibnamefont {Leung}}, \bibinfo {author} {\bibfnamefont {K.~R.}\ \bibnamefont
  {Brown}},\ and\ \bibinfo {author} {\bibfnamefont {J.}~\bibnamefont {Kim}},\
  }\bibfield  {title} {\bibinfo {title} {High-{Fidelity} {Two}-{Qubit} {Gates}
  {Using} a {Microelectromechanical}-{System}-{Based} {Beam} {Steering}
  {System} for {Individual} {Qubit} {Addressing}},\ }\href
  {https://doi.org/10.1103/PhysRevLett.125.150505} {\bibfield  {journal}
  {\bibinfo  {journal} {Phys. Rev. Lett.}\ }\textbf {\bibinfo {volume} {125}},\
  \bibinfo {pages} {150505} (\bibinfo {year} {2020})}\BibitemShut {NoStop}%
\bibitem [{\citenamefont {Wang}\ \emph {et~al.}(2017)\citenamefont {Wang},
  \citenamefont {Um}, \citenamefont {Zhang}, \citenamefont {An}, \citenamefont
  {Lyu}, \citenamefont {Zhang}, \citenamefont {Duan}, \citenamefont {Yum},\
  and\ \citenamefont {Kim}}]{wang_single-qubit_2017}%
  \BibitemOpen
  \bibfield  {author} {\bibinfo {author} {\bibfnamefont {Y.}~\bibnamefont
  {Wang}}, \bibinfo {author} {\bibfnamefont {M.}~\bibnamefont {Um}}, \bibinfo
  {author} {\bibfnamefont {J.}~\bibnamefont {Zhang}}, \bibinfo {author}
  {\bibfnamefont {S.}~\bibnamefont {An}}, \bibinfo {author} {\bibfnamefont
  {M.}~\bibnamefont {Lyu}}, \bibinfo {author} {\bibfnamefont {J.-N.}\
  \bibnamefont {Zhang}}, \bibinfo {author} {\bibfnamefont {L.-M.}\ \bibnamefont
  {Duan}}, \bibinfo {author} {\bibfnamefont {D.}~\bibnamefont {Yum}},\ and\
  \bibinfo {author} {\bibfnamefont {K.}~\bibnamefont {Kim}},\ }\bibfield
  {title} {\bibinfo {title} {Single-qubit quantum memory exceeding ten-minute
  coherence time},\ }\href {https://doi.org/10.1038/s41566-017-0007-1}
  {\bibfield  {journal} {\bibinfo  {journal} {Nat. Photonics}\ }\textbf
  {\bibinfo {volume} {11}},\ \bibinfo {pages} {646} (\bibinfo {year}
  {2017})}\BibitemShut {NoStop}%
\bibitem [{\citenamefont {Wang}\ \emph {et~al.}(2021)\citenamefont {Wang},
  \citenamefont {Luan}, \citenamefont {Qiao}, \citenamefont {Um}, \citenamefont
  {Zhang}, \citenamefont {Wang}, \citenamefont {Yuan}, \citenamefont {Gu},
  \citenamefont {Zhang},\ and\ \citenamefont {Kim}}]{wang_single_2021}%
  \BibitemOpen
  \bibfield  {author} {\bibinfo {author} {\bibfnamefont {P.}~\bibnamefont
  {Wang}}, \bibinfo {author} {\bibfnamefont {C.-Y.}\ \bibnamefont {Luan}},
  \bibinfo {author} {\bibfnamefont {M.}~\bibnamefont {Qiao}}, \bibinfo {author}
  {\bibfnamefont {M.}~\bibnamefont {Um}}, \bibinfo {author} {\bibfnamefont
  {J.}~\bibnamefont {Zhang}}, \bibinfo {author} {\bibfnamefont
  {Y.}~\bibnamefont {Wang}}, \bibinfo {author} {\bibfnamefont {X.}~\bibnamefont
  {Yuan}}, \bibinfo {author} {\bibfnamefont {M.}~\bibnamefont {Gu}}, \bibinfo
  {author} {\bibfnamefont {J.}~\bibnamefont {Zhang}},\ and\ \bibinfo {author}
  {\bibfnamefont {K.}~\bibnamefont {Kim}},\ }\bibfield  {title} {\bibinfo
  {title} {Single ion qubit with estimated coherence time exceeding one hour},\
  }\href {https://doi.org/10.1038/s41467-020-20330-w} {\bibfield  {journal}
  {\bibinfo  {journal} {Nat. Commun.}\ }\textbf {\bibinfo {volume} {12}},\
  \bibinfo {pages} {233} (\bibinfo {year} {2021})}\BibitemShut {NoStop}%
\bibitem [{\citenamefont {Leibfried}\ \emph {et~al.}(2003)\citenamefont
  {Leibfried}, \citenamefont {Blatt}, \citenamefont {Monroe},\ and\
  \citenamefont {Wineland}}]{leibfried_quantum_2003}%
  \BibitemOpen
  \bibfield  {author} {\bibinfo {author} {\bibfnamefont {D.}~\bibnamefont
  {Leibfried}}, \bibinfo {author} {\bibfnamefont {R.}~\bibnamefont {Blatt}},
  \bibinfo {author} {\bibfnamefont {C.}~\bibnamefont {Monroe}},\ and\ \bibinfo
  {author} {\bibfnamefont {D.}~\bibnamefont {Wineland}},\ }\bibfield  {title}
  {\bibinfo {title} {Quantum dynamics of single trapped ions},\ }\href
  {https://doi.org/10.1103/RevModPhys.75.281} {\bibfield  {journal} {\bibinfo
  {journal} {Rev. Mod. Phys.}\ }\textbf {\bibinfo {volume} {75}},\ \bibinfo
  {pages} {281} (\bibinfo {year} {2003})}\BibitemShut {NoStop}%
\bibitem [{\citenamefont {Wineland}\ \emph {et~al.}(2003)\citenamefont
  {Wineland}, \citenamefont {Barrett}, \citenamefont {Britton}, \citenamefont
  {Chiaverini}, \citenamefont {DeMarco}, \citenamefont {Itano}, \citenamefont
  {Jelenkovi{\'c}}, \citenamefont {Langer}, \citenamefont {Leibfried},
  \citenamefont {Meyer}, \citenamefont {Rosenband},\ and\ \citenamefont
  {Schätz}}]{wineland_quantum_2003}%
  \BibitemOpen
  \bibfield  {author} {\bibinfo {author} {\bibfnamefont {D.~J.}\ \bibnamefont
  {Wineland}}, \bibinfo {author} {\bibfnamefont {M.}~\bibnamefont {Barrett}},
  \bibinfo {author} {\bibfnamefont {J.}~\bibnamefont {Britton}}, \bibinfo
  {author} {\bibfnamefont {J.}~\bibnamefont {Chiaverini}}, \bibinfo {author}
  {\bibfnamefont {B.}~\bibnamefont {DeMarco}}, \bibinfo {author} {\bibfnamefont
  {W.~M.}\ \bibnamefont {Itano}}, \bibinfo {author} {\bibfnamefont
  {B.}~\bibnamefont {Jelenkovi{\'c}}}, \bibinfo {author} {\bibfnamefont
  {C.}~\bibnamefont {Langer}}, \bibinfo {author} {\bibfnamefont
  {D.}~\bibnamefont {Leibfried}}, \bibinfo {author} {\bibfnamefont
  {V.}~\bibnamefont {Meyer}}, \bibinfo {author} {\bibfnamefont
  {T.}~\bibnamefont {Rosenband}},\ and\ \bibinfo {author} {\bibfnamefont
  {T.}~\bibnamefont {Schätz}},\ }\bibfield  {title} {\bibinfo {title} {Quantum
  information processing with trapped ions},\ }\href
  {https://doi.org/10.1098/rsta.2003.1205} {\bibfield  {journal} {\bibinfo
  {journal} {Philos. Trans. R. Soc. London, Ser. A}\ }\textbf {\bibinfo
  {volume} {361}},\ \bibinfo {pages} {1349} (\bibinfo {year}
  {2003})}\BibitemShut {NoStop}%
\bibitem [{\citenamefont {Flühmann}\ and\ \citenamefont
  {Home}(2020)}]{fluhmann_direct_2020}%
  \BibitemOpen
  \bibfield  {author} {\bibinfo {author} {\bibfnamefont {C.}~\bibnamefont
  {Flühmann}}\ and\ \bibinfo {author} {\bibfnamefont {J.~P.}\ \bibnamefont
  {Home}},\ }\bibfield  {title} {\bibinfo {title} {Direct
  {Characteristic}-{Function} {Tomography} of {Quantum} {States} of the
  {Trapped}-{Ion} {Motional} {Oscillator}},\ }\href
  {https://doi.org/10.1103/PhysRevLett.125.043602} {\bibfield  {journal}
  {\bibinfo  {journal} {Phys. Rev. Lett.}\ }\textbf {\bibinfo {volume} {125}},\
  \bibinfo {pages} {043602} (\bibinfo {year} {2020})}\BibitemShut {NoStop}%
\bibitem [{\citenamefont {Pedernales}\ \emph {et~al.}(2015)\citenamefont
  {Pedernales}, \citenamefont {Lizuain}, \citenamefont {Felicetti},
  \citenamefont {Romero}, \citenamefont {Lamata},\ and\ \citenamefont
  {Solano}}]{pedernales_quantum_2015}%
  \BibitemOpen
  \bibfield  {author} {\bibinfo {author} {\bibfnamefont {J.~S.}\ \bibnamefont
  {Pedernales}}, \bibinfo {author} {\bibfnamefont {I.}~\bibnamefont {Lizuain}},
  \bibinfo {author} {\bibfnamefont {S.}~\bibnamefont {Felicetti}}, \bibinfo
  {author} {\bibfnamefont {G.}~\bibnamefont {Romero}}, \bibinfo {author}
  {\bibfnamefont {L.}~\bibnamefont {Lamata}},\ and\ \bibinfo {author}
  {\bibfnamefont {E.}~\bibnamefont {Solano}},\ }\bibfield  {title} {\bibinfo
  {title} {Quantum {Rabi} {Model} with {Trapped} {Ions}},\ }\href
  {https://doi.org/10.1038/srep15472} {\bibfield  {journal} {\bibinfo
  {journal} {Sci. Rep.}\ }\textbf {\bibinfo {volume} {5}},\ \bibinfo {pages}
  {15472} (\bibinfo {year} {2015})}\BibitemShut {NoStop}%
\bibitem [{\citenamefont {Lv}\ \emph {et~al.}(2018)\citenamefont {Lv},
  \citenamefont {An}, \citenamefont {Liu}, \citenamefont {Zhang}, \citenamefont
  {Pedernales}, \citenamefont {Lamata}, \citenamefont {Solano},\ and\
  \citenamefont {Kim}}]{lv_quantum_2018}%
  \BibitemOpen
  \bibfield  {author} {\bibinfo {author} {\bibfnamefont {D.}~\bibnamefont
  {Lv}}, \bibinfo {author} {\bibfnamefont {S.}~\bibnamefont {An}}, \bibinfo
  {author} {\bibfnamefont {Z.}~\bibnamefont {Liu}}, \bibinfo {author}
  {\bibfnamefont {J.-N.}\ \bibnamefont {Zhang}}, \bibinfo {author}
  {\bibfnamefont {J.~S.}\ \bibnamefont {Pedernales}}, \bibinfo {author}
  {\bibfnamefont {L.}~\bibnamefont {Lamata}}, \bibinfo {author} {\bibfnamefont
  {E.}~\bibnamefont {Solano}},\ and\ \bibinfo {author} {\bibfnamefont
  {K.}~\bibnamefont {Kim}},\ }\bibfield  {title} {\bibinfo {title} {Quantum
  {Simulation} of the {Quantum} {Rabi} {Model} in a {Trapped} {Ion}},\ }\href
  {https://doi.org/10.1103/PhysRevX.8.021027} {\bibfield  {journal} {\bibinfo
  {journal} {Phys. Rev. X}\ }\textbf {\bibinfo {volume} {8}},\ \bibinfo {pages}
  {021027} (\bibinfo {year} {2018})}\BibitemShut {NoStop}%
\bibitem [{\citenamefont {Cai}\ \emph {et~al.}(2021)\citenamefont {Cai},
  \citenamefont {Liu}, \citenamefont {Zhao}, \citenamefont {Wu}, \citenamefont
  {Mei}, \citenamefont {Jiang}, \citenamefont {He}, \citenamefont {Zhang},
  \citenamefont {Zhou},\ and\ \citenamefont {Duan}}]{cai_observation_2021}%
  \BibitemOpen
  \bibfield  {author} {\bibinfo {author} {\bibfnamefont {M.-L.}\ \bibnamefont
  {Cai}}, \bibinfo {author} {\bibfnamefont {Z.-D.}\ \bibnamefont {Liu}},
  \bibinfo {author} {\bibfnamefont {W.-D.}\ \bibnamefont {Zhao}}, \bibinfo
  {author} {\bibfnamefont {Y.-K.}\ \bibnamefont {Wu}}, \bibinfo {author}
  {\bibfnamefont {Q.-X.}\ \bibnamefont {Mei}}, \bibinfo {author} {\bibfnamefont
  {Y.}~\bibnamefont {Jiang}}, \bibinfo {author} {\bibfnamefont
  {L.}~\bibnamefont {He}}, \bibinfo {author} {\bibfnamefont {X.}~\bibnamefont
  {Zhang}}, \bibinfo {author} {\bibfnamefont {Z.-C.}\ \bibnamefont {Zhou}},\
  and\ \bibinfo {author} {\bibfnamefont {L.-M.}\ \bibnamefont {Duan}},\
  }\bibfield  {title} {\bibinfo {title} {Observation of a quantum phase
  transition in the quantum {Rabi} model with a single trapped ion},\ }\href
  {https://doi.org/10.1038/s41467-021-21425-8} {\bibfield  {journal} {\bibinfo
  {journal} {Nat. Commun.}\ }\textbf {\bibinfo {volume} {12}},\ \bibinfo
  {pages} {1126} (\bibinfo {year} {2021})}\BibitemShut {NoStop}%
\bibitem [{\citenamefont {Zhao}\ \emph {et~al.}(2025)\citenamefont {Zhao},
  \citenamefont {Bin}, \citenamefont {Hou}, \citenamefont {Li}, \citenamefont
  {Li}, \citenamefont {Lin}, \citenamefont {Lü},\ and\ \citenamefont
  {Du}}]{zhao_experimental_2025}%
  \BibitemOpen
  \bibfield  {author} {\bibinfo {author} {\bibfnamefont {X.}~\bibnamefont
  {Zhao}}, \bibinfo {author} {\bibfnamefont {Q.}~\bibnamefont {Bin}}, \bibinfo
  {author} {\bibfnamefont {W.}~\bibnamefont {Hou}}, \bibinfo {author}
  {\bibfnamefont {Y.}~\bibnamefont {Li}}, \bibinfo {author} {\bibfnamefont
  {Y.}~\bibnamefont {Li}}, \bibinfo {author} {\bibfnamefont {Y.}~\bibnamefont
  {Lin}}, \bibinfo {author} {\bibfnamefont {X.-Y.}\ \bibnamefont {Lü}},\ and\
  \bibinfo {author} {\bibfnamefont {J.}~\bibnamefont {Du}},\ }\bibfield
  {title} {\bibinfo {title} {Experimental {Observation} of
  {Parity}-{Symmetry}-{Protected} {Phenomena} in the {Quantum} {Rabi} {Model}
  with a {Trapped} {Ion}},\ }\href
  {https://doi.org/10.1103/PhysRevLett.134.193604} {\bibfield  {journal}
  {\bibinfo  {journal} {Phys. Rev. Lett.}\ }\textbf {\bibinfo {volume} {134}},\
  \bibinfo {pages} {193604} (\bibinfo {year} {2025})}\BibitemShut {NoStop}%
\bibitem [{\citenamefont {Dutta}\ \emph {et~al.}(2012)\citenamefont {Dutta},
  \citenamefont {Mukherjee},\ and\ \citenamefont
  {Sengupta}}]{dutta_nonequilibrium_2012}%
  \BibitemOpen
  \bibfield  {author} {\bibinfo {author} {\bibfnamefont {T.}~\bibnamefont
  {Dutta}}, \bibinfo {author} {\bibfnamefont {M.}~\bibnamefont {Mukherjee}},\
  and\ \bibinfo {author} {\bibfnamefont {K.}~\bibnamefont {Sengupta}},\
  }\bibfield  {title} {\bibinfo {title} {Nonequilibrium phonon dynamics in
  trapped-ion systems},\ }\href {https://doi.org/10.1103/PhysRevA.85.063401}
  {\bibfield  {journal} {\bibinfo  {journal} {Phys. Rev. A}\ }\textbf {\bibinfo
  {volume} {85}},\ \bibinfo {pages} {063401} (\bibinfo {year}
  {2012})}\BibitemShut {NoStop}%
\bibitem [{\citenamefont {Mei}\ \emph {et~al.}(2022)\citenamefont {Mei},
  \citenamefont {Li}, \citenamefont {Wu}, \citenamefont {Cai}, \citenamefont
  {Wang}, \citenamefont {Yao}, \citenamefont {Zhou},\ and\ \citenamefont
  {Duan}}]{mei_experimental_2022}%
  \BibitemOpen
  \bibfield  {author} {\bibinfo {author} {\bibfnamefont {Q.-X.}\ \bibnamefont
  {Mei}}, \bibinfo {author} {\bibfnamefont {B.-W.}\ \bibnamefont {Li}},
  \bibinfo {author} {\bibfnamefont {Y.-K.}\ \bibnamefont {Wu}}, \bibinfo
  {author} {\bibfnamefont {M.-L.}\ \bibnamefont {Cai}}, \bibinfo {author}
  {\bibfnamefont {Y.}~\bibnamefont {Wang}}, \bibinfo {author} {\bibfnamefont
  {L.}~\bibnamefont {Yao}}, \bibinfo {author} {\bibfnamefont {Z.-C.}\
  \bibnamefont {Zhou}},\ and\ \bibinfo {author} {\bibfnamefont {L.-M.}\
  \bibnamefont {Duan}},\ }\bibfield  {title} {\bibinfo {title} {Experimental
  {Realization} of the {Rabi}-{Hubbard} {Model} with {Trapped} {Ions}},\ }\href
  {https://doi.org/10.1103/PhysRevLett.128.160504} {\bibfield  {journal}
  {\bibinfo  {journal} {Phys. Rev. Lett.}\ }\textbf {\bibinfo {volume} {128}},\
  \bibinfo {pages} {160504} (\bibinfo {year} {2022})}\BibitemShut {NoStop}%
\bibitem [{\citenamefont {Sun}\ \emph {et~al.}(2025)\citenamefont {Sun},
  \citenamefont {Kang}, \citenamefont {Nuomin}, \citenamefont {Schwartz},
  \citenamefont {Beratan}, \citenamefont {Brown},\ and\ \citenamefont
  {Kim}}]{sun_quantum_2025}%
  \BibitemOpen
  \bibfield  {author} {\bibinfo {author} {\bibfnamefont {K.}~\bibnamefont
  {Sun}}, \bibinfo {author} {\bibfnamefont {M.}~\bibnamefont {Kang}}, \bibinfo
  {author} {\bibfnamefont {H.}~\bibnamefont {Nuomin}}, \bibinfo {author}
  {\bibfnamefont {G.}~\bibnamefont {Schwartz}}, \bibinfo {author}
  {\bibfnamefont {D.~N.}\ \bibnamefont {Beratan}}, \bibinfo {author}
  {\bibfnamefont {K.~R.}\ \bibnamefont {Brown}},\ and\ \bibinfo {author}
  {\bibfnamefont {J.}~\bibnamefont {Kim}},\ }\bibfield  {title} {\bibinfo
  {title} {Quantum simulation of spin-boson models with structured bath},\
  }\href {https://doi.org/10.1038/s41467-025-59296-y} {\bibfield  {journal}
  {\bibinfo  {journal} {Nat. Commun.}\ }\textbf {\bibinfo {volume} {16}},\
  \bibinfo {pages} {4042} (\bibinfo {year} {2025})}\BibitemShut {NoStop}%
\bibitem [{\citenamefont {So}\ \emph {et~al.}(2024)\citenamefont {So},
  \citenamefont {Duraisamy~Suganthi}, \citenamefont {Menon}, \citenamefont
  {Zhu}, \citenamefont {Zhuravel}, \citenamefont {Pu}, \citenamefont {Wolynes},
  \citenamefont {Onuchic},\ and\ \citenamefont {Pagano}}]{so_trapped-ion_2024}%
  \BibitemOpen
  \bibfield  {author} {\bibinfo {author} {\bibfnamefont {V.}~\bibnamefont
  {So}}, \bibinfo {author} {\bibfnamefont {M.}~\bibnamefont
  {Duraisamy~Suganthi}}, \bibinfo {author} {\bibfnamefont {A.}~\bibnamefont
  {Menon}}, \bibinfo {author} {\bibfnamefont {M.}~\bibnamefont {Zhu}}, \bibinfo
  {author} {\bibfnamefont {R.}~\bibnamefont {Zhuravel}}, \bibinfo {author}
  {\bibfnamefont {H.}~\bibnamefont {Pu}}, \bibinfo {author} {\bibfnamefont
  {P.~G.}\ \bibnamefont {Wolynes}}, \bibinfo {author} {\bibfnamefont {J.~N.}\
  \bibnamefont {Onuchic}},\ and\ \bibinfo {author} {\bibfnamefont
  {G.}~\bibnamefont {Pagano}},\ }\bibfield  {title} {\bibinfo {title}
  {Trapped-ion quantum simulation of electron transfer models with tunable
  dissipation},\ }\href {https://doi.org/10.1126/sciadv.ads8011} {\bibfield
  {journal} {\bibinfo  {journal} {Sci. Adv.}\ }\textbf {\bibinfo {volume}
  {10}},\ \bibinfo {pages} {eads8011} (\bibinfo {year} {2024})}\BibitemShut
  {NoStop}%
\bibitem [{\citenamefont {Casanova}\ \emph {et~al.}(2010)\citenamefont
  {Casanova}, \citenamefont {Romero}, \citenamefont {Lizuain}, \citenamefont
  {García-Ripoll},\ and\ \citenamefont {Solano}}]{casanova_deep_2010}%
  \BibitemOpen
  \bibfield  {author} {\bibinfo {author} {\bibfnamefont {J.}~\bibnamefont
  {Casanova}}, \bibinfo {author} {\bibfnamefont {G.}~\bibnamefont {Romero}},
  \bibinfo {author} {\bibfnamefont {I.}~\bibnamefont {Lizuain}}, \bibinfo
  {author} {\bibfnamefont {J.~J.}\ \bibnamefont {García-Ripoll}},\ and\
  \bibinfo {author} {\bibfnamefont {E.}~\bibnamefont {Solano}},\ }\bibfield
  {title} {\bibinfo {title} {Deep {Strong} {Coupling} {Regime} of the
  {Jaynes}-{Cummings} {Model}},\ }\href
  {https://doi.org/10.1103/PhysRevLett.105.263603} {\bibfield  {journal}
  {\bibinfo  {journal} {Phys. Rev. Lett.}\ }\textbf {\bibinfo {volume} {105}},\
  \bibinfo {pages} {263603} (\bibinfo {year} {2010})}\BibitemShut {NoStop}%
\bibitem [{\citenamefont {Batra}\ and\ \citenamefont
  {Sheet}(2020)}]{batra_physics_2020}%
  \BibitemOpen
  \bibfield  {author} {\bibinfo {author} {\bibfnamefont {N.}~\bibnamefont
  {Batra}}\ and\ \bibinfo {author} {\bibfnamefont {G.}~\bibnamefont {Sheet}},\
  }\bibfield  {title} {\bibinfo {title} {Physics with {Coffee} and {Doughnuts}:
  {Understanding} the {Physics} {Behind} {Topological} {Insulators} {Through}
  {Su}-{Schrieffer}-{Heeger} {Model}},\ }\href
  {https://doi.org/10.1007/s12045-020-0995-x} {\bibfield  {journal} {\bibinfo
  {journal} {Resonance}\ }\textbf {\bibinfo {volume} {25}},\ \bibinfo {pages}
  {765} (\bibinfo {year} {2020})}\BibitemShut {NoStop}%
\bibitem [{\citenamefont {Thiang}(2023)}]{thiang_topological_2023}%
  \BibitemOpen
  \bibfield  {author} {\bibinfo {author} {\bibfnamefont {G.~C.}\ \bibnamefont
  {Thiang}},\ }\bibfield  {title} {\bibinfo {title} {Topological edge states of
  {1D} chains and index theory},\ }\href {https://doi.org/10.1063/5.0150870}
  {\bibfield  {journal} {\bibinfo  {journal} {J. Math. Phys.}\ }\textbf
  {\bibinfo {volume} {64}},\ \bibinfo {pages} {061901} (\bibinfo {year}
  {2023})}\BibitemShut {NoStop}%
\bibitem [{\citenamefont {Essin}\ and\ \citenamefont
  {Gurarie}(2011)}]{Essin11a}%
  \BibitemOpen
  \bibfield  {author} {\bibinfo {author} {\bibfnamefont {A.~M.}\ \bibnamefont
  {Essin}}\ and\ \bibinfo {author} {\bibfnamefont {V.}~\bibnamefont
  {Gurarie}},\ }\bibfield  {title} {\bibinfo {title} {Bulk-boundary
  correspondence of topological insulators from their respective {Green}'s
  functions},\ }\href {https://doi.org/10.1103/physrevb.84.125132} {\bibfield
  {journal} {\bibinfo  {journal} {Phys. Rev. B}\ }\textbf {\bibinfo {volume}
  {84}},\ \bibinfo {pages} {125132} (\bibinfo {year} {2011})}\BibitemShut
  {NoStop}%
\bibitem [{\citenamefont {Revelle}(2020)}]{revelle_phoenix_2020}%
  \BibitemOpen
  \bibfield  {author} {\bibinfo {author} {\bibfnamefont {M.~C.}\ \bibnamefont
  {Revelle}},\ }\href {https://doi.org/10.48550/ARXIV.2009.02398} {\bibinfo
  {title} {Phoenix and {Peregrine} {Ion} {Traps}}} (\bibinfo {year} {2020}),\
  \Eprint {https://arxiv.org/abs/2009.02398} {arXiv:2009.02398} \BibitemShut
  {NoStop}%
\bibitem [{\citenamefont {Clark}\ \emph {et~al.}(2021)\citenamefont {Clark},
  \citenamefont {Lobser}, \citenamefont {Revelle}, \citenamefont {Yale},
  \citenamefont {Bossert}, \citenamefont {Burch}, \citenamefont {Chow},
  \citenamefont {Hogle}, \citenamefont {Ivory}, \citenamefont {Pehr},
  \citenamefont {Salzbrenner}, \citenamefont {Stick}, \citenamefont {Sweatt},
  \citenamefont {Wilson}, \citenamefont {Winrow},\ and\ \citenamefont
  {Maunz}}]{clark_engineering_2021}%
  \BibitemOpen
  \bibfield  {author} {\bibinfo {author} {\bibfnamefont {S.~M.}\ \bibnamefont
  {Clark}}, \bibinfo {author} {\bibfnamefont {D.}~\bibnamefont {Lobser}},
  \bibinfo {author} {\bibfnamefont {M.~C.}\ \bibnamefont {Revelle}}, \bibinfo
  {author} {\bibfnamefont {C.~G.}\ \bibnamefont {Yale}}, \bibinfo {author}
  {\bibfnamefont {D.}~\bibnamefont {Bossert}}, \bibinfo {author} {\bibfnamefont
  {A.~D.}\ \bibnamefont {Burch}}, \bibinfo {author} {\bibfnamefont {M.~N.}\
  \bibnamefont {Chow}}, \bibinfo {author} {\bibfnamefont {C.~W.}\ \bibnamefont
  {Hogle}}, \bibinfo {author} {\bibfnamefont {M.}~\bibnamefont {Ivory}},
  \bibinfo {author} {\bibfnamefont {J.}~\bibnamefont {Pehr}}, \bibinfo {author}
  {\bibfnamefont {B.}~\bibnamefont {Salzbrenner}}, \bibinfo {author}
  {\bibfnamefont {D.}~\bibnamefont {Stick}}, \bibinfo {author} {\bibfnamefont
  {W.}~\bibnamefont {Sweatt}}, \bibinfo {author} {\bibfnamefont {J.~M.}\
  \bibnamefont {Wilson}}, \bibinfo {author} {\bibfnamefont {E.}~\bibnamefont
  {Winrow}},\ and\ \bibinfo {author} {\bibfnamefont {P.}~\bibnamefont
  {Maunz}},\ }\bibfield  {title} {\bibinfo {title} {Engineering the {Quantum}
  {Scientific} {Computing} {Open} {User} {Testbed}},\ }\href
  {https://doi.org/10.1109/TQE.2021.3096480} {\bibfield  {journal} {\bibinfo
  {journal} {IEEE Trans. Quantum Eng.}\ }\textbf {\bibinfo {volume} {2}},\
  \bibinfo {eid} {3102832} (\bibinfo {year} {2021})}\BibitemShut {NoStop}%
\bibitem [{\citenamefont {Haljan}\ \emph
  {et~al.}(2005{\natexlab{a}})\citenamefont {Haljan}, \citenamefont {Brickman},
  \citenamefont {Deslauriers}, \citenamefont {Lee},\ and\ \citenamefont
  {Monroe}}]{haljan_spin-dependent_2005}%
  \BibitemOpen
  \bibfield  {author} {\bibinfo {author} {\bibfnamefont {P.~C.}\ \bibnamefont
  {Haljan}}, \bibinfo {author} {\bibfnamefont {K.-A.}\ \bibnamefont
  {Brickman}}, \bibinfo {author} {\bibfnamefont {L.}~\bibnamefont
  {Deslauriers}}, \bibinfo {author} {\bibfnamefont {P.~J.}\ \bibnamefont
  {Lee}},\ and\ \bibinfo {author} {\bibfnamefont {C.}~\bibnamefont {Monroe}},\
  }\bibfield  {title} {\bibinfo {title} {Spin-{Dependent} {Forces} on {Trapped}
  {Ions} for {Phase}-{Stable} {Quantum} {Gates} and {Entangled} {States} of
  {Spin} and {Motion}},\ }\href {https://doi.org/10.1103/PhysRevLett.94.153602}
  {\bibfield  {journal} {\bibinfo  {journal} {Phys. Rev. Lett.}\ }\textbf
  {\bibinfo {volume} {94}},\ \bibinfo {pages} {153602} (\bibinfo {year}
  {2005}{\natexlab{a}})}\BibitemShut {NoStop}%
\bibitem [{\citenamefont {Haljan}\ \emph
  {et~al.}(2005{\natexlab{b}})\citenamefont {Haljan}, \citenamefont {Lee},
  \citenamefont {Brickman}, \citenamefont {Acton}, \citenamefont
  {Deslauriers},\ and\ \citenamefont {Monroe}}]{haljan_entanglement_2005}%
  \BibitemOpen
  \bibfield  {author} {\bibinfo {author} {\bibfnamefont {P.~C.}\ \bibnamefont
  {Haljan}}, \bibinfo {author} {\bibfnamefont {P.~J.}\ \bibnamefont {Lee}},
  \bibinfo {author} {\bibfnamefont {K.-A.}\ \bibnamefont {Brickman}}, \bibinfo
  {author} {\bibfnamefont {M.}~\bibnamefont {Acton}}, \bibinfo {author}
  {\bibfnamefont {L.}~\bibnamefont {Deslauriers}},\ and\ \bibinfo {author}
  {\bibfnamefont {C.}~\bibnamefont {Monroe}},\ }\bibfield  {title} {\bibinfo
  {title} {Entanglement of trapped-ion clock states},\ }\href
  {https://doi.org/10.1103/PhysRevA.72.062316} {\bibfield  {journal} {\bibinfo
  {journal} {Phys. Rev. A}\ }\textbf {\bibinfo {volume} {72}},\ \bibinfo
  {pages} {062316} (\bibinfo {year} {2005}{\natexlab{b}})}\BibitemShut
  {NoStop}%
\bibitem [{\citenamefont {Lobser}\ \emph {et~al.}(2023)\citenamefont {Lobser},
  \citenamefont {Goldberg}, \citenamefont {Landahl}, \citenamefont {Maunz},
  \citenamefont {Morrison}, \citenamefont {Rudinger}, \citenamefont {Russo},
  \citenamefont {Ruzic}, \citenamefont {Stick}, \citenamefont {Van Der~Wall},\
  and\ \citenamefont {Clark}}]{lobser_jaqalpaw_2023}%
  \BibitemOpen
  \bibfield  {author} {\bibinfo {author} {\bibfnamefont {D.}~\bibnamefont
  {Lobser}}, \bibinfo {author} {\bibfnamefont {J.}~\bibnamefont {Goldberg}},
  \bibinfo {author} {\bibfnamefont {A.~J.}\ \bibnamefont {Landahl}}, \bibinfo
  {author} {\bibfnamefont {P.}~\bibnamefont {Maunz}}, \bibinfo {author}
  {\bibfnamefont {B.~C.~A.}\ \bibnamefont {Morrison}}, \bibinfo {author}
  {\bibfnamefont {K.}~\bibnamefont {Rudinger}}, \bibinfo {author}
  {\bibfnamefont {A.}~\bibnamefont {Russo}}, \bibinfo {author} {\bibfnamefont
  {B.}~\bibnamefont {Ruzic}}, \bibinfo {author} {\bibfnamefont
  {D.}~\bibnamefont {Stick}}, \bibinfo {author} {\bibfnamefont
  {J.}~\bibnamefont {Van Der~Wall}},\ and\ \bibinfo {author} {\bibfnamefont
  {S.~M.}\ \bibnamefont {Clark}},\ }\href
  {https://doi.org/10.48550/ARXIV.2305.02311} {\bibinfo {title} {{JaqalPaw}:
  {A} {Guide} to {Defining} {Pulses} and {Waveforms} for {Jaqal}}} (\bibinfo
  {year} {2023}),\ \Eprint {https://arxiv.org/abs/2305.02311}
  {arXiv:2305.02311} \BibitemShut {NoStop}%
\bibitem [{\citenamefont {Born}\ and\ \citenamefont
  {Fock}(1928)}]{born_beweis_1928}%
  \BibitemOpen
  \bibfield  {author} {\bibinfo {author} {\bibfnamefont {M.}~\bibnamefont
  {Born}}\ and\ \bibinfo {author} {\bibfnamefont {V.}~\bibnamefont {Fock}},\
  }\bibfield  {title} {\bibinfo {title} {Beweis des {Adiabatensatzes}},\ }\href
  {https://doi.org/10.1007/BF01343193} {\bibfield  {journal} {\bibinfo
  {journal} {Z. Phys.}\ }\textbf {\bibinfo {volume} {51}},\ \bibinfo {pages}
  {165} (\bibinfo {year} {1928})}\BibitemShut {NoStop}%
\bibitem [{\citenamefont {Islam}\ \emph {et~al.}(2015)\citenamefont {Islam},
  \citenamefont {Ma}, \citenamefont {Preiss}, \citenamefont {Eric~Tai},
  \citenamefont {Lukin}, \citenamefont {Rispoli},\ and\ \citenamefont
  {Greiner}}]{islam_measuring_2015}%
  \BibitemOpen
  \bibfield  {author} {\bibinfo {author} {\bibfnamefont {R.}~\bibnamefont
  {Islam}}, \bibinfo {author} {\bibfnamefont {R.}~\bibnamefont {Ma}}, \bibinfo
  {author} {\bibfnamefont {P.~M.}\ \bibnamefont {Preiss}}, \bibinfo {author}
  {\bibfnamefont {M.}~\bibnamefont {Eric~Tai}}, \bibinfo {author}
  {\bibfnamefont {A.}~\bibnamefont {Lukin}}, \bibinfo {author} {\bibfnamefont
  {M.}~\bibnamefont {Rispoli}},\ and\ \bibinfo {author} {\bibfnamefont
  {M.}~\bibnamefont {Greiner}},\ }\bibfield  {title} {\bibinfo {title}
  {Measuring entanglement entropy in a quantum many-body system},\ }\href
  {https://doi.org/10.1038/nature15750} {\bibfield  {journal} {\bibinfo
  {journal} {Nature}\ }\textbf {\bibinfo {volume} {528}},\ \bibinfo {pages}
  {77} (\bibinfo {year} {2015})}\BibitemShut {NoStop}%
\bibitem [{\citenamefont {Wang}\ \emph {et~al.}(2024)\citenamefont {Wang},
  \citenamefont {Wu}, \citenamefont {Jiang}, \citenamefont {Cai}, \citenamefont
  {Li}, \citenamefont {Mei}, \citenamefont {Qi}, \citenamefont {Zhou},\ and\
  \citenamefont {Duan}}]{wang_realizing_2024}%
  \BibitemOpen
  \bibfield  {author} {\bibinfo {author} {\bibfnamefont {Y.}~\bibnamefont
  {Wang}}, \bibinfo {author} {\bibfnamefont {Y.-K.}\ \bibnamefont {Wu}},
  \bibinfo {author} {\bibfnamefont {Y.}~\bibnamefont {Jiang}}, \bibinfo
  {author} {\bibfnamefont {M.-L.}\ \bibnamefont {Cai}}, \bibinfo {author}
  {\bibfnamefont {B.-W.}\ \bibnamefont {Li}}, \bibinfo {author} {\bibfnamefont
  {Q.-X.}\ \bibnamefont {Mei}}, \bibinfo {author} {\bibfnamefont {B.-X.}\
  \bibnamefont {Qi}}, \bibinfo {author} {\bibfnamefont {Z.-C.}\ \bibnamefont
  {Zhou}},\ and\ \bibinfo {author} {\bibfnamefont {L.-M.}\ \bibnamefont
  {Duan}},\ }\bibfield  {title} {\bibinfo {title} {Realizing {Synthetic}
  {Dimensions} and {Artificial} {Magnetic} {Flux} in a {Trapped}-{Ion}
  {Quantum} {Simulator}},\ }\href
  {https://doi.org/10.1103/PhysRevLett.132.130601} {\bibfield  {journal}
  {\bibinfo  {journal} {Phys. Rev. Lett.}\ }\textbf {\bibinfo {volume} {132}},\
  \bibinfo {pages} {130601} (\bibinfo {year} {2024})}\BibitemShut {NoStop}%
\bibitem [{\citenamefont {Zhang}\ \emph {et~al.}(2025)\citenamefont {Zhang},
  \citenamefont {Huang}, \citenamefont {Chu}, \citenamefont {Qiu},
  \citenamefont {Sun}, \citenamefont {Tao}, \citenamefont {Zhang},
  \citenamefont {Zhang}, \citenamefont {Zhou}, \citenamefont {Chen},
  \citenamefont {Liu}, \citenamefont {Liu}, \citenamefont {Zhong},
  \citenamefont {Miao}, \citenamefont {Niu},\ and\ \citenamefont
  {Yu}}]{zhang_synthetic_2025}%
  \BibitemOpen
  \bibfield  {author} {\bibinfo {author} {\bibfnamefont {J.}~\bibnamefont
  {Zhang}}, \bibinfo {author} {\bibfnamefont {W.}~\bibnamefont {Huang}},
  \bibinfo {author} {\bibfnamefont {J.}~\bibnamefont {Chu}}, \bibinfo {author}
  {\bibfnamefont {J.}~\bibnamefont {Qiu}}, \bibinfo {author} {\bibfnamefont
  {X.}~\bibnamefont {Sun}}, \bibinfo {author} {\bibfnamefont {Z.}~\bibnamefont
  {Tao}}, \bibinfo {author} {\bibfnamefont {J.}~\bibnamefont {Zhang}}, \bibinfo
  {author} {\bibfnamefont {L.}~\bibnamefont {Zhang}}, \bibinfo {author}
  {\bibfnamefont {Y.}~\bibnamefont {Zhou}}, \bibinfo {author} {\bibfnamefont
  {Y.}~\bibnamefont {Chen}}, \bibinfo {author} {\bibfnamefont {Y.}~\bibnamefont
  {Liu}}, \bibinfo {author} {\bibfnamefont {S.}~\bibnamefont {Liu}}, \bibinfo
  {author} {\bibfnamefont {Y.}~\bibnamefont {Zhong}}, \bibinfo {author}
  {\bibfnamefont {J.-J.}\ \bibnamefont {Miao}}, \bibinfo {author}
  {\bibfnamefont {J.}~\bibnamefont {Niu}},\ and\ \bibinfo {author}
  {\bibfnamefont {D.}~\bibnamefont {Yu}},\ }\bibfield  {title} {\bibinfo
  {title} {Synthetic {Multidimensional} {Aharonov}-{Bohm} {Cages} in {Fock}
  {State} {Lattices}},\ }\href {https://doi.org/10.1103/PhysRevLett.134.070601}
  {\bibfield  {journal} {\bibinfo  {journal} {Phys. Rev. Lett.}\ }\textbf
  {\bibinfo {volume} {134}},\ \bibinfo {pages} {070601} (\bibinfo {year}
  {2025})}\BibitemShut {NoStop}%
\bibitem [{\citenamefont {Kienzler}\ \emph {et~al.}(2015)\citenamefont
  {Kienzler}, \citenamefont {Lo}, \citenamefont {Keitch}, \citenamefont
  {De~Clercq}, \citenamefont {Leupold}, \citenamefont {Lindenfelser},
  \citenamefont {Marinelli}, \citenamefont {Negnevitsky},\ and\ \citenamefont
  {Home}}]{kienzler_quantum_2015}%
  \BibitemOpen
  \bibfield  {author} {\bibinfo {author} {\bibfnamefont {D.}~\bibnamefont
  {Kienzler}}, \bibinfo {author} {\bibfnamefont {H.-Y.}\ \bibnamefont {Lo}},
  \bibinfo {author} {\bibfnamefont {B.}~\bibnamefont {Keitch}}, \bibinfo
  {author} {\bibfnamefont {L.}~\bibnamefont {De~Clercq}}, \bibinfo {author}
  {\bibfnamefont {F.}~\bibnamefont {Leupold}}, \bibinfo {author} {\bibfnamefont
  {F.}~\bibnamefont {Lindenfelser}}, \bibinfo {author} {\bibfnamefont
  {M.}~\bibnamefont {Marinelli}}, \bibinfo {author} {\bibfnamefont
  {V.}~\bibnamefont {Negnevitsky}},\ and\ \bibinfo {author} {\bibfnamefont
  {J.~P.}\ \bibnamefont {Home}},\ }\bibfield  {title} {\bibinfo {title}
  {Quantum harmonic oscillator state synthesis by reservoir engineering},\
  }\href {https://doi.org/10.1126/science.1261033} {\bibfield  {journal}
  {\bibinfo  {journal} {Science}\ }\textbf {\bibinfo {volume} {347}},\ \bibinfo
  {pages} {53} (\bibinfo {year} {2015})}\BibitemShut {NoStop}%
\bibitem [{\citenamefont {Balian}\ and\ \citenamefont
  {Brezin}(1969)}]{balian_nonunitary_1969}%
  \BibitemOpen
  \bibfield  {author} {\bibinfo {author} {\bibfnamefont {R.}~\bibnamefont
  {Balian}}\ and\ \bibinfo {author} {\bibfnamefont {E.}~\bibnamefont
  {Brezin}},\ }\bibfield  {title} {\bibinfo {title} {Nonunitary bogoliubov
  transformations and extension of {Wick}’s theorem},\ }\href
  {https://doi.org/10.1007/BF02710281} {\bibfield  {journal} {\bibinfo
  {journal} {Nuovo Cimento B}\ }\textbf {\bibinfo {volume} {64}},\ \bibinfo
  {pages} {37} (\bibinfo {year} {1969})}\BibitemShut {NoStop}%
\bibitem [{\citenamefont {Hayes}\ \emph {et~al.}(2010)\citenamefont {Hayes},
  \citenamefont {Matsukevich}, \citenamefont {Maunz}, \citenamefont {Hucul},
  \citenamefont {Quraishi}, \citenamefont {Olmschenk}, \citenamefont
  {Campbell}, \citenamefont {Mizrahi}, \citenamefont {Senko},\ and\
  \citenamefont {Monroe}}]{hayes_entanglement_2010}%
  \BibitemOpen
  \bibfield  {author} {\bibinfo {author} {\bibfnamefont {D.}~\bibnamefont
  {Hayes}}, \bibinfo {author} {\bibfnamefont {D.~N.}\ \bibnamefont
  {Matsukevich}}, \bibinfo {author} {\bibfnamefont {P.}~\bibnamefont {Maunz}},
  \bibinfo {author} {\bibfnamefont {D.}~\bibnamefont {Hucul}}, \bibinfo
  {author} {\bibfnamefont {Q.}~\bibnamefont {Quraishi}}, \bibinfo {author}
  {\bibfnamefont {S.}~\bibnamefont {Olmschenk}}, \bibinfo {author}
  {\bibfnamefont {W.}~\bibnamefont {Campbell}}, \bibinfo {author}
  {\bibfnamefont {J.}~\bibnamefont {Mizrahi}}, \bibinfo {author} {\bibfnamefont
  {C.}~\bibnamefont {Senko}},\ and\ \bibinfo {author} {\bibfnamefont
  {C.}~\bibnamefont {Monroe}},\ }\bibfield  {title} {\bibinfo {title}
  {Entanglement of {Atomic} {Qubits} {Using} an {Optical} {Frequency} {Comb}},\
  }\href {https://doi.org/10.1103/PhysRevLett.104.140501} {\bibfield  {journal}
  {\bibinfo  {journal} {Phys. Rev. Lett.}\ }\textbf {\bibinfo {volume} {104}},\
  \bibinfo {pages} {140501} (\bibinfo {year} {2010})}\BibitemShut {NoStop}%
\bibitem [{\citenamefont {Olmschenk}\ \emph {et~al.}(2007)\citenamefont
  {Olmschenk}, \citenamefont {Younge}, \citenamefont {Moehring}, \citenamefont
  {Matsukevich}, \citenamefont {Maunz},\ and\ \citenamefont
  {Monroe}}]{olmschenk_manipulation_2007}%
  \BibitemOpen
  \bibfield  {author} {\bibinfo {author} {\bibfnamefont {S.}~\bibnamefont
  {Olmschenk}}, \bibinfo {author} {\bibfnamefont {K.~C.}\ \bibnamefont
  {Younge}}, \bibinfo {author} {\bibfnamefont {D.~L.}\ \bibnamefont
  {Moehring}}, \bibinfo {author} {\bibfnamefont {D.~N.}\ \bibnamefont
  {Matsukevich}}, \bibinfo {author} {\bibfnamefont {P.}~\bibnamefont {Maunz}},\
  and\ \bibinfo {author} {\bibfnamefont {C.}~\bibnamefont {Monroe}},\
  }\bibfield  {title} {\bibinfo {title} {Manipulation and detection of a
  trapped {$\mathrm{Yb}^{+}$} hyperfine qubit},\ }\href
  {https://doi.org/10.1103/PhysRevA.76.052314} {\bibfield  {journal} {\bibinfo
  {journal} {Phys. Rev. A}\ }\textbf {\bibinfo {volume} {76}},\ \bibinfo
  {pages} {052314} (\bibinfo {year} {2007})}\BibitemShut {NoStop}%
\bibitem [{\citenamefont {Lim}\ \emph {et~al.}(2025)\citenamefont {Lim},
  \citenamefont {Baek}, \citenamefont {Whitlow}, \citenamefont {D'Onofrio},
  \citenamefont {Chen}, \citenamefont {Phiri}, \citenamefont {Crain},
  \citenamefont {Brown}, \citenamefont {Kim},\ and\ \citenamefont
  {Kim}}]{lim_design_2025}%
  \BibitemOpen
  \bibfield  {author} {\bibinfo {author} {\bibfnamefont {S.}~\bibnamefont
  {Lim}}, \bibinfo {author} {\bibfnamefont {S.}~\bibnamefont {Baek}}, \bibinfo
  {author} {\bibfnamefont {J.}~\bibnamefont {Whitlow}}, \bibinfo {author}
  {\bibfnamefont {M.}~\bibnamefont {D'Onofrio}}, \bibinfo {author}
  {\bibfnamefont {T.}~\bibnamefont {Chen}}, \bibinfo {author} {\bibfnamefont
  {S.}~\bibnamefont {Phiri}}, \bibinfo {author} {\bibfnamefont
  {S.}~\bibnamefont {Crain}}, \bibinfo {author} {\bibfnamefont {K.~R.}\
  \bibnamefont {Brown}}, \bibinfo {author} {\bibfnamefont {J.}~\bibnamefont
  {Kim}},\ and\ \bibinfo {author} {\bibfnamefont {J.}~\bibnamefont {Kim}},\
  }\bibfield  {title} {\bibinfo {title} {Design and characterization of
  individual addressing optics based on multi-channel acousto-optic modulator
  for {$^{171}\mathrm{Yb}^{+}$} qubits},\ }\href
  {https://doi.org/10.1016/j.optlastec.2024.111436} {\bibfield  {journal}
  {\bibinfo  {journal} {Opt. Laser Technol.}\ }\textbf {\bibinfo {volume}
  {180}},\ \bibinfo {pages} {111436} (\bibinfo {year} {2025})}\BibitemShut
  {NoStop}%
\bibitem [{\citenamefont {Debnath}\ \emph {et~al.}(2016)\citenamefont
  {Debnath}, \citenamefont {Linke}, \citenamefont {Figgatt}, \citenamefont
  {Landsman}, \citenamefont {Wright},\ and\ \citenamefont
  {Monroe}}]{debnath_demonstration_2016}%
  \BibitemOpen
  \bibfield  {author} {\bibinfo {author} {\bibfnamefont {S.}~\bibnamefont
  {Debnath}}, \bibinfo {author} {\bibfnamefont {N.~M.}\ \bibnamefont {Linke}},
  \bibinfo {author} {\bibfnamefont {C.}~\bibnamefont {Figgatt}}, \bibinfo
  {author} {\bibfnamefont {K.~A.}\ \bibnamefont {Landsman}}, \bibinfo {author}
  {\bibfnamefont {K.}~\bibnamefont {Wright}},\ and\ \bibinfo {author}
  {\bibfnamefont {C.}~\bibnamefont {Monroe}},\ }\bibfield  {title} {\bibinfo
  {title} {Demonstration of a small programmable quantum computer with atomic
  qubits},\ }\href {https://doi.org/10.1038/nature18648} {\bibfield  {journal}
  {\bibinfo  {journal} {Nature}\ }\textbf {\bibinfo {volume} {536}},\ \bibinfo
  {pages} {63} (\bibinfo {year} {2016})}\BibitemShut {NoStop}%
\bibitem [{\citenamefont {Wright}\ \emph {et~al.}(2019)\citenamefont {Wright},
  \citenamefont {Beck}, \citenamefont {Debnath}, \citenamefont {Amini},
  \citenamefont {Nam}, \citenamefont {Grzesiak}, \citenamefont {Chen},
  \citenamefont {Pisenti}, \citenamefont {Chmielewski}, \citenamefont
  {Collins}, \citenamefont {Hudek}, \citenamefont {Mizrahi}, \citenamefont
  {Wong-Campos}, \citenamefont {Allen}, \citenamefont {Apisdorf}, \citenamefont
  {Solomon}, \citenamefont {Williams}, \citenamefont {Ducore}, \citenamefont
  {Blinov}, \citenamefont {Kreikemeier}, \citenamefont {Chaplin}, \citenamefont
  {Keesan}, \citenamefont {Monroe},\ and\ \citenamefont
  {Kim}}]{wright_benchmarking_2019}%
  \BibitemOpen
  \bibfield  {author} {\bibinfo {author} {\bibfnamefont {K.}~\bibnamefont
  {Wright}}, \bibinfo {author} {\bibfnamefont {K.~M.}\ \bibnamefont {Beck}},
  \bibinfo {author} {\bibfnamefont {S.}~\bibnamefont {Debnath}}, \bibinfo
  {author} {\bibfnamefont {J.~M.}\ \bibnamefont {Amini}}, \bibinfo {author}
  {\bibfnamefont {Y.}~\bibnamefont {Nam}}, \bibinfo {author} {\bibfnamefont
  {N.}~\bibnamefont {Grzesiak}}, \bibinfo {author} {\bibfnamefont {J.-S.}\
  \bibnamefont {Chen}}, \bibinfo {author} {\bibfnamefont {N.~C.}\ \bibnamefont
  {Pisenti}}, \bibinfo {author} {\bibfnamefont {M.}~\bibnamefont
  {Chmielewski}}, \bibinfo {author} {\bibfnamefont {C.}~\bibnamefont
  {Collins}}, \bibinfo {author} {\bibfnamefont {K.~M.}\ \bibnamefont {Hudek}},
  \bibinfo {author} {\bibfnamefont {J.}~\bibnamefont {Mizrahi}}, \bibinfo
  {author} {\bibfnamefont {J.~D.}\ \bibnamefont {Wong-Campos}}, \bibinfo
  {author} {\bibfnamefont {S.}~\bibnamefont {Allen}}, \bibinfo {author}
  {\bibfnamefont {J.}~\bibnamefont {Apisdorf}}, \bibinfo {author}
  {\bibfnamefont {P.}~\bibnamefont {Solomon}}, \bibinfo {author} {\bibfnamefont
  {M.}~\bibnamefont {Williams}}, \bibinfo {author} {\bibfnamefont {A.~M.}\
  \bibnamefont {Ducore}}, \bibinfo {author} {\bibfnamefont {A.}~\bibnamefont
  {Blinov}}, \bibinfo {author} {\bibfnamefont {S.~M.}\ \bibnamefont
  {Kreikemeier}}, \bibinfo {author} {\bibfnamefont {V.}~\bibnamefont
  {Chaplin}}, \bibinfo {author} {\bibfnamefont {M.}~\bibnamefont {Keesan}},
  \bibinfo {author} {\bibfnamefont {C.}~\bibnamefont {Monroe}},\ and\ \bibinfo
  {author} {\bibfnamefont {J.}~\bibnamefont {Kim}},\ }\bibfield  {title}
  {\bibinfo {title} {Benchmarking an 11-qubit quantum computer},\ }\href
  {https://doi.org/10.1038/s41467-019-13534-2} {\bibfield  {journal} {\bibinfo
  {journal} {Nat. Commun.}\ }\textbf {\bibinfo {volume} {10}},\ \bibinfo
  {pages} {5464} (\bibinfo {year} {2019})}\BibitemShut {NoStop}%
\bibitem [{\citenamefont {Lee}\ \emph {et~al.}(2016)\citenamefont {Lee},
  \citenamefont {Smith}, \citenamefont {Richerme}, \citenamefont {Neyenhuis},
  \citenamefont {Hess}, \citenamefont {Zhang},\ and\ \citenamefont
  {Monroe}}]{lee_engineering_2016}%
  \BibitemOpen
  \bibfield  {author} {\bibinfo {author} {\bibfnamefont {A.~C.}\ \bibnamefont
  {Lee}}, \bibinfo {author} {\bibfnamefont {J.}~\bibnamefont {Smith}}, \bibinfo
  {author} {\bibfnamefont {P.}~\bibnamefont {Richerme}}, \bibinfo {author}
  {\bibfnamefont {B.}~\bibnamefont {Neyenhuis}}, \bibinfo {author}
  {\bibfnamefont {P.~W.}\ \bibnamefont {Hess}}, \bibinfo {author}
  {\bibfnamefont {J.}~\bibnamefont {Zhang}},\ and\ \bibinfo {author}
  {\bibfnamefont {C.}~\bibnamefont {Monroe}},\ }\bibfield  {title} {\bibinfo
  {title} {Engineering {Large} {Stark} {Shifts} for {Control} of {Individual}
  {Clock} {State} {Qubits}},\ }\href
  {https://doi.org/10.1103/PhysRevA.94.042308} {\bibfield  {journal} {\bibinfo
  {journal} {Phys. Rev. A}\ }\textbf {\bibinfo {volume} {94}},\ \bibinfo
  {pages} {042308} (\bibinfo {year} {2016})}\BibitemShut {NoStop}%
\bibitem [{\citenamefont {Yale}\ \emph {et~al.}(2025)\citenamefont {Yale},
  \citenamefont {Burch}, \citenamefont {Chow}, \citenamefont {Ruzic},
  \citenamefont {Lobser}, \citenamefont {McFarland}, \citenamefont {Revelle},\
  and\ \citenamefont {Clark}}]{yale_realization_2025}%
  \BibitemOpen
  \bibfield  {author} {\bibinfo {author} {\bibfnamefont {C.~G.}\ \bibnamefont
  {Yale}}, \bibinfo {author} {\bibfnamefont {A.~D.}\ \bibnamefont {Burch}},
  \bibinfo {author} {\bibfnamefont {M.~N.~H.}\ \bibnamefont {Chow}}, \bibinfo
  {author} {\bibfnamefont {B.~P.}\ \bibnamefont {Ruzic}}, \bibinfo {author}
  {\bibfnamefont {D.~S.}\ \bibnamefont {Lobser}}, \bibinfo {author}
  {\bibfnamefont {B.~K.}\ \bibnamefont {McFarland}}, \bibinfo {author}
  {\bibfnamefont {M.~C.}\ \bibnamefont {Revelle}},\ and\ \bibinfo {author}
  {\bibfnamefont {S.~M.}\ \bibnamefont {Clark}},\ }\bibfield  {title} {\bibinfo
  {title} {Realization and {Calibration} of {Continuously} {Parameterized}
  {Two}-{Qubit} {Gates} on a {Trapped}-{Ion} {Quantum} {Processor}},\ }\href
  {https://doi.org/10.1109/TQE.2025.3600216} {\bibfield  {journal} {\bibinfo
  {journal} {IEEE Trans. Quantum Eng.}\ }\textbf {\bibinfo {volume} {6}},\
  \bibinfo {eid} {3102117} (\bibinfo {year} {2025})}\BibitemShut {NoStop}%
\bibitem [{\citenamefont {Johansson}\ \emph {et~al.}(2013)\citenamefont
  {Johansson}, \citenamefont {Nation},\ and\ \citenamefont
  {Nori}}]{johansson_qutip_2013}%
  \BibitemOpen
  \bibfield  {author} {\bibinfo {author} {\bibfnamefont {J.~R.}\ \bibnamefont
  {Johansson}}, \bibinfo {author} {\bibfnamefont {P.~D.}\ \bibnamefont
  {Nation}},\ and\ \bibinfo {author} {\bibfnamefont {F.}~\bibnamefont {Nori}},\
  }\bibfield  {title} {\bibinfo {title} {{QuTiP} 2: {A} {Python} framework for
  the dynamics of open quantum systems},\ }\href
  {https://doi.org/10.1016/j.cpc.2012.11.019} {\bibfield  {journal} {\bibinfo
  {journal} {Comput. Phys. Commun.}\ }\textbf {\bibinfo {volume} {184}},\
  \bibinfo {pages} {1234} (\bibinfo {year} {2013})}\BibitemShut {NoStop}%
\bibitem [{\citenamefont {Lo}\ \emph {et~al.}(2015)\citenamefont {Lo},
  \citenamefont {Kienzler}, \citenamefont {De~Clercq}, \citenamefont
  {Marinelli}, \citenamefont {Negnevitsky}, \citenamefont {Keitch},\ and\
  \citenamefont {Home}}]{lo_spinmotion_2015}%
  \BibitemOpen
  \bibfield  {author} {\bibinfo {author} {\bibfnamefont {H.-Y.}\ \bibnamefont
  {Lo}}, \bibinfo {author} {\bibfnamefont {D.}~\bibnamefont {Kienzler}},
  \bibinfo {author} {\bibfnamefont {L.}~\bibnamefont {De~Clercq}}, \bibinfo
  {author} {\bibfnamefont {M.}~\bibnamefont {Marinelli}}, \bibinfo {author}
  {\bibfnamefont {V.}~\bibnamefont {Negnevitsky}}, \bibinfo {author}
  {\bibfnamefont {B.~C.}\ \bibnamefont {Keitch}},\ and\ \bibinfo {author}
  {\bibfnamefont {J.~P.}\ \bibnamefont {Home}},\ }\bibfield  {title} {\bibinfo
  {title} {Spin--motion entanglement and state diagnosis with squeezed
  oscillator wavepackets},\ }\href {https://doi.org/10.1038/nature14458}
  {\bibfield  {journal} {\bibinfo  {journal} {Nature}\ }\textbf {\bibinfo
  {volume} {521}},\ \bibinfo {pages} {336} (\bibinfo {year}
  {2015})}\BibitemShut {NoStop}%
\end{thebibliography}
\end{document}